\documentclass[pr,twocolumn,superscriptaddress]{revtex4}
\usepackage{amsmath}
\usepackage{amssymb}
\usepackage{amsthm}
\usepackage{amsfonts}
\usepackage{listings}
\usepackage{longtable}
\usepackage{enumerate}
\usepackage{latexsym}
\usepackage{color}
\usepackage{bm}
\usepackage{hyperref}

\usepackage{array,etoolbox}
\preto\tabular{\setcounter{magicrownumbers}{0}}
\newcounter{magicrownumbers}
\newcommand\rownumber{\stepcounter{magicrownumbers}\arabic{magicrownumbers}     }

\hypersetup{
 pdfnewwindow=true, colorlinks=true,
 linkcolor=blue, anchorcolor=blue,
 citecolor=blue, filecolor=blue,
 menucolor=blue, urlcolor=blue}

\newcommand{\beginsupplement}{%
        \setcounter{table}{0}
        \renewcommand{\thetable}{MSG\arabic{table}}%
        \setcounter{figure}{0}
        \renewcommand{\thefigure}{S\arabic{figure}}%
     }

\usepackage{psfrag}

\usepackage{bm}
\usepackage{graphicx}
\usepackage{subfigure}
\usepackage{url}

\newcommand{\webirvsp}{\href{https://github.com/zjwang11/irvsp/blob/master/src_irvsp_v2.tar.gz}{\ttfamily IRVSP}}

\newcommand{\webPh}{\href{https://phonopy.github.io/phonopy/}{\ttfamily Phonopy}}

\newcommand{\bB}{{\bf B}}
\newcommand{\bk}{{\bf k}}
\newcommand{\bL}{\mathbf L}
\newcommand{\bv}{{\bf v}}
\newcommand{\bG}{{\bf G}}
\newcommand{\bM}{{\bf M}}
\newcommand{\bq}{{\bf q}}
\newcommand{\vecr}{\vec r}

\newcommand{\ket}[1]{\left|#1\right\rangle}

\def\ie{{\it i.e.},\ }

\input{epsf}

\begin{document}

\tolerance 10000

\newcommand{\vk}{{\bf k}}

\draft

\title{Magnetic band representations, Fu-Kane-like symmetry indicators \\
and magnetic topological materials}

\author{Jiacheng Gao}
\affiliation{Beijing National Laboratory for Condensed Matter Physics,
and Institute of Physics, Chinese Academy of Sciences, Beijing 100190, China}
\affiliation{University of Chinese Academy of Sciences, Beijing 100049, China}

\author{Zhaopeng Guo}
\affiliation{Beijing National Laboratory for Condensed Matter Physics,
and Institute of Physics, Chinese Academy of Sciences, Beijing 100190, China}
\affiliation{University of Chinese Academy of Sciences, Beijing 100049, China}

\author{Hongming Weng}
\affiliation{Beijing National Laboratory for Condensed Matter Physics,
and Institute of Physics, Chinese Academy of Sciences, Beijing 100190, China}
\affiliation{University of Chinese Academy of Sciences, Beijing 100049, China}

\author{Zhijun Wang}
\email{wzj@iphy.ac.cn}
\affiliation{Beijing National Laboratory for Condensed Matter Physics,
and Institute of Physics, Chinese Academy of Sciences, Beijing 100190, China}
\affiliation{University of Chinese Academy of Sciences, Beijing 100049, China}

{
\begin{abstract}
To realize novel topological phases and to pursue potential applications in low-energy consumption spintronics, the study of magnetic topological materials is of great interest. Starting from the theory of nonmagnetic topological quantum chemistry [\emph{Bradlyn et al., Nature 547, 298 (2017)}], we have obtained irreducible (co)representations and compatibility relations (CRs) in momentum space, and we constructed a complete list of magnetic band (co)representations (MBRs) in real space for other MSGs with anti-unitary symmetries (\ie type-III and type-IV MSGs). The results are consistent with the magnetic topological quantum chemistry [\emph{Elcoro et al., Nat. Comm. 12, 5965 (2021)}]. Using the CRs and MBRs, we reproduce the symmetry-based classifications for MSGs, and we obtain a set of Fu-Kane-like formulas of symmetry indicators (SIs) in both spinless (bosonic) and spinful (fermionic) systems, which are implemented in an automatic code -- {\ttfamily TopMat} -- to diagnose topological magnetic materials.
The magnetic topological materials, whose occupied states can not be decomposed into a sum of MBRs, are consistent with nonzero SIs. 
Lastly, using our online code, we have performed spin-polarized calculations for magnetic compounds in the materials database and find many magnetic topological candidates.
\end{abstract}

\maketitle
\section{Introduction}
In the last few decades, the topological phases of matter have attracted a great deal of interest in the field of condensed-matter physics. New phenomena such as the integer and fractional quantum Hall effects \cite{PhysRevLett.45.494,PhysRevLett.50.1395,PhysRevLett.48.1559}, time-reversal invariant two- and three-dimensional topological insulators (TIs)~\cite{PhysRevLett.95.226801,Bernevig1757,PhysRevLett.98.106803,xia2009observation,zhang2009}, topological crystalline insulators (TCIs)~\cite{PhysRevLett.106.106802,hsieh2012topological,wang2016hourglass,Ma2017,Wieder246,Schindlereaat0346}, and topological semimetals~\cite{wan2011, xu2011, wang2013, weng2015, Xu613,lv2015observation} have been discovered. The topologically nontrivial materials exhibit robust transport properties such as the quantized Hall and magnetoelectric effects, surface states, and Fermi arcs. Theoretical works reveal that the topology of noninteracting electrons in 3D crystals relies on the structure of Bloch states as a function of momentum.
Most recently, the symmetry eigenvalues or irreducible representations (irreps) of all 230 space groups are used to characterize topological materials  through the theories of symmetry-based indicators~\cite{po2017symmetry,PhysRevX.8.031070,song2018,PhysRevX.7.041069} and topological quantum chemistry (TQC)~\cite{bradlyn2017tqc}. As a result, high-throughput screening for topological materials has been performed in nonmagnetic materials~\cite{tang2019comprehensive,zhang2019catalogue,vergniory2019complete}.

Pervious studies~\cite{slager2013space,bradlyn2017tqc,po2017symmetry,song2018} have concentrated on nonmagnetic space groups, which are known as 230 type-II MSGs with time reversal ($\cal T$).
In fact, there are 1651 MSGs, each of which generally contains a unitary part and an antiunitary part ($\bM\equiv \bG+A\bG$, where $\bM$ is a MSG, $\bG$ is its unitary part, and $A$ is an antiunitary symmetry). These MSGs can also be categorized into four classes: 230 ordinary crystallographic space groups without any antiunitary symmetry (type I with $A=\varnothing$), 230 type-II (with $A={\cal T}$), 674 type-III and 517 type-IV groups (with $A={\cal T}R$). In the type-IV group $R$ is a pure translation, and other cases belong to the type-III group.
In the nonmagnetic TQC work, the irreps and compatibility relations (CRs) in momentum space, and the (elementary) band representations (BRs) are enumerated for type-I and type-II MSGs. To extend the TQC theory to all MSGs, the enumerations of irreducible corepresentations (coirreps), CRs and MBRs for the type-III and type-IV MSGs are needed.
Even though the symmetry-based classifications of all MSGs were finished in Ref.~\cite{watanabe2018structure}, they do not give the physical meaning of the SIs. Recently, the authors of Ref.~\cite{xu2020high,elcoro2021magnetic} developed magnetic topological quantum chemistry (MTQC) and obtained the physically meaningful SIs for MSGs. They have released the full set of MTQC, including the corepresentations, their compatibility relations, and the magnetic band corepresentations accordingly~\cite{elcoro2021magnetic}.
On the other hand, the mappings from SIs to topological invariants in MSGs have also been investigated in Refs.~\cite{peng2021topological,PhysRevB.103.245127,PhysRevB.103.195145}. 

In this work, starting from the CRs and BRs of the nonmagnetic TQC, we have reconstructed MBRs and the Fu-Kane-like formulas of SIs for MSGs. Although they were obtained in the MTQC during the long preparation of this work, it is important to have independent works to generate them and apply them in the DFT calculations due to the complexity of the problem. Therefore, we have developed an automatic code -- {\ttfamily TopMat}~\cite{web} -- to diagnose magnetic topological materials, and we performed a theoretical search for topological magnetic materials. In Section \ref{sec:tqc1651}, we review nonmagnetic TQC theory in type-I (and type-II with $\cal T$) MSGs, and then derive coirreps, CRs, and MBRs of type-III and type-IV MSGs.
In Sec.~\ref{sec:si1651}, we reproduce the symmetry-based classifications for MSGs, and we obtain the Fu-Kane-like formulas of SIs by Smith decomposition. An online code -- {\ttfamily TopMat} -- is developed to compute them for any (non)magnetic material with the input ``{\ttfamily tqc.data}" (from \webirvsp~\cite{gao2021irvsp}). 
Finally, in Sec.~\ref{sec:cal}, based on spin-polarized density functional theory (DFT) calculations, we have performed a theoretical search for topological magnetic compounds in the materials database by using our homemade codes. Some good candidates are presented.


\section{MTQC in 1651 MSGs}
\label{sec:tqc1651}
In topologically trivial band structures, the Wannier functions are exponentially localized and respect the crystal symmetries.  A set of bands arising from localized, symmetric Wannier functions form a representation of the crystal symmetry group, and they are referred to as BR in TQC theory. TQC theory contains two concrete aspects. i) The CRs: based on them, one can get the information of irreps at any $k$-points in the entire Brillouin zone (BZ) from those at maximal high-symmetry $k$-points (HSKPs) only, which makes it possible for the high-throughput screening for topological materials. ii) A full list of (elementary) BRs: they are given by not only the irreps in momentum space, but also by a specific set of orbitals (\ie $\rho@q$) in real space. The materials, whose occupied states can not be expressed as a sum of BRs, are classified to be topological. On the other hand, these BRs also indicate the orbital character in real space, \ie the average charge center and site-symmetry character.
Therefore, the application of TQC has allowed for the discovery of both topologically nontrivial materials with novel phenomena~\cite{tang2019comprehensive,zhang2019catalogue,vergniory2019complete,xu2020high} and topologically trivial unconventional materials (or obstructed atomic limits) with interesting properties~\cite{gao2022unconventional,xu2021filling,xu2021three,li2021obstructed}.

\subsection{TQC in MSGs without antiunitary symmetry}
The CRs in momentum space and the BRs in real space are obtained and established for 230 type-I MSGs in Refs.~\cite{bradlyn2017tqc,PhysRevE.96.023310,elcoro2017double}.
We briefly review space-group operators, assuming the reader is a physicist familiar with group theory. 
In 3D crystals, space-group operators, $h\equiv\{R_h|\bv_h\}$, consist of two parts: a rotation part $R_h$ and a translation part $\bv_h$. The product of two operations is defined as $\{R_s|\bv_s\}\{R_t|\bv_t\}=\{R_sR_t|R_s\bv_t+\bv_s\}$. 
The lattice translation $\{E|\bL\}$ acting on a Bloch state $\phi_\bk$ gives a phase factor of $e^{-i\bk\cdot \bL}$.

\subsubsection{Compatibility relations in momentum space}
We start by identifying the maximal $k$-vectors in the first BZ~\cite{aroyo2014brillouin,tasci2012introduction}. The little group of $\bk$ is defined as $LG(\bk):\{h|h\bk=\bk+\bB,~h\in \bG\}$ with $\bB$ integer reciprocal lattice translations. In a 3D BZ, all adjacent $\bk_i$ points connecting to the maximal $k$-vector $\bk_0$ satisfy $LG(\bk_i)\subset LG(\bk_0)$. 
The group theory tells that the irreducible representation of $LG(\bk_0)$ is also a representation of $LG(\bk_i)$, which may be reducible or not. These relations are known as the CRs~\cite{PhysRevE.96.023310}.

\subsubsection{Band representations in real space}
For any position $\bq$ in the unit cell of a crystal, the set of symmetry operations $s\in \bG$ that leave $\bq$ fixed (absolutely, not up to integer lattice translations $\bL$) is called the stabilizer group, or the site-symmetry group $G_\bq$. By definition, a site-symmetry group is isomorphic to a point group. 
Thus, we assume that the local orbits $\ket{W_j(\vecr-\bq)}~(j=1,\dots,m)$ are transformed as basis of the $m$-dimensional representation $\rho$ [\ie $\Delta^{\rho}(s)$ is the $m\times m$ matrix representation of symmetry operation $s$ in irrep $\rho$].

The set of positions $\{\bq_\alpha=g_\alpha\bq~|~\bq_\alpha\neq \bq_\beta+\bL\};~g_1=\{E|000\};~\alpha,\beta=1,\dots,n$) are classified by a Wyckoff position of multiplicity $n$. Thus, one can find that there are $m\times n$ orbitals $\ket{ W^{\alpha}_{j}(\vecr-\bq_\alpha)}\equiv\ket{g_\alpha W_j(\vecr-\bq)}$ in a unit cell.
Considering the duplicates due to the lattice translations, the space group $\bf G$ is spanned by the set of all the orbitals ($\rho@q$). In other words, they form a representation of $\bf G$ in real space. After Fourier transforms, we can derive the corresponding matrix representation of symmetry operation $h$ (at any $k$) as below:
\begin{eqnarray}
& c^\dagger_{\bk,j\alpha}: a_{\bk,j\alpha}(\vecr)=\frac{1}{\sqrt{N}}\sum_\bL e^{i\bk\cdot (\bq_\alpha+\bL)} W^{\alpha}_{j}(\vecr-\bq_\alpha-\bL)\notag \\
&  h c_{\bk, j\alpha}^\dagger= e^{-i(hk)\cdot \bL_{\beta \alpha}} \sum_{j'} \Delta^{\rho}_{j'j}(s) c_{R_h\bk,j'\beta}^\dagger \label{reph} \label{eq:mpk}\\
& {\rm with}~s=g_{\beta}^{-1}\{E|-\bL_{\beta\alpha}\}hg_\alpha \notag
\end{eqnarray}
Here, $\beta$ and $\bL_{\beta\alpha}$ are determined by $h \bq_\alpha \equiv \bq_\beta+\bL_{\beta\alpha}$.  Thus, the matrix representations of $\bf G$ are given in Eq.~(\ref{eq:mpk}) on the basis of $\{\ket{a_{\bk,j\alpha}(\vecr)},\ket{a_{\bk_2,j\alpha}(\vecr)},\dots,\ket{a_{\bk_l,j\alpha}(\vecr)}\}$ with $j=1,\dots,m$ and $\alpha=1,\dots,n$.
The set of symmetry related $k$-vectors $\{\bk_\gamma=R_{h_\gamma}\bk~|~\bk_\gamma\neq \bk_\delta+\bB\};~h_1=\{E|000\};~\gamma,\delta=1,\dots,l$) are classified by $k$-stars.
The little group of $LG(\bf k)$ is presented in the basis of $\{\ket{a_{\bk,j\alpha}(\vecr)}\}$. By assigning the representations to the irreps of $k$ little group~\cite{elcoro2017double}, one can obtain a set of $k$-irreps of the BR $\rho@q$, especially at the maximal $k$-vectors. 

\begin{table*}[!htb]
\footnotesize
\caption{The table of magnetic topological candidates shows the ICSD number, magnetic space group and type, the converged energy, and the nonzero SIs.
}\label{table:mat}
\begin{ruledtabular}
\begin{tabular}{ p{4cm} p{1.5cm} p{1cm} p{2cm} p{2cm} }
\#ICSD & \#MSG\ (OG)  & Type & Energy (eV/atom) & SIs \\
\hline
44084 SG223 V$_{3}$As & 1608 &  III &  -5.9877 & $z_{2}$=1 \\
\hline
58096 SG223 V$_{3}$Sb & 1608 &  III &  -5.7167 & $z_{2}$=1 \\
\hline
428795 SG164 EuMg$_{2}$Bi$_{2}$ & 1319 &  III &  -4.9684 & $z_{2}$=0, $z_{12}$=9 \\
 & 1321 &  IV &  -4.9687 & $z_{2}$=1 \\
\hline
67671 SG194 InMnO$_{3}$
 & 1497 &  III &  -6.5824 & $z_{3}$=2 \\
 & 1501 &  III &  -6.5832 & $z_{3}$=0, $z_{6}$=4 \\
\hline
260109 SG194 EuIn$_{2}$As$_{2}$
 & 1499 &  III &  -5.5771 & $z_{2}$=1 \\
 & 1501 &  III &  -5.5768 & $z_{3}$=2, $z_{6}$=3 \\
\hline
97965 SG148 La$_{2}$CuRuO$_{6}$ & 1247 &  I &  -6.9781 & $z_{2}$=1, $z_{4}$=3 \\
\hline
104953 SG221 Mn$_{3}$Pt & 1600 &  IV &  -7.1117 & $z_{4}$=1 \\
\hline
20449 SG11
YClO$_{2}$
 & 65 &  IV &  -5.9291 & $z_{2}$=1 \\
\hline
20492  SG11
GdClO$_{2}$ & 65 &  IV &  -8.2142 & $z_{2}$=1 \\
\hline
76845  SG221 Mn$_3$GaC & 1600 &  IV &  -6.5658 & $z_{4}$=2 \\
\hline
76070 SG221 Mn$_{3}$ZnN & 1600 &  IV &  -6.2377 & $z_{4}$=2 \\
\end{tabular}
\end{ruledtabular}
\end{table*} 

\subsection{TQC in MSGs with antiunitary symmetry}
To construct BRs in MSGs with antiunitary symmetry (AS), one needs to get the relations between coirreps in momentum space and AS-related orbitals in real space. 
To get an appropriate description of a MSG $\bM$, one should start with its unitary group $\bG$. 
Saying the eigenstates $\ket{\phi_{j=1,\dots,m}}$ transform as a $m$-dimensional irrep $\rho$ of $\bG$, one can take $\ket{\phi_j}$ and $A\ket{\phi_j}$ as the bases for the magnetic group $\bM$. Thus, these bases are dubbed ``corepresentations" of $\bM$, denoted as $D$ below.
To analyze the reducibility of a corepresentation of a MSG, one can apply the ``Herring rule"~\cite{cornwell1984group,streitwolf1971group,PhysRevB.91.115317},
\begin{eqnarray}
&\frac{1}{|\bG|}\sum_{B\in{A\bG}}{\chi_{\rho}(B^2)}
=\begin{cases}
\begin{array}{ccc}
1  &&\text{ case (a)} \\
-1 &&\text{ case (b)} \\
0  &&\text{ case (c)}
\end{array}\end{cases} \label{eq:herring}
\end{eqnarray}
where $\chi_\rho$ is the character of irrep $\rho$, $|\bG|$ is the number of elements in $\bG$. The Herring rule classify $\rho$ into three classes: $\rho^{(a)}$, $\rho^{(b)}$, and $\rho^{(c)}$. In case (a), the corepresentation $D$ is reducible, $D=\rho^{(a)}_i\oplus \rho^{(a)}_i$, while in cases (b) and (c), $D$ is irreducible, resulting in the double degeneracy of $\rho^{(b)}_i\rho^{(b)}_i$ and $\rho^{(c)}_i \rho^{(c)}_j$, respectively. Two related irreps of case (c) are satisfied in the condition [$\chi_{\rho_j}(R)\equiv \chi^*_{\rho_i}(A^{-1}RA)$].
Thus, the coirreps of the magnetic group $\bM$ are labeled by the combined irreps of unitary group $\bG$: $\rho^{(a)}_i$, $\rho^{(b)}_i\rho^{(b)}_i$, and $\rho^{(c)}_i \rho^{(c)}_j$.

\begin{figure*}[!th]
 \includegraphics[width=0.9\textwidth]{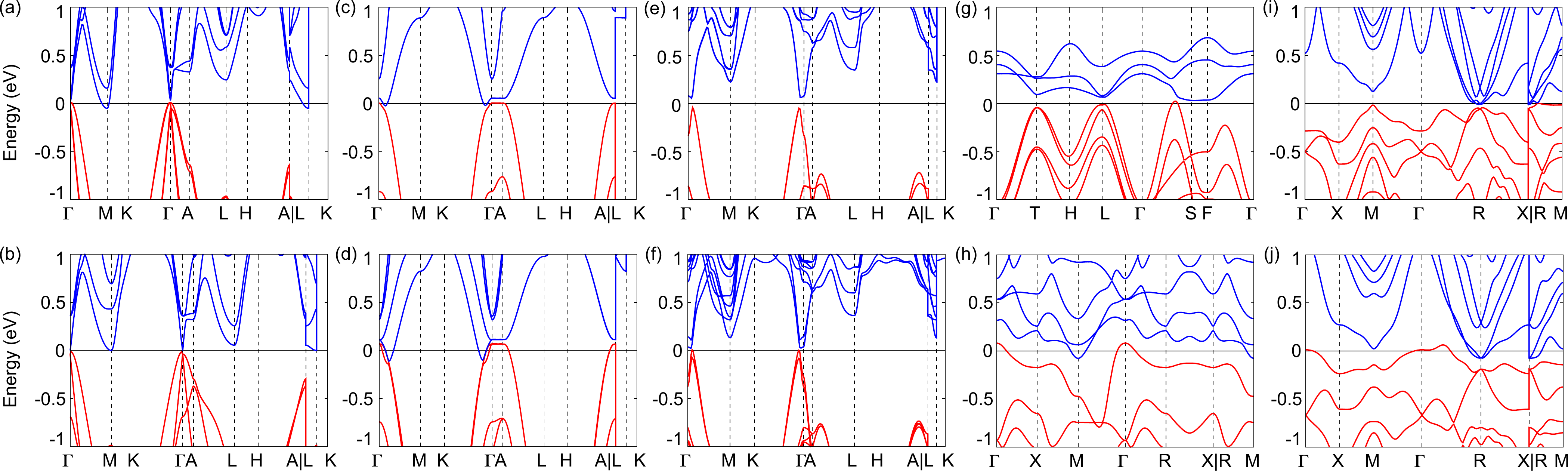}
    \caption{(color online).
Band structures of magnetic topological candidates:
(a)MSG1319 EuMg$_2$Bi$_2$,
(b)MSG1321 EuMg$_2$Bi$_2$,
(c)MSG1497 InMnO$_3$,
(d)MSG1501 InMnO$_3$,
(e)MSG1499 EuIn$_2$As$_2$,
(f)MSG1501 EuIn$_2$As$_2$,
(g)MSG1247 La$_2$CuRuO$_6$, 
(h)MSG1600 Mn$_3$Pt,
(i)MSG1608 V$_3$As,
(j)MSG1608 V$_3$Sb.
    }\label{fig:dft1}
\end{figure*}

\subsubsection{Irreducible corepresentations in momentum space}
First, we need to identify its unitary group $\bG$.
After considering the Herring rule in Eq.~(\ref{eq:herring}), we can label all the coirreps of $\bM$ by the combined irreps of $\bG$. In this work, all the coirreps for MSGs are solved. They are presented explicitly in the Appendix.
Thus, from the CRs of type-I MSG obtained in the previous work~\cite{bradlyn2017tqc},
the CRs for coirreps in type III and type-IV MSG can be conveniently worked out.
	To calculate the corresponding coirreps in the DFT calculcations, the unitary group of a MSG is crucial, which can be obtained {\href{http://tm.iphy.ac.cn/TopMat_1651msg.html}{\ttfamily online}} (see MSG tables in the Appendix). 
With the program \webirvsp~and its unitary group, one can get the coirreps for a magnetic compound.

\begin{figure*}[!t]
    \centering
    \includegraphics[width=0.9\textwidth]{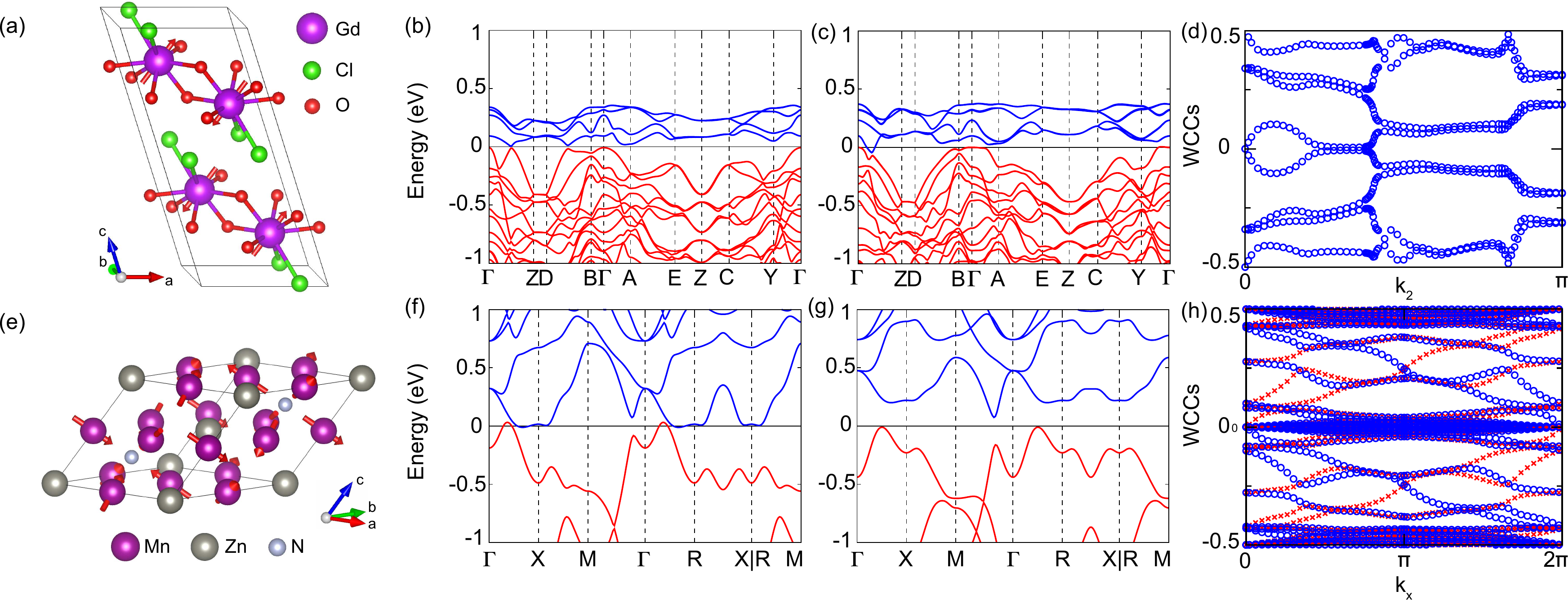}
    \caption{(color online).
	Crystal structures and magnetic configurations (denoted by red arrows) of (a) GdClO$_2$ and (e) Mn$_3$ZnN.
Band structures of magnetic topological candidates:
(b)MSG65 YClO$_2$,
(c)MSG65 GdClO$_2$,
(f)MSG1600 Mn$_3$GaC,
(g)MSG1600 Mn$_3$ZnN.
(d) WCCs of the k$_1$k$_2$ plane in MSG65 GdClO$_2$.
(h) WCCs of the k$_x$k$_y$ plane in MSG1600 Mn$_3$ZnN.
    }\label{fig:dft2}
\end{figure*}

\subsubsection{Magnetic band representations in real space}
At a magnetic Wyckoff position $\bq$ in a MSG~\cite{gallego2012magnetic}, if its site-symmetry group is antiunitary, we have to figure out the degeneracy of the irreps of its unitary part by the Herring rule in Eq.~(\ref{eq:herring}) (isomorphic to magnetic point groups). Otherwise, we have to add other AS-related orbitals to form a MBR in real space. Thus, we can get the MBRs for MSGs from the BRs of type-I MSGs.
After solving the matching of the orbitals in real space, MBRs are obtained and presented explicitly in the Appendix.

\section{Symmetry-based indicators of MSGs}
\label{sec:si1651}
With the CRs and coirreps for all MSGs, a MBR can be expressed as a vector, consisting of the numbers of different coirreps at maximal HSKPs. They form a vector space $\{BR\}$, in which the linearly independent ones are regarded as elementary (M)BRs.
On the other hand, the coirreps of band structures, satisfying the CRs, form another vector space $\{CR\}$.  These band structures can be classified by the quotient group of $\frac{\{CR\}}{\{BR\}}$~\cite{po2017symmetry}. We derive the symmetry-based classifications of MSGs by Smith decomposition~\cite{song2020twisted, PhysRevB.102.035110}, which are consistent with Refs.~\cite{watanabe2018structure,xu2020high,elcoro2021magnetic}. The Fu-Kane-like formulas of the SIs are obtained in the tables of the Appendix and computed by an online code ({\ttfamily TopMat}) for any magnetic material. 

The occupied bands of a material can also be expressed as a vector. 
If it is a sum of BRs, it is consistent with the trivial case with zero SIs. Otherwise, it's topological with nonzero SIs.
In fact, the BR decomposition approach is equivalent to the SI description.
Unlike the symmetry-based classifications for type-II MSG, the classifications of other MSG may yield Weyl semimetal phase. For example, an odd-number inversion-based $z_4$ and an odd-number S$_4$-based $z_2$ indicate a set of Weyl nodes at some generic momenta~\cite{PhysRevB.83.245132,nie2020magnetic,gao2021high}.
For more information, one can find the physically meaningful SIs for MSGs in Refs.~\cite{xu2020high,elcoro2021magnetic,peng2021topological,PhysRevB.103.245127,PhysRevB.103.195145}.

\section{Searching for magnetic topological materials}
\label{sec:cal}
To search for magnetic topological materials, we have performed spin-polarized DFT calculations with Hubbard-$U$ calculations (\ie $U=3$ and 7 eV for $d$ and $f$ electrons respectively). To investigate different magnetic configurations, non-collinear magnetism and spin-orbit coupling are considered in our calculations.
We propose an automatic process to search for topological magnetic materials. It is implemented in {\ttfamily TopMat} ({\href{http://tm.iphy.ac.cn/TopMat\_1651msg.html}{http://tm.iphy.ac.cn/TopMat\_1651msg.html}}), and it is user-friendly for the DFT researchers. The searching process can be easily reproduced online by others for any magnetic material.

\subsection{An automatic process to search for topological materials}
\label{sec:workflow}
i) We generate {\ttfamily POSCAR} (SG \#A) of a crystallographic structure with magnetic atoms from the materials database. Then, we run \webPh~to standardize the structure and generate {\ttfamily PPOSCAR}.

ii) The MSGs in the Opechowski-Guccione (OG) setting~\cite{OGref} (A.1.$X$,A.2.$X$,$\dots$) for different magnetic configurations can be generated by our code. The unitary-part group (SG \#B) and the magnetic configuration in a MSG are given explicitly by {\ttfamily TopMat}. For compounds with local magnetic moments on magnetic atoms (negnecting type-II MSGs), we perform the DFT calculations to obtain the total energies and Bloch states on maximal HSKPs. By comparing their total energies (per atom), one can tell the ground-state configuration and metastable configurations.

iii) By using \webirvsp~({\ttfamily irvsp -sg \#B}), we compute the (co)irreps for occupied bands and generate {\ttfamily tqc.data}. Then, we use the online code {\ttfamily TopMat}~to solve the CRs and compute the SIs. One can find the Fu-Kane-like SIs in the MSG tables of the Appendix.
If the band structure doesn't satisfy CRs, it is a symmetry-protected magnetic metal with crossing points near the $E_F$.
Otherwise, it could be a topological magnetic insulator if it  has nonzero SIs.

\subsection{Magnetic topological candidates}

Based on the DFT calculations, we find some magnetic topological candidates in Table~\ref{table:mat} with nonzero SIs. Their band structures are shown in Figs.~\ref{fig:dft1} and \ref{fig:dft2}. To understand detailed nontrivial topology, such as antiferromagnetic topological insulator (AFM TI), axion insulator, and magnetic topological crystalline insulator (MTCI), one needs to do extra work, such as the Wannier charge centers (WCCs) of 1D Wilson loops. In the following, we have investigated GdClO$_2$ and Mn$_3$ZnN compounds in detail. Their crystal structures and magnetic configurations are generated by the workflow described in Section~\ref{sec:workflow}.

The SI of MSG65 type-IV GdClO$_2$ is computed to be $z_2=1$, indicating nontrivial nature. Its magnetic band structure is presented in Fig.~\ref{fig:dft2}(c). In the configuration of MSG 65 ($P_{2c}2_1/m'$), the AS operator is $\{{\cal T}|0 0 \frac{1}{2}\}$. Thus, the $k_1k_2$ 2D plane has a time-reversal $\mathbb Z_2$ classification. Note that all energy bands are doubly degenerate in the bulk due to the presence of inversion symmetry $\{{\cal I}|0 0 0\}$ with the relation $[\{{\cal T}|0 0 \frac{1}{2}\}\{{\cal I}|0 0 0\}]^2=-1$. 
The WCCs on $k_1$-directed Wilson loops are computed and plotted as a function of $k_2$ in Fig.~\ref{fig:dft2}(d). At $\overline\Gamma~(k_2=0)$ and $\overline {\rm Y}~(k_2=\pi)$, all the WCCs are twofold, and they switch partners during the revolution, indicating $z_2=1$ with nontrivial $\mathbb Z_2$ topology.
Thus, MSG65 type-IV GdClO$_2$ belongs to the AFM TI phase.
On $\{{\cal T}|0 0 \frac{1}{2}\}$-preserving (100)-surface, the topological gapless surface states are expected.

The SI of MSG1600 type-IV Mn$_3$ZnN is computed to be $z_4=2$. Its insulating band structure is shown in Fig.~\ref{fig:dft2}(g). In the magnetic configuration of MSG 1600 ($P_Fm\bar 3 m'$), an AS operator is $\{{\cal T}| 0 0 \frac{1}{2}\}$ [with respect to the (cubic) conventional cell of MSG structure]. A mirror symmetry $\{m_z|0 0 0\}$ is preserved in the $k_xk_y$ plane, on which the mirror Chern number is well defined. The WCCs of $k_x$-directed Wilson loops in the $k_z=0$ plane are obtained in Fig.~\ref{fig:dft2}(h). The crossings and circles denotes different $m_z$ eigenvalues. The evolutions of the WCCs indicate that mirror Chern numbers are $C^{\pm}_{m_z}=\pm 2$. Thus, we conclude that MSG1600 Mn$_3$ZnN belongs to MTCI phase.

\section{Conclusions}
From the nonmagnetic TQC, we have constructed the CRs and MBRs for type-III and type-IV MSGs, which are consistent with the MTQC. Then we obtain the Fu-Kane-like formulas of SIs for MSGs directly by Smith decomposition. Next, we propose an automatic process to search for magnetic topological materials, which is implemented in the online code {\ttfamily TopMat}~to check CRs and compute SIs with the help of \webirvsp. In the end, based on the DFT calculations, we found many magnetic topological candidates.


\ \\
\noindent \textbf{Acknowledgments}
This work was supported by the National Natural Science Foundation of China (Grants No. 11974395, No. 12188101), the Strategic Priority Research Program of Chinese Academy of Sciences (Grant No. XDB33000000), the China Postdoctoral Science Foundation funded project (Grant No. 2021M703461), and the Center for Materials Genome.

\bibliography{main}
}

  \ \\
  \clearpage
 \begin{widetext}

 \beginsupplement{}
 \setcounter{section}{0}
 \renewcommand{\thesubsection}{\arabic{subsection}}
 \renewcommand{\thesubsubsection}{\alph{subsubsection}}


 \clearpage
 \section*{The Appendices of "Magnetic band representations, Fu-Kane-like symmetry indicators and magnetic topological materials"} 
 \label{tabindex}
 The CRs and BRs in type-I and type-II MSGs originate from the nonmagnetic TQC work~\cite{bradlyn2017tqc}.
 From them, we then derive the CRs and magnetic BRs in type-III and type-IV MSGs. The results are consistent with the MTQC work~\cite{xu2020high,elcoro2021magnetic}.
 The notations of Wyckoff positions, $k$-points, irreducible representations (irrep) of $k$-little groups, and MSGs are employed on the \href{https://www.cryst.ehu.es}{BCS server}~\cite{PhysRevE.96.023310,elcoro2017double,aroyo2014brillouin,tasci2012introduction,gallego2012magnetic}.
 After getting the coirreps in type-III and type-IV MSGs, their CR conditions are derived from those of the unitary part of the MSGs.
The Fu-Kane-like formulas of SIs are obtained from the Smith decomposition to solve the quotient group of $\{CR\}/\{BR\}$.  They may be different from the physically meaningful ones~\cite{elcoro2021magnetic}, since they are directly generated by codes, without further modification. 
The nonzero SIs indicate that the band structure of a given magnetic material is not a sum of BRs, which is topological. In the end, an automatic process implemented in the code {\ttfamily TopMat}~ is presented to search for magnetic topological materials with the help of the code \webirvsp~\cite{gao2021irvsp}.
 \begin{longtable}{p{4cm}p{4cm}p{4cm}p{4cm}}
    \hyperref[sup:sg1]   {SG 1}   (Appendix\ref{sup:sg1})    &
    \hyperref[sup:sg2]   {SG 2}   (Appendix\ref{sup:sg2})    &
    \hyperref[sup:sg3]   {SG 3}   (Appendix\ref{sup:sg3})    &
    \hyperref[sup:sg4]   {SG 4}   (Appendix\ref{sup:sg4})    \\
    \hyperref[sup:sg5]   {SG 5}   (Appendix\ref{sup:sg5})    &  
    \hyperref[sup:sg6]   {SG 6}   (Appendix\ref{sup:sg6})    &  
    \hyperref[sup:sg7]   {SG 7}   (Appendix\ref{sup:sg7})    &  
    \hyperref[sup:sg8]   {SG 8}   (Appendix\ref{sup:sg8})    \\ 
    \hyperref[sup:sg9]   {SG 9}   (Appendix\ref{sup:sg9})    &  
    \hyperref[sup:sg10]  {SG 10}  (Appendix\ref{sup:sg10})   &  
    \hyperref[sup:sg11]  {SG 11}  (Appendix\ref{sup:sg11})   &  
    \hyperref[sup:sg12]  {SG 12}  (Appendix\ref{sup:sg12})   \\ 
    \hyperref[sup:sg13]  {SG 13}  (Appendix\ref{sup:sg13})   &  
    \hyperref[sup:sg14]  {SG 14}  (Appendix\ref{sup:sg14})   &  
    \hyperref[sup:sg15]  {SG 15}  (Appendix\ref{sup:sg15})   &  
    \hyperref[sup:sg16]  {SG 16}  (Appendix\ref{sup:sg16})   \\ 
    \hyperref[sup:sg17]  {SG 17}  (Appendix\ref{sup:sg17})   &  
    \hyperref[sup:sg18]  {SG 18}  (Appendix\ref{sup:sg18})   &  
    \hyperref[sup:sg19]  {SG 19}  (Appendix\ref{sup:sg19})   &  
    \hyperref[sup:sg20]  {SG 20}  (Appendix\ref{sup:sg20})   \\ 
    \hyperref[sup:sg21]  {SG 21}  (Appendix\ref{sup:sg21})   &  
    \hyperref[sup:sg22]  {SG 22}  (Appendix\ref{sup:sg22})   &  
    \hyperref[sup:sg23]  {SG 23}  (Appendix\ref{sup:sg23})   &  
    \hyperref[sup:sg24]  {SG 24}  (Appendix\ref{sup:sg24})   \\ 
    \hyperref[sup:sg25]  {SG 25}  (Appendix\ref{sup:sg25})   &  
    \hyperref[sup:sg26]  {SG 26}  (Appendix\ref{sup:sg26})   &  
    \hyperref[sup:sg27]  {SG 27}  (Appendix\ref{sup:sg27})   &  
    \hyperref[sup:sg28]  {SG 28}  (Appendix\ref{sup:sg28})   \\ 
    \hyperref[sup:sg29]  {SG 29}  (Appendix\ref{sup:sg29})   &  
    \hyperref[sup:sg30]  {SG 30}  (Appendix\ref{sup:sg30})   &  
    \hyperref[sup:sg31]  {SG 31}  (Appendix\ref{sup:sg31})   &  
    \hyperref[sup:sg32]  {SG 32}  (Appendix\ref{sup:sg32})   \\ 
    \hyperref[sup:sg33]  {SG 33}  (Appendix\ref{sup:sg33})   &  
    \hyperref[sup:sg34]  {SG 34}  (Appendix\ref{sup:sg34})   &  
    \hyperref[sup:sg35]  {SG 35}  (Appendix\ref{sup:sg35})   &  
    \hyperref[sup:sg36]  {SG 36}  (Appendix\ref{sup:sg36})   \\ 
    \hyperref[sup:sg37]  {SG 37}  (Appendix\ref{sup:sg37})   &  
    \hyperref[sup:sg38]  {SG 38}  (Appendix\ref{sup:sg38})   &  
    \hyperref[sup:sg39]  {SG 39}  (Appendix\ref{sup:sg39})   &  
    \hyperref[sup:sg40]  {SG 40}  (Appendix\ref{sup:sg40})   \\ 
    \hyperref[sup:sg41]  {SG 41}  (Appendix\ref{sup:sg41})   &  
    \hyperref[sup:sg42]  {SG 42}  (Appendix\ref{sup:sg42})   &  
    \hyperref[sup:sg43]  {SG 43}  (Appendix\ref{sup:sg43})   &  
    \hyperref[sup:sg44]  {SG 44}  (Appendix\ref{sup:sg44})   \\ 
    \hyperref[sup:sg45]  {SG 45}  (Appendix\ref{sup:sg45})   &  
    \hyperref[sup:sg46]  {SG 46}  (Appendix\ref{sup:sg46})   &  
    \hyperref[sup:sg47]  {SG 47}  (Appendix\ref{sup:sg47})   &  
    \hyperref[sup:sg48]  {SG 48}  (Appendix\ref{sup:sg48})   \\ 
    \hyperref[sup:sg49]  {SG 49}  (Appendix\ref{sup:sg49})   &  
    \hyperref[sup:sg50]  {SG 50}  (Appendix\ref{sup:sg50})   &  
    \hyperref[sup:sg51]  {SG 51}  (Appendix\ref{sup:sg51})   &  
    \hyperref[sup:sg52]  {SG 52}  (Appendix\ref{sup:sg52})   \\ 
    \hyperref[sup:sg53]  {SG 53}  (Appendix\ref{sup:sg53})   &  
    \hyperref[sup:sg54]  {SG 54}  (Appendix\ref{sup:sg54})   &  
    \hyperref[sup:sg55]  {SG 55}  (Appendix\ref{sup:sg55})   &  
    \hyperref[sup:sg56]  {SG 56}  (Appendix\ref{sup:sg56})   \\ 
    \hyperref[sup:sg57]  {SG 57}  (Appendix\ref{sup:sg57})   &  
    \hyperref[sup:sg58]  {SG 58}  (Appendix\ref{sup:sg58})   &  
    \hyperref[sup:sg59]  {SG 59}  (Appendix\ref{sup:sg59})   &  
    \hyperref[sup:sg60]  {SG 60}  (Appendix\ref{sup:sg60})   \\ 
    \hyperref[sup:sg61]  {SG 61}  (Appendix\ref{sup:sg61})   &  
    \hyperref[sup:sg62]  {SG 62}  (Appendix\ref{sup:sg62})   &  
    \hyperref[sup:sg63]  {SG 63}  (Appendix\ref{sup:sg63})   &  
    \hyperref[sup:sg64]  {SG 64}  (Appendix\ref{sup:sg64})   \\ 
    \hyperref[sup:sg65]  {SG 65}  (Appendix\ref{sup:sg65})   &  
    \hyperref[sup:sg66]  {SG 66}  (Appendix\ref{sup:sg66})   &  
    \hyperref[sup:sg67]  {SG 67}  (Appendix\ref{sup:sg67})   &  
    \hyperref[sup:sg68]  {SG 68}  (Appendix\ref{sup:sg68})   \\ 
    \hyperref[sup:sg69]  {SG 69}  (Appendix\ref{sup:sg69})   &  
    \hyperref[sup:sg70]  {SG 70}  (Appendix\ref{sup:sg70})   &  
    \hyperref[sup:sg71]  {SG 71}  (Appendix\ref{sup:sg71})   &  
    \hyperref[sup:sg72]  {SG 72}  (Appendix\ref{sup:sg72})   \\ 
    \hyperref[sup:sg73]  {SG 73}  (Appendix\ref{sup:sg73})   &  
    \hyperref[sup:sg74]  {SG 74}  (Appendix\ref{sup:sg74})   &  
    \hyperref[sup:sg75]  {SG 75}  (Appendix\ref{sup:sg75})   &  
    \hyperref[sup:sg76]  {SG 76}  (Appendix\ref{sup:sg76})   \\ 
    \hyperref[sup:sg77]  {SG 77}  (Appendix\ref{sup:sg77})   &  
    \hyperref[sup:sg78]  {SG 78}  (Appendix\ref{sup:sg78})   &  
    \hyperref[sup:sg79]  {SG 79}  (Appendix\ref{sup:sg79})   &  
    \hyperref[sup:sg80]  {SG 80}  (Appendix\ref{sup:sg80})   \\ 
    \hyperref[sup:sg81]  {SG 81}  (Appendix\ref{sup:sg81})   &  
    \hyperref[sup:sg82]  {SG 82}  (Appendix\ref{sup:sg82})   &  
    \hyperref[sup:sg83]  {SG 83}  (Appendix\ref{sup:sg83})   &  
    \hyperref[sup:sg84]  {SG 84}  (Appendix\ref{sup:sg84})   \\ 
    \hyperref[sup:sg85]  {SG 85}  (Appendix\ref{sup:sg85})   &  
    \hyperref[sup:sg86]  {SG 86}  (Appendix\ref{sup:sg86})   &  
    \hyperref[sup:sg87]  {SG 87}  (Appendix\ref{sup:sg87})   &  
    \hyperref[sup:sg88]  {SG 88}  (Appendix\ref{sup:sg88})   \\ 
    \hyperref[sup:sg89]  {SG 89}  (Appendix\ref{sup:sg89})   &  
    \hyperref[sup:sg90]  {SG 90}  (Appendix\ref{sup:sg90})   &  
    \hyperref[sup:sg91]  {SG 91}  (Appendix\ref{sup:sg91})   &  
    \hyperref[sup:sg92]  {SG 92}  (Appendix\ref{sup:sg92})   \\ 
    \hyperref[sup:sg93]  {SG 93}  (Appendix\ref{sup:sg93})   &  
    \hyperref[sup:sg94]  {SG 94}  (Appendix\ref{sup:sg94})   &  
    \hyperref[sup:sg95]  {SG 95}  (Appendix\ref{sup:sg95})   &  
    \hyperref[sup:sg96]  {SG 96}  (Appendix\ref{sup:sg96})   \\ 
    \hyperref[sup:sg97]  {SG 97}  (Appendix\ref{sup:sg97})   &  
    \hyperref[sup:sg98]  {SG 98}  (Appendix\ref{sup:sg98})   &  
    \hyperref[sup:sg99]  {SG 99}  (Appendix\ref{sup:sg99})   &  
    \hyperref[sup:sg100] {SG 100} (Appendix\ref{sup:sg100})  \\ 
    \hyperref[sup:sg101] {SG 101} (Appendix\ref{sup:sg101})  &  
    \hyperref[sup:sg102] {SG 102} (Appendix\ref{sup:sg102})  &  
    \hyperref[sup:sg103] {SG 103} (Appendix\ref{sup:sg103})  &  
    \hyperref[sup:sg104] {SG 104} (Appendix\ref{sup:sg104})  \\ 
    \hyperref[sup:sg105] {SG 105} (Appendix\ref{sup:sg105})  &  
    \hyperref[sup:sg106] {SG 106} (Appendix\ref{sup:sg106})  &  
    \hyperref[sup:sg107] {SG 107} (Appendix\ref{sup:sg107})  &  
    \hyperref[sup:sg108] {SG 108} (Appendix\ref{sup:sg108})  \\ 
    \hyperref[sup:sg109] {SG 109} (Appendix\ref{sup:sg109})  &  
    \hyperref[sup:sg110] {SG 110} (Appendix\ref{sup:sg110})  &  
    \hyperref[sup:sg111] {SG 111} (Appendix\ref{sup:sg111})  &  
    \hyperref[sup:sg112] {SG 112} (Appendix\ref{sup:sg112})  \\ 
    \hyperref[sup:sg113] {SG 113} (Appendix\ref{sup:sg113})  &  
    \hyperref[sup:sg114] {SG 114} (Appendix\ref{sup:sg114})  &  
    \hyperref[sup:sg115] {SG 115} (Appendix\ref{sup:sg115})  &  
    \hyperref[sup:sg116] {SG 116} (Appendix\ref{sup:sg116})  \\ 
    \hyperref[sup:sg117] {SG 117} (Appendix\ref{sup:sg117})  &  
    \hyperref[sup:sg118] {SG 118} (Appendix\ref{sup:sg118})  &  
    \hyperref[sup:sg119] {SG 119} (Appendix\ref{sup:sg119})  &  
    \hyperref[sup:sg120] {SG 120} (Appendix\ref{sup:sg120})  \\ 
    \hyperref[sup:sg121] {SG 121} (Appendix\ref{sup:sg121})  &  
    \hyperref[sup:sg122] {SG 122} (Appendix\ref{sup:sg122})  &  
    \hyperref[sup:sg123] {SG 123} (Appendix\ref{sup:sg123})  &  
    \hyperref[sup:sg124] {SG 124} (Appendix\ref{sup:sg124})  \\ 
    \hyperref[sup:sg125] {SG 125} (Appendix\ref{sup:sg125})  &  
    \hyperref[sup:sg126] {SG 126} (Appendix\ref{sup:sg126})  &  
    \hyperref[sup:sg127] {SG 127} (Appendix\ref{sup:sg127})  &  
    \hyperref[sup:sg128] {SG 128} (Appendix\ref{sup:sg128})  \\ 
    \hyperref[sup:sg129] {SG 129} (Appendix\ref{sup:sg129})  &  
    \hyperref[sup:sg130] {SG 130} (Appendix\ref{sup:sg130})  &  
    \hyperref[sup:sg131] {SG 131} (Appendix\ref{sup:sg131})  &  
    \hyperref[sup:sg132] {SG 132} (Appendix\ref{sup:sg132})  \\ 
    \hyperref[sup:sg133] {SG 133} (Appendix\ref{sup:sg133})  &  
    \hyperref[sup:sg134] {SG 134} (Appendix\ref{sup:sg134})  &  
    \hyperref[sup:sg135] {SG 135} (Appendix\ref{sup:sg135})  &  
    \hyperref[sup:sg136] {SG 136} (Appendix\ref{sup:sg136})  \\ 
    \hyperref[sup:sg137] {SG 137} (Appendix\ref{sup:sg137})  &  
    \hyperref[sup:sg138] {SG 138} (Appendix\ref{sup:sg138})  &  
    \hyperref[sup:sg139] {SG 139} (Appendix\ref{sup:sg139})  &  
    \hyperref[sup:sg140] {SG 140} (Appendix\ref{sup:sg140})  \\ 
    \hyperref[sup:sg141] {SG 141} (Appendix\ref{sup:sg141})  &  
    \hyperref[sup:sg142] {SG 142} (Appendix\ref{sup:sg142})  &  
    \hyperref[sup:sg143] {SG 143} (Appendix\ref{sup:sg143})  &  
    \hyperref[sup:sg144] {SG 144} (Appendix\ref{sup:sg144})  \\ 
    \hyperref[sup:sg145] {SG 145} (Appendix\ref{sup:sg145})  &  
    \hyperref[sup:sg146] {SG 146} (Appendix\ref{sup:sg146})  &  
    \hyperref[sup:sg147] {SG 147} (Appendix\ref{sup:sg147})  &  
    \hyperref[sup:sg148] {SG 148} (Appendix\ref{sup:sg148})  \\ 
    \hyperref[sup:sg149] {SG 149} (Appendix\ref{sup:sg149})  &  
    \hyperref[sup:sg150] {SG 150} (Appendix\ref{sup:sg150})  &  
    \hyperref[sup:sg151] {SG 151} (Appendix\ref{sup:sg151})  &  
    \hyperref[sup:sg152] {SG 152} (Appendix\ref{sup:sg152})  \\ 
    \hyperref[sup:sg153] {SG 153} (Appendix\ref{sup:sg153})  &  
    \hyperref[sup:sg154] {SG 154} (Appendix\ref{sup:sg154})  &  
    \hyperref[sup:sg155] {SG 155} (Appendix\ref{sup:sg155})  &  
    \hyperref[sup:sg156] {SG 156} (Appendix\ref{sup:sg156})  \\ 
    \hyperref[sup:sg157] {SG 157} (Appendix\ref{sup:sg157})  &  
    \hyperref[sup:sg158] {SG 158} (Appendix\ref{sup:sg158})  &  
    \hyperref[sup:sg159] {SG 159} (Appendix\ref{sup:sg159})  &  
    \hyperref[sup:sg160] {SG 160} (Appendix\ref{sup:sg160})  \\ 
    \hyperref[sup:sg161] {SG 161} (Appendix\ref{sup:sg161})  &  
    \hyperref[sup:sg162] {SG 162} (Appendix\ref{sup:sg162})  &  
    \hyperref[sup:sg163] {SG 163} (Appendix\ref{sup:sg163})  &  
    \hyperref[sup:sg164] {SG 164} (Appendix\ref{sup:sg164})  \\ 
    \hyperref[sup:sg165] {SG 165} (Appendix\ref{sup:sg165})  &  
    \hyperref[sup:sg166] {SG 166} (Appendix\ref{sup:sg166})  &  
    \hyperref[sup:sg167] {SG 167} (Appendix\ref{sup:sg167})  &  
    \hyperref[sup:sg168] {SG 168} (Appendix\ref{sup:sg168})  \\ 
    \hyperref[sup:sg169] {SG 169} (Appendix\ref{sup:sg169})  &  
    \hyperref[sup:sg170] {SG 170} (Appendix\ref{sup:sg170})  &  
    \hyperref[sup:sg171] {SG 171} (Appendix\ref{sup:sg171})  &  
    \hyperref[sup:sg172] {SG 172} (Appendix\ref{sup:sg172})  \\ 
    \hyperref[sup:sg173] {SG 173} (Appendix\ref{sup:sg173})  &  
    \hyperref[sup:sg174] {SG 174} (Appendix\ref{sup:sg174})  &  
    \hyperref[sup:sg175] {SG 175} (Appendix\ref{sup:sg175})  &  
    \hyperref[sup:sg176] {SG 176} (Appendix\ref{sup:sg176})  \\ 
    \hyperref[sup:sg177] {SG 177} (Appendix\ref{sup:sg177})  &  
    \hyperref[sup:sg178] {SG 178} (Appendix\ref{sup:sg178})  &  
    \hyperref[sup:sg179] {SG 179} (Appendix\ref{sup:sg179})  &  
    \hyperref[sup:sg180] {SG 180} (Appendix\ref{sup:sg180})  \\ 
    \hyperref[sup:sg181] {SG 181} (Appendix\ref{sup:sg181})  &  
    \hyperref[sup:sg182] {SG 182} (Appendix\ref{sup:sg182})  &  
    \hyperref[sup:sg183] {SG 183} (Appendix\ref{sup:sg183})  &  
    \hyperref[sup:sg184] {SG 184} (Appendix\ref{sup:sg184})  \\ 
    \hyperref[sup:sg185] {SG 185} (Appendix\ref{sup:sg185})  &  
    \hyperref[sup:sg186] {SG 186} (Appendix\ref{sup:sg186})  &  
    \hyperref[sup:sg187] {SG 187} (Appendix\ref{sup:sg187})  &  
    \hyperref[sup:sg188] {SG 188} (Appendix\ref{sup:sg188})  \\ 
    \hyperref[sup:sg189] {SG 189} (Appendix\ref{sup:sg189})  &  
    \hyperref[sup:sg190] {SG 190} (Appendix\ref{sup:sg190})  &  
    \hyperref[sup:sg191] {SG 191} (Appendix\ref{sup:sg191})  &  
    \hyperref[sup:sg192] {SG 192} (Appendix\ref{sup:sg192})  \\ 
    \hyperref[sup:sg193] {SG 193} (Appendix\ref{sup:sg193})  &  
    \hyperref[sup:sg194] {SG 194} (Appendix\ref{sup:sg194})  &  
    \hyperref[sup:sg195] {SG 195} (Appendix\ref{sup:sg195})  &  
    \hyperref[sup:sg196] {SG 196} (Appendix\ref{sup:sg196})  \\ 
    \hyperref[sup:sg197] {SG 197} (Appendix\ref{sup:sg197})  &  
    \hyperref[sup:sg198] {SG 198} (Appendix\ref{sup:sg198})  &  
    \hyperref[sup:sg199] {SG 199} (Appendix\ref{sup:sg199})  &  
    \hyperref[sup:sg200] {SG 200} (Appendix\ref{sup:sg200})  \\ 
    \hyperref[sup:sg201] {SG 201} (Appendix\ref{sup:sg201})  &  
    \hyperref[sup:sg202] {SG 202} (Appendix\ref{sup:sg202})  &  
    \hyperref[sup:sg203] {SG 203} (Appendix\ref{sup:sg203})  &  
    \hyperref[sup:sg204] {SG 204} (Appendix\ref{sup:sg204})  \\ 
    \hyperref[sup:sg205] {SG 205} (Appendix\ref{sup:sg205})  &  
    \hyperref[sup:sg206] {SG 206} (Appendix\ref{sup:sg206})  &  
    \hyperref[sup:sg207] {SG 207} (Appendix\ref{sup:sg207})  &  
    \hyperref[sup:sg208] {SG 208} (Appendix\ref{sup:sg208})  \\ 
    \hyperref[sup:sg209] {SG 209} (Appendix\ref{sup:sg209})  &  
    \hyperref[sup:sg210] {SG 210} (Appendix\ref{sup:sg210})  &  
    \hyperref[sup:sg211] {SG 211} (Appendix\ref{sup:sg211})  &  
    \hyperref[sup:sg212] {SG 212} (Appendix\ref{sup:sg212})  \\ 
    \hyperref[sup:sg213] {SG 213} (Appendix\ref{sup:sg213})  &  
    \hyperref[sup:sg214] {SG 214} (Appendix\ref{sup:sg214})  &  
    \hyperref[sup:sg215] {SG 215} (Appendix\ref{sup:sg215})  &  
    \hyperref[sup:sg216] {SG 216} (Appendix\ref{sup:sg216})  \\ 
    \hyperref[sup:sg217] {SG 217} (Appendix\ref{sup:sg217})  &  
    \hyperref[sup:sg218] {SG 218} (Appendix\ref{sup:sg218})  &  
    \hyperref[sup:sg219] {SG 219} (Appendix\ref{sup:sg219})  &  
    \hyperref[sup:sg220] {SG 220} (Appendix\ref{sup:sg220})  \\ 
    \hyperref[sup:sg221] {SG 221} (Appendix\ref{sup:sg221})  &  
    \hyperref[sup:sg222] {SG 222} (Appendix\ref{sup:sg222})  &  
    \hyperref[sup:sg223] {SG 223} (Appendix\ref{sup:sg223})  &  
    \hyperref[sup:sg224] {SG 224} (Appendix\ref{sup:sg224})  \\ 
    \hyperref[sup:sg225] {SG 225} (Appendix\ref{sup:sg225})  &  
    \hyperref[sup:sg226] {SG 226} (Appendix\ref{sup:sg226})  &  
    \hyperref[sup:sg227] {SG 227} (Appendix\ref{sup:sg227})  &  
    \hyperref[sup:sg228] {SG 228} (Appendix\ref{sup:sg228})  \\ 
    \hyperref[sup:sg229] {SG 229} (Appendix\ref{sup:sg229})  & 
    \hyperref[sup:sg230] {SG 230} (Appendix\ref{sup:sg230})  &&\\
   \end{longtable}
\setcounter{table}{0}

\subsection{SG 1}  \label{sup:sg1}
TABLE SG1. The crystalline space group is what the crystal has if the magnetic order is neglected. Once condidering magnetic order, the MSGs, magnetic type, and the symmetry-indicator classifications are given below. For each MSG, the detailed information is given in the corresponding MSG table. \hyperref[tabindex]{See the list of tables.}
\setcounter{table}{-1}
\begin{longtable}{p{1.5cm}p{2cm}|p{1.5cm}p{2cm}|c|c|c|c}
\hline
\hline
   \multicolumn{2}{c|}{OG setting} & \multicolumn{2}{c|}{BNS setting} & MSG type & Detailed Inf. & integer spin & half-integer spin\\
\hline
\endfirsthead
\multicolumn{8}{c} {{ TABLE SG\ref{sup:sg1} -- continued}}\\
\hline
   \multicolumn{2}{c|}{OG setting} & \multicolumn{2}{c|}{BNS setting} & MSG type & Detailed Inf. & integer spin & half-integer spin\\
\hline
\endhead
\hline
\endfoot
\hline
\hline
\endlastfoot
1.1.\rownumber & $P1$ & $P1$ & \#1.1 & I & Table~\ref{tab:msg1.1} & $\varnothing$ & $\varnothing$\\
1.2.\rownumber & $P11'$ & $P11'$ & \#1.2 & II & Table~\ref{tab:msg1.2} & $\varnothing$ & $\varnothing$\\
1.3.\rownumber & $P_{2s}1$ & $P_{S}1$ & \#1.3 & IV & Table~\ref{tab:msg1.3} & $\varnothing$ & $\varnothing$\\
\end{longtable}
\input{src_tex_mtable/msg1.1.tex}
\input{src_tex_mtable/msg1.2.tex}
\input{src_tex_mtable/msg1.3.tex}
\subsection{SG 2}  \label{sup:sg2}
TABLE SG2. The crystalline space group is what the crystal has if the magnetic order is neglected. Once condidering magnetic order, the MSGs, magnetic type, and the symmetry-indicator classifications are given below. For each MSG, the detailed information is given in the corresponding MSG table. \hyperref[tabindex]{See the list of tables.}
\setcounter{table}{2}
\begin{longtable}{p{1.5cm}p{2cm}|p{1.5cm}p{2cm}|c|c|c|c}
\hline
\hline
   \multicolumn{2}{c|}{OG setting} & \multicolumn{2}{c|}{BNS setting} & MSG type & Detailed Inf. & integer spin & half-integer spin\\
\hline
\endfirsthead
\multicolumn{8}{c} {{ TABLE SG\ref{sup:sg2} -- continued}}\\
\hline
   \multicolumn{2}{c|}{OG setting} & \multicolumn{2}{c|}{BNS setting} & MSG type & Detailed Inf. & integer spin & half-integer spin\\
\hline
\endhead
\hline
\endfoot
\hline
\hline
\endlastfoot
2.1.\rownumber & $P$-$1$ & $P$-$1$ & \#2.4 & I & Table~\ref{tab:msg2.4} & $Z_{2}$$\times$$Z_{2}$$\times$$Z_{2}$$\times$$Z_{4}$ & $Z_{2}$$\times$$Z_{2}$$\times$$Z_{2}$$\times$$Z_{4}$\\
2.2.\rownumber & $P$-$11'$ & $P$-$11'$ & \#2.5 & II & Table~\ref{tab:msg2.5} & $Z_{2}$$\times$$Z_{2}$$\times$$Z_{2}$$\times$$Z_{4}$ & $Z_{2}$$\times$$Z_{2}$$\times$$Z_{2}$$\times$$Z_{4}$\\
2.3.\rownumber & $P$-$1'$ & $P$-$1'$ & \#2.6 & III & Table~\ref{tab:msg2.6} & $\varnothing$ & $\varnothing$\\
2.4.\rownumber & $P_{2s}$-$1$ & $P_{S}$-$1$ & \#2.7 & IV & Table~\ref{tab:msg2.7} & $Z_{2}$$\times$$Z_{2}$$\times$$Z_{4}$ & $Z_{2}$\\
\end{longtable}
\input{src_tex_mtable/msg2.4.tex}
\input{src_tex_mtable/msg2.5.tex}
\input{src_tex_mtable/msg2.6.tex}
\input{src_tex_mtable/msg2.7.tex}
\subsection{SG 3}  \label{sup:sg3}
TABLE SG3. The crystalline space group is what the crystal has if the magnetic order is neglected. Once condidering magnetic order, the MSGs, magnetic type, and the symmetry-indicator classifications are given below. For each MSG, the detailed information is given in the corresponding MSG table. \hyperref[tabindex]{See the list of tables.}
\setcounter{table}{6}
\begin{longtable}{p{1.5cm}p{2cm}|p{1.5cm}p{2cm}|c|c|c|c}
\hline
\hline
   \multicolumn{2}{c|}{OG setting} & \multicolumn{2}{c|}{BNS setting} & MSG type & Detailed Inf. & integer spin & half-integer spin\\
\hline
\endfirsthead
\multicolumn{8}{c} {{ TABLE SG\ref{sup:sg3} -- continued}}\\
\hline
   \multicolumn{2}{c|}{OG setting} & \multicolumn{2}{c|}{BNS setting} & MSG type & Detailed Inf. & integer spin & half-integer spin\\
\hline
\endhead
\hline
\endfoot
\hline
\hline
\endlastfoot
3.1.\rownumber & $P2$ & $P2$ & \#3.1 & I & Table~\ref{tab:msg3.1} & $Z_{2}$ & $Z_{2}$\\
3.2.\rownumber & $P21'$ & $P21'$ & \#3.2 & II & Table~\ref{tab:msg3.2} & $Z_{2}$ & $\varnothing$\\
3.3.\rownumber & $P2'$ & $P2'$ & \#3.3 & III & Table~\ref{tab:msg3.3} & $\varnothing$ & $\varnothing$\\
3.4.\rownumber & $P_{2a}2$ & $P_{a}2$ & \#3.4 & IV & Table~\ref{tab:msg3.4} & $Z_{2}$ & $Z_{2}$\\
3.5.\rownumber & $P_{2b}2$ & $P_{b}2$ & \#3.5 & IV & Table~\ref{tab:msg3.5} & $Z_{2}$ & $\varnothing$\\
3.6.\rownumber & $P_{C}2$ & $C_{a}2$ & \#5.17 & IV & Table~\ref{tab:msg5.17} & $Z_{2}$ & $\varnothing$\\
3.7.\rownumber & $P_{2b}2'$ & $P_{b}2_{1}$ & \#4.11 & IV & Table~\ref{tab:msg4.11} & $\varnothing$ & $\varnothing$\\
\hline
&&$P_{C}2$&\#3.6&IV&OG~\ref{sup:sg5}.5.23&\\
\end{longtable}
\input{src_tex_mtable/msg3.1.tex}
\input{src_tex_mtable/msg3.2.tex}
\input{src_tex_mtable/msg3.3.tex}
\input{src_tex_mtable/msg3.4.tex}
\input{src_tex_mtable/msg3.5.tex}
\input{src_tex_mtable/msg5.17.tex}
\input{src_tex_mtable/msg4.11.tex}
\subsection{SG 4}  \label{sup:sg4}
TABLE SG4. The crystalline space group is what the crystal has if the magnetic order is neglected. Once condidering magnetic order, the MSGs, magnetic type, and the symmetry-indicator classifications are given below. For each MSG, the detailed information is given in the corresponding MSG table. \hyperref[tabindex]{See the list of tables.}
\setcounter{table}{13}
\begin{longtable}{p{1.5cm}p{2cm}|p{1.5cm}p{2cm}|c|c|c|c}
\hline
\hline
   \multicolumn{2}{c|}{OG setting} & \multicolumn{2}{c|}{BNS setting} & MSG type & Detailed Inf. & integer spin & half-integer spin\\
\hline
\endfirsthead
\multicolumn{8}{c} {{ TABLE SG\ref{sup:sg4} -- continued}}\\
\hline
   \multicolumn{2}{c|}{OG setting} & \multicolumn{2}{c|}{BNS setting} & MSG type & Detailed Inf. & integer spin & half-integer spin\\
\hline
\endhead
\hline
\endfoot
\hline
\hline
\endlastfoot
4.1.\rownumber & $P2_{1}$ & $P2_{1}$ & \#4.7 & I & Table~\ref{tab:msg4.7} & $\varnothing$ & $\varnothing$\\
4.2.\rownumber & $P2_{1}1'$ & $P2_{1}1'$ & \#4.8 & II & Table~\ref{tab:msg4.8} & $\varnothing$ & $\varnothing$\\
4.3.\rownumber & $P2_{1}'$ & $P2_{1}'$ & \#4.9 & III & Table~\ref{tab:msg4.9} & $\varnothing$ & $\varnothing$\\
4.4.\rownumber & $P_{2a}2_{1}$ & $P_{a}2_{1}$ & \#4.10 & IV & Table~\ref{tab:msg4.10} & $\varnothing$ & $\varnothing$\\
\hline
&&$P_{b}2_{1}$&\#4.11&IV&OG~\ref{sup:sg3}.7.14&\\
&&$P_{C}2_{1}$&\#4.12&IV&OG~\ref{sup:sg5}.6.24&\\
\end{longtable}
\input{src_tex_mtable/msg4.7.tex}
\input{src_tex_mtable/msg4.8.tex}
\input{src_tex_mtable/msg4.9.tex}
\input{src_tex_mtable/msg4.10.tex}
\subsection{SG 5}  \label{sup:sg5}
TABLE SG5. The crystalline space group is what the crystal has if the magnetic order is neglected. Once condidering magnetic order, the MSGs, magnetic type, and the symmetry-indicator classifications are given below. For each MSG, the detailed information is given in the corresponding MSG table. \hyperref[tabindex]{See the list of tables.}
\setcounter{table}{17}
\begin{longtable}{p{1.5cm}p{2cm}|p{1.5cm}p{2cm}|c|c|c|c}
\hline
\hline
   \multicolumn{2}{c|}{OG setting} & \multicolumn{2}{c|}{BNS setting} & MSG type & Detailed Inf. & integer spin & half-integer spin\\
\hline
\endfirsthead
\multicolumn{8}{c} {{ TABLE SG\ref{sup:sg5} -- continued}}\\
\hline
   \multicolumn{2}{c|}{OG setting} & \multicolumn{2}{c|}{BNS setting} & MSG type & Detailed Inf. & integer spin & half-integer spin\\
\hline
\endhead
\hline
\endfoot
\hline
\hline
\endlastfoot
5.1.\rownumber & $C2$ & $C2$ & \#5.13 & I & Table~\ref{tab:msg5.13} & $\varnothing$ & $\varnothing$\\
5.2.\rownumber & $C21'$ & $C21'$ & \#5.14 & II & Table~\ref{tab:msg5.14} & $\varnothing$ & $\varnothing$\\
5.3.\rownumber & $C2'$ & $C2'$ & \#5.15 & III & Table~\ref{tab:msg5.15} & $\varnothing$ & $\varnothing$\\
5.4.\rownumber & $C_{2c}2$ & $C_{c}2$ & \#5.16 & IV & Table~\ref{tab:msg5.16} & $\varnothing$ & $\varnothing$\\
5.5.\rownumber & $C_{P}2$ & $P_{C}2$ & \#3.6 & IV & Table~\ref{tab:msg3.6} & $\varnothing$ & $\varnothing$\\
5.6.\rownumber & $C_{P}2'$ & $P_{C}2_{1}$ & \#4.12 & IV & Table~\ref{tab:msg4.12} & $\varnothing$ & $\varnothing$\\
\hline
&&$C_{a}2$&\#5.17&IV&OG~\ref{sup:sg3}.6.13&\\
\end{longtable}
\input{src_tex_mtable/msg5.13.tex}
\input{src_tex_mtable/msg5.14.tex}
\input{src_tex_mtable/msg5.15.tex}
\input{src_tex_mtable/msg5.16.tex}
\input{src_tex_mtable/msg3.6.tex}
\input{src_tex_mtable/msg4.12.tex}
\subsection{SG 6}  \label{sup:sg6}
TABLE SG6. The crystalline space group is what the crystal has if the magnetic order is neglected. Once condidering magnetic order, the MSGs, magnetic type, and the symmetry-indicator classifications are given below. For each MSG, the detailed information is given in the corresponding MSG table. \hyperref[tabindex]{See the list of tables.}
\setcounter{table}{23}
\begin{longtable}{p{1.5cm}p{2cm}|p{1.5cm}p{2cm}|c|c|c|c}
\hline
\hline
   \multicolumn{2}{c|}{OG setting} & \multicolumn{2}{c|}{BNS setting} & MSG type & Detailed Inf. & integer spin & half-integer spin\\
\hline
\endfirsthead
\multicolumn{8}{c} {{ TABLE SG\ref{sup:sg6} -- continued}}\\
\hline
   \multicolumn{2}{c|}{OG setting} & \multicolumn{2}{c|}{BNS setting} & MSG type & Detailed Inf. & integer spin & half-integer spin\\
\hline
\endhead
\hline
\endfoot
\hline
\hline
\endlastfoot
6.1.\rownumber & $Pm$ & $Pm$ & \#6.18 & I & Table~\ref{tab:msg6.18} & $\varnothing$ & $\varnothing$\\
6.2.\rownumber & $Pm1'$ & $Pm1'$ & \#6.19 & II & Table~\ref{tab:msg6.19} & $\varnothing$ & $\varnothing$\\
6.3.\rownumber & $Pm'$ & $Pm'$ & \#6.20 & III & Table~\ref{tab:msg6.20} & $\varnothing$ & $\varnothing$\\
6.4.\rownumber & $P_{2a}m$ & $P_{a}m$ & \#6.21 & IV & Table~\ref{tab:msg6.21} & $\varnothing$ & $\varnothing$\\
6.5.\rownumber & $P_{2b}m$ & $P_{b}m$ & \#6.22 & IV & Table~\ref{tab:msg6.22} & $\varnothing$ & $\varnothing$\\
6.6.\rownumber & $P_{C}m$ & $C_{a}m$ & \#8.36 & IV & Table~\ref{tab:msg8.36} & $\varnothing$ & $\varnothing$\\
6.7.\rownumber & $P_{2c}m'$ & $P_{c}c$ & \#7.28 & IV & Table~\ref{tab:msg7.28} & $\varnothing$ & $\varnothing$\\
\hline
&&$P_{C}m$&\#6.23&IV&OG~\ref{sup:sg8}.5.42&\\
\end{longtable}
\input{src_tex_mtable/msg6.18.tex}
\input{src_tex_mtable/msg6.19.tex}
\input{src_tex_mtable/msg6.20.tex}
\input{src_tex_mtable/msg6.21.tex}
\input{src_tex_mtable/msg6.22.tex}
\input{src_tex_mtable/msg8.36.tex}
\input{src_tex_mtable/msg7.28.tex}
\subsection{SG 7}  \label{sup:sg7}
TABLE SG7. The crystalline space group is what the crystal has if the magnetic order is neglected. Once condidering magnetic order, the MSGs, magnetic type, and the symmetry-indicator classifications are given below. For each MSG, the detailed information is given in the corresponding MSG table. \hyperref[tabindex]{See the list of tables.}
\setcounter{table}{30}
\begin{longtable}{p{1.5cm}p{2cm}|p{1.5cm}p{2cm}|c|c|c|c}
\hline
\hline
   \multicolumn{2}{c|}{OG setting} & \multicolumn{2}{c|}{BNS setting} & MSG type & Detailed Inf. & integer spin & half-integer spin\\
\hline
\endfirsthead
\multicolumn{8}{c} {{ TABLE SG\ref{sup:sg7} -- continued}}\\
\hline
   \multicolumn{2}{c|}{OG setting} & \multicolumn{2}{c|}{BNS setting} & MSG type & Detailed Inf. & integer spin & half-integer spin\\
\hline
\endhead
\hline
\endfoot
\hline
\hline
\endlastfoot
7.1.\rownumber & $Pc$ & $Pc$ & \#7.24 & I & Table~\ref{tab:msg7.24} & $\varnothing$ & $\varnothing$\\
7.2.\rownumber & $Pc1'$ & $Pc1'$ & \#7.25 & II & Table~\ref{tab:msg7.25} & $\varnothing$ & $\varnothing$\\
7.3.\rownumber & $Pc'$ & $Pc'$ & \#7.26 & III & Table~\ref{tab:msg7.26} & $\varnothing$ & $\varnothing$\\
7.4.\rownumber & $P_{2a}c$ & $P_{a}c$ & \#7.27 & IV & Table~\ref{tab:msg7.27} & $\varnothing$ & $\varnothing$\\
7.5.\rownumber & $P_{2b}c$ & $P_{b}c$ & \#7.29 & IV & Table~\ref{tab:msg7.29} & $\varnothing$ & $\varnothing$\\
7.6.\rownumber & $P_{C}c$ & $C_{a}c$ & \#9.41 & IV & Table~\ref{tab:msg9.41} & $Z_{2}$ & $\varnothing$\\
\hline
&&$P_{c}c$&\#7.28&IV&OG~\ref{sup:sg6}.7.31&\\
&&$P_{C}c$&\#7.30&IV&OG~\ref{sup:sg9}.4.48&\\
&&$P_{A}c$&\#7.31&IV&OG~\ref{sup:sg8}.7.44&\\
\end{longtable}
\input{src_tex_mtable/msg7.24.tex}
\input{src_tex_mtable/msg7.25.tex}
\input{src_tex_mtable/msg7.26.tex}
\input{src_tex_mtable/msg7.27.tex}
\input{src_tex_mtable/msg7.29.tex}
\input{src_tex_mtable/msg9.41.tex}
\subsection{SG 8}  \label{sup:sg8}
TABLE SG8. The crystalline space group is what the crystal has if the magnetic order is neglected. Once condidering magnetic order, the MSGs, magnetic type, and the symmetry-indicator classifications are given below. For each MSG, the detailed information is given in the corresponding MSG table. \hyperref[tabindex]{See the list of tables.}
\setcounter{table}{36}
\begin{longtable}{p{1.5cm}p{2cm}|p{1.5cm}p{2cm}|c|c|c|c}
\hline
\hline
   \multicolumn{2}{c|}{OG setting} & \multicolumn{2}{c|}{BNS setting} & MSG type & Detailed Inf. & integer spin & half-integer spin\\
\hline
\endfirsthead
\multicolumn{8}{c} {{ TABLE SG\ref{sup:sg8} -- continued}}\\
\hline
   \multicolumn{2}{c|}{OG setting} & \multicolumn{2}{c|}{BNS setting} & MSG type & Detailed Inf. & integer spin & half-integer spin\\
\hline
\endhead
\hline
\endfoot
\hline
\hline
\endlastfoot
8.1.\rownumber & $Cm$ & $Cm$ & \#8.32 & I & Table~\ref{tab:msg8.32} & $\varnothing$ & $\varnothing$\\
8.2.\rownumber & $Cm1'$ & $Cm1'$ & \#8.33 & II & Table~\ref{tab:msg8.33} & $\varnothing$ & $\varnothing$\\
8.3.\rownumber & $Cm'$ & $Cm'$ & \#8.34 & III & Table~\ref{tab:msg8.34} & $\varnothing$ & $\varnothing$\\
8.4.\rownumber & $C_{2c}m$ & $C_{c}m$ & \#8.35 & IV & Table~\ref{tab:msg8.35} & $\varnothing$ & $\varnothing$\\
8.5.\rownumber & $C_{P}m$ & $P_{C}m$ & \#6.23 & IV & Table~\ref{tab:msg6.23} & $\varnothing$ & $\varnothing$\\
8.6.\rownumber & $C_{2c}m'$ & $C_{c}c$ & \#9.40 & IV & Table~\ref{tab:msg9.40} & $\varnothing$ & $\varnothing$\\
8.7.\rownumber & $C_{P}m'$ & $P_{A}c$ & \#7.31 & IV & Table~\ref{tab:msg7.31} & $\varnothing$ & $\varnothing$\\
\hline
&&$C_{a}m$&\#8.36&IV&OG~\ref{sup:sg6}.6.30&\\
\end{longtable}
\input{src_tex_mtable/msg8.32.tex}
\input{src_tex_mtable/msg8.33.tex}
\input{src_tex_mtable/msg8.34.tex}
\input{src_tex_mtable/msg8.35.tex}
\input{src_tex_mtable/msg6.23.tex}
\input{src_tex_mtable/msg9.40.tex}
\input{src_tex_mtable/msg7.31.tex}
\subsection{SG 9}  \label{sup:sg9}
TABLE SG9. The crystalline space group is what the crystal has if the magnetic order is neglected. Once condidering magnetic order, the MSGs, magnetic type, and the symmetry-indicator classifications are given below. For each MSG, the detailed information is given in the corresponding MSG table. \hyperref[tabindex]{See the list of tables.}
\setcounter{table}{43}
\begin{longtable}{p{1.5cm}p{2cm}|p{1.5cm}p{2cm}|c|c|c|c}
\hline
\hline
   \multicolumn{2}{c|}{OG setting} & \multicolumn{2}{c|}{BNS setting} & MSG type & Detailed Inf. & integer spin & half-integer spin\\
\hline
\endfirsthead
\multicolumn{8}{c} {{ TABLE SG\ref{sup:sg9} -- continued}}\\
\hline
   \multicolumn{2}{c|}{OG setting} & \multicolumn{2}{c|}{BNS setting} & MSG type & Detailed Inf. & integer spin & half-integer spin\\
\hline
\endhead
\hline
\endfoot
\hline
\hline
\endlastfoot
9.1.\rownumber & $Cc$ & $Cc$ & \#9.37 & I & Table~\ref{tab:msg9.37} & $\varnothing$ & $\varnothing$\\
9.2.\rownumber & $Cc1'$ & $Cc1'$ & \#9.38 & II & Table~\ref{tab:msg9.38} & $\varnothing$ & $\varnothing$\\
9.3.\rownumber & $Cc'$ & $Cc'$ & \#9.39 & III & Table~\ref{tab:msg9.39} & $\varnothing$ & $\varnothing$\\
9.4.\rownumber & $C_{P}c$ & $P_{C}c$ & \#7.30 & IV & Table~\ref{tab:msg7.30} & $\varnothing$ & $\varnothing$\\
\hline
&&$C_{c}c$&\#9.40&IV&OG~\ref{sup:sg8}.6.43&\\
&&$C_{a}c$&\#9.41&IV&OG~\ref{sup:sg7}.6.37&\\
\end{longtable}
\input{src_tex_mtable/msg9.37.tex}
\input{src_tex_mtable/msg9.38.tex}
\input{src_tex_mtable/msg9.39.tex}
\input{src_tex_mtable/msg7.30.tex}
\subsection{SG 10}  \label{sup:sg10}
TABLE SG10. The crystalline space group is what the crystal has if the magnetic order is neglected. Once condidering magnetic order, the MSGs, magnetic type, and the symmetry-indicator classifications are given below. For each MSG, the detailed information is given in the corresponding MSG table. \hyperref[tabindex]{See the list of tables.}
\setcounter{table}{47}
\begin{longtable}{p{1.5cm}p{2cm}|p{1.5cm}p{2cm}|c|c|c|c}
\hline
\hline
   \multicolumn{2}{c|}{OG setting} & \multicolumn{2}{c|}{BNS setting} & MSG type & Detailed Inf. & integer spin & half-integer spin\\
\hline
\endfirsthead
\multicolumn{8}{c} {{ TABLE SG\ref{sup:sg10} -- continued}}\\
\hline
   \multicolumn{2}{c|}{OG setting} & \multicolumn{2}{c|}{BNS setting} & MSG type & Detailed Inf. & integer spin & half-integer spin\\
\hline
\endhead
\hline
\endfoot
\hline
\hline
\endlastfoot
10.1.\rownumber & $P2/m$ & $P2/m$ & \#10.42 & I & Table~\ref{tab:msg10.42} & $Z_{2}$$\times$$Z_{2}$$\times$$Z_{2}$ & $Z_{2}$$\times$$Z_{2}$$\times$$Z_{2}$\\
10.2.\rownumber & $P2/m1'$ & $P2/m1'$ & \#10.43 & II & Table~\ref{tab:msg10.43} & $Z_{2}$$\times$$Z_{2}$$\times$$Z_{2}$ & $Z_{2}$$\times$$Z_{2}$$\times$$Z_{2}$$\times$$Z_{4}$\\
10.3.\rownumber & $P2'/m$ & $P2'/m$ & \#10.44 & III & Table~\ref{tab:msg10.44} & $\varnothing$ & $\varnothing$\\
10.4.\rownumber & $P2/m'$ & $P2/m'$ & \#10.45 & III & Table~\ref{tab:msg10.45} & $Z_{2}$ & $\varnothing$\\
10.5.\rownumber & $P2'/m'$ & $P2'/m'$ & \#10.46 & III & Table~\ref{tab:msg10.46} & $Z_{2}$$\times$$Z_{2}$$\times$$Z_{2}$$\times$$Z_{4}$ & $Z_{2}$$\times$$Z_{2}$$\times$$Z_{2}$$\times$$Z_{4}$\\
10.6.\rownumber & $P_{2a}2/m$ & $P_{a}2/m$ & \#10.47 & IV & Table~\ref{tab:msg10.47} & $Z_{2}$$\times$$Z_{2}$$\times$$Z_{2}$ & $Z_{2}$$\times$$Z_{2}$\\
10.7.\rownumber & $P_{2b}2/m$ & $P_{b}2/m$ & \#10.48 & IV & Table~\ref{tab:msg10.48} & $Z_{2}$$\times$$Z_{2}$ & $Z_{2}$\\
10.8.\rownumber & $P_{C}2/m$ & $C_{a}2/m$ & \#12.64 & IV & Table~\ref{tab:msg12.64} & $Z_{2}$$\times$$Z_{2}$ & $Z_{2}$\\
10.9.\rownumber & $P_{2b}2'/m$ & $P_{b}2_{1}/m$ & \#11.56 & IV & Table~\ref{tab:msg11.56} & $Z_{2}$ & $Z_{2}$\\
10.10.\rownumber & $P_{2c}2/m'$ & $P_{c}2/c$ & \#13.72 & IV & Table~\ref{tab:msg13.72} & $Z_{2}$$\times$$Z_{2}$ & $Z_{2}$\\
\hline
&&$P_{C}2/m$&\#10.49&IV&OG~\ref{sup:sg12}.7.72&\\
\end{longtable}
\input{src_tex_mtable/msg10.42.tex}
\input{src_tex_mtable/msg10.43.tex}
\input{src_tex_mtable/msg10.44.tex}
\input{src_tex_mtable/msg10.45.tex}
\input{src_tex_mtable/msg10.46.tex}
\input{src_tex_mtable/msg10.47.tex}
\input{src_tex_mtable/msg10.48.tex}
\input{src_tex_mtable/msg12.64.tex}
\input{src_tex_mtable/msg11.56.tex}
\input{src_tex_mtable/msg13.72.tex}
\subsection{SG 11}  \label{sup:sg11}
TABLE SG11. The crystalline space group is what the crystal has if the magnetic order is neglected. Once condidering magnetic order, the MSGs, magnetic type, and the symmetry-indicator classifications are given below. For each MSG, the detailed information is given in the corresponding MSG table. \hyperref[tabindex]{See the list of tables.}
\setcounter{table}{57}
\begin{longtable}{p{1.5cm}p{2cm}|p{1.5cm}p{2cm}|c|c|c|c}
\hline
\hline
   \multicolumn{2}{c|}{OG setting} & \multicolumn{2}{c|}{BNS setting} & MSG type & Detailed Inf. & integer spin & half-integer spin\\
\hline
\endfirsthead
\multicolumn{8}{c} {{ TABLE SG\ref{sup:sg11} -- continued}}\\
\hline
   \multicolumn{2}{c|}{OG setting} & \multicolumn{2}{c|}{BNS setting} & MSG type & Detailed Inf. & integer spin & half-integer spin\\
\hline
\endhead
\hline
\endfoot
\hline
\hline
\endlastfoot
11.1.\rownumber & $P2_{1}/m$ & $P2_{1}/m$ & \#11.50 & I & Table~\ref{tab:msg11.50} & $Z_{2}$ & $Z_{2}$\\
11.2.\rownumber & $P2_{1}/m1'$ & $P2_{1}/m1'$ & \#11.51 & II & Table~\ref{tab:msg11.51} & $Z_{2}$ & $Z_{2}$$\times$$Z_{2}$$\times$$Z_{4}$\\
11.3.\rownumber & $P2_{1}'/m$ & $P2_{1}'/m$ & \#11.52 & III & Table~\ref{tab:msg11.52} & $\varnothing$ & $\varnothing$\\
11.4.\rownumber & $P2_{1}/m'$ & $P2_{1}/m'$ & \#11.53 & III & Table~\ref{tab:msg11.53} & $\varnothing$ & $\varnothing$\\
11.5.\rownumber & $P2_{1}'/m'$ & $P2_{1}'/m'$ & \#11.54 & III & Table~\ref{tab:msg11.54} & $Z_{2}$$\times$$Z_{2}$$\times$$Z_{4}$ & $Z_{2}$$\times$$Z_{2}$$\times$$Z_{4}$\\
11.6.\rownumber & $P_{2a}2_{1}/m$ & $P_{a}2_{1}/m$ & \#11.55 & IV & Table~\ref{tab:msg11.55} & $Z_{2}$ & $Z_{2}$\\
11.7.\rownumber & $P_{2c}2_{1}/m'$ & $P_{c}2_{1}/c$ & \#14.82 & IV & Table~\ref{tab:msg14.82} & $Z_{2}$ & $Z_{2}$\\
\hline
&&$P_{b}2_{1}/m$&\#11.56&IV&OG~\ref{sup:sg10}.9.57&\\
&&$P_{C}2_{1}/m$&\#11.57&IV&OG~\ref{sup:sg12}.9.74&\\
\end{longtable}
\input{src_tex_mtable/msg11.50.tex}
\input{src_tex_mtable/msg11.51.tex}
\input{src_tex_mtable/msg11.52.tex}
\input{src_tex_mtable/msg11.53.tex}
\input{src_tex_mtable/msg11.54.tex}
\input{src_tex_mtable/msg11.55.tex}
\input{src_tex_mtable/msg14.82.tex}
\subsection{SG 12}  \label{sup:sg12}
TABLE SG12. The crystalline space group is what the crystal has if the magnetic order is neglected. Once condidering magnetic order, the MSGs, magnetic type, and the symmetry-indicator classifications are given below. For each MSG, the detailed information is given in the corresponding MSG table. \hyperref[tabindex]{See the list of tables.}
\setcounter{table}{64}
\begin{longtable}{p{1.5cm}p{2cm}|p{1.5cm}p{2cm}|c|c|c|c}
\hline
\hline
   \multicolumn{2}{c|}{OG setting} & \multicolumn{2}{c|}{BNS setting} & MSG type & Detailed Inf. & integer spin & half-integer spin\\
\hline
\endfirsthead
\multicolumn{8}{c} {{ TABLE SG\ref{sup:sg12} -- continued}}\\
\hline
   \multicolumn{2}{c|}{OG setting} & \multicolumn{2}{c|}{BNS setting} & MSG type & Detailed Inf. & integer spin & half-integer spin\\
\hline
\endhead
\hline
\endfoot
\hline
\hline
\endlastfoot
12.1.\rownumber & $C2/m$ & $C2/m$ & \#12.58 & I & Table~\ref{tab:msg12.58} & $Z_{2}$$\times$$Z_{2}$ & $Z_{2}$$\times$$Z_{2}$\\
12.2.\rownumber & $C2/m1'$ & $C2/m1'$ & \#12.59 & II & Table~\ref{tab:msg12.59} & $Z_{2}$$\times$$Z_{2}$ & $Z_{2}$$\times$$Z_{2}$$\times$$Z_{4}$\\
12.3.\rownumber & $C2'/m$ & $C2'/m$ & \#12.60 & III & Table~\ref{tab:msg12.60} & $\varnothing$ & $\varnothing$\\
12.4.\rownumber & $C2/m'$ & $C2/m'$ & \#12.61 & III & Table~\ref{tab:msg12.61} & $\varnothing$ & $\varnothing$\\
12.5.\rownumber & $C2'/m'$ & $C2'/m'$ & \#12.62 & III & Table~\ref{tab:msg12.62} & $Z_{2}$$\times$$Z_{2}$$\times$$Z_{4}$ & $Z_{2}$$\times$$Z_{2}$$\times$$Z_{4}$\\
12.6.\rownumber & $C_{2c}2/m$ & $C_{c}2/m$ & \#12.63 & IV & Table~\ref{tab:msg12.63} & $Z_{2}$$\times$$Z_{2}$ & $Z_{2}$\\
12.7.\rownumber & $C_{P}2/m$ & $P_{C}2/m$ & \#10.49 & IV & Table~\ref{tab:msg10.49} & $Z_{2}$$\times$$Z_{2}$ & $Z_{2}$\\
12.8.\rownumber & $C_{2c}2/m'$ & $C_{c}2/c$ & \#15.90 & IV & Table~\ref{tab:msg15.90} & $Z_{2}$$\times$$Z_{2}$ & $Z_{2}$\\
12.9.\rownumber & $C_{P}2'/m$ & $P_{C}2_{1}/m$ & \#11.57 & IV & Table~\ref{tab:msg11.57} & $Z_{2}$ & $Z_{2}$\\
12.10.\rownumber & $C_{P}2/m'$ & $P_{A}2/c$ & \#13.73 & IV & Table~\ref{tab:msg13.73} & $Z_{2}$ & $Z_{2}$\\
12.11.\rownumber & $C_{P}2'/m'$ & $P_{A}2_{1}/c$ & \#14.83 & IV & Table~\ref{tab:msg14.83} & $Z_{2}$ & $Z_{2}$\\
\hline
&&$C_{a}2/m$&\#12.64&IV&OG~\ref{sup:sg10}.8.56&\\
\end{longtable}
\input{src_tex_mtable/msg12.58.tex}
\input{src_tex_mtable/msg12.59.tex}
\input{src_tex_mtable/msg12.60.tex}
\input{src_tex_mtable/msg12.61.tex}
\input{src_tex_mtable/msg12.62.tex}
\input{src_tex_mtable/msg12.63.tex}
\input{src_tex_mtable/msg10.49.tex}
\input{src_tex_mtable/msg15.90.tex}
\input{src_tex_mtable/msg11.57.tex}
\input{src_tex_mtable/msg13.73.tex}
\input{src_tex_mtable/msg14.83.tex}
\subsection{SG 13}  \label{sup:sg13}
TABLE SG13. The crystalline space group is what the crystal has if the magnetic order is neglected. Once condidering magnetic order, the MSGs, magnetic type, and the symmetry-indicator classifications are given below. For each MSG, the detailed information is given in the corresponding MSG table. \hyperref[tabindex]{See the list of tables.}
\setcounter{table}{75}
\begin{longtable}{p{1.5cm}p{2cm}|p{1.5cm}p{2cm}|c|c|c|c}
\hline
\hline
   \multicolumn{2}{c|}{OG setting} & \multicolumn{2}{c|}{BNS setting} & MSG type & Detailed Inf. & integer spin & half-integer spin\\
\hline
\endfirsthead
\multicolumn{8}{c} {{ TABLE SG\ref{sup:sg13} -- continued}}\\
\hline
   \multicolumn{2}{c|}{OG setting} & \multicolumn{2}{c|}{BNS setting} & MSG type & Detailed Inf. & integer spin & half-integer spin\\
\hline
\endhead
\hline
\endfoot
\hline
\hline
\endlastfoot
13.1.\rownumber & $P2/c$ & $P2/c$ & \#13.65 & I & Table~\ref{tab:msg13.65} & $Z_{2}$$\times$$Z_{2}$ & $Z_{2}$$\times$$Z_{2}$\\
13.2.\rownumber & $P2/c1'$ & $P2/c1'$ & \#13.66 & II & Table~\ref{tab:msg13.66} & $Z_{2}$$\times$$Z_{2}$ & $Z_{2}$$\times$$Z_{2}$$\times$$Z_{4}$\\
13.3.\rownumber & $P2'/c$ & $P2'/c$ & \#13.67 & III & Table~\ref{tab:msg13.67} & $\varnothing$ & $\varnothing$\\
13.4.\rownumber & $P2/c'$ & $P2/c'$ & \#13.68 & III & Table~\ref{tab:msg13.68} & $Z_{2}$ & $\varnothing$\\
13.5.\rownumber & $P2'/c'$ & $P2'/c'$ & \#13.69 & III & Table~\ref{tab:msg13.69} & $Z_{2}$$\times$$Z_{2}$$\times$$Z_{4}$ & $Z_{2}$$\times$$Z_{2}$$\times$$Z_{4}$\\
13.6.\rownumber & $P_{2a}2/c$ & $P_{a}2/c$ & \#13.70 & IV & Table~\ref{tab:msg13.70} & $Z_{2}$$\times$$Z_{2}$ & $Z_{2}$\\
13.7.\rownumber & $P_{2b}2/c$ & $P_{b}2/c$ & \#13.71 & IV & Table~\ref{tab:msg13.71} & $Z_{2}$$\times$$Z_{2}$ & $Z_{2}$\\
13.8.\rownumber & $P_{C}2/c$ & $C_{a}2/c$ & \#15.91 & IV & Table~\ref{tab:msg15.91} & $Z_{2}$$\times$$Z_{2}$$\times$$Z_{2}$ & $Z_{2}$\\
13.9.\rownumber & $P_{2b}2'/c$ & $P_{b}2_{1}/c$ & \#14.81 & IV & Table~\ref{tab:msg14.81} & $Z_{2}$ & $Z_{2}$\\
\hline
&&$P_{c}2/c$&\#13.72&IV&OG~\ref{sup:sg10}.10.58&\\
&&$P_{A}2/c$&\#13.73&IV&OG~\ref{sup:sg12}.10.75&\\
&&$P_{C}2/c$&\#13.74&IV&OG~\ref{sup:sg15}.6.97&\\
\end{longtable}
\input{src_tex_mtable/msg13.65.tex}
\input{src_tex_mtable/msg13.66.tex}
\input{src_tex_mtable/msg13.67.tex}
\input{src_tex_mtable/msg13.68.tex}
\input{src_tex_mtable/msg13.69.tex}
\input{src_tex_mtable/msg13.70.tex}
\input{src_tex_mtable/msg13.71.tex}
\input{src_tex_mtable/msg15.91.tex}
\input{src_tex_mtable/msg14.81.tex}
\subsection{SG 14}  \label{sup:sg14}
TABLE SG14. The crystalline space group is what the crystal has if the magnetic order is neglected. Once condidering magnetic order, the MSGs, magnetic type, and the symmetry-indicator classifications are given below. For each MSG, the detailed information is given in the corresponding MSG table. \hyperref[tabindex]{See the list of tables.}
\setcounter{table}{84}
\begin{longtable}{p{1.5cm}p{2cm}|p{1.5cm}p{2cm}|c|c|c|c}
\hline
\hline
   \multicolumn{2}{c|}{OG setting} & \multicolumn{2}{c|}{BNS setting} & MSG type & Detailed Inf. & integer spin & half-integer spin\\
\hline
\endfirsthead
\multicolumn{8}{c} {{ TABLE SG\ref{sup:sg14} -- continued}}\\
\hline
   \multicolumn{2}{c|}{OG setting} & \multicolumn{2}{c|}{BNS setting} & MSG type & Detailed Inf. & integer spin & half-integer spin\\
\hline
\endhead
\hline
\endfoot
\hline
\hline
\endlastfoot
14.1.\rownumber & $P2_{1}/c$ & $P2_{1}/c$ & \#14.75 & I & Table~\ref{tab:msg14.75} & $Z_{2}$ & $Z_{2}$\\
14.2.\rownumber & $P2_{1}/c1'$ & $P2_{1}/c1'$ & \#14.76 & II & Table~\ref{tab:msg14.76} & $Z_{2}$ & $Z_{2}$$\times$$Z_{4}$\\
14.3.\rownumber & $P2_{1}'/c$ & $P2_{1}'/c$ & \#14.77 & III & Table~\ref{tab:msg14.77} & $\varnothing$ & $\varnothing$\\
14.4.\rownumber & $P2_{1}/c'$ & $P2_{1}/c'$ & \#14.78 & III & Table~\ref{tab:msg14.78} & $\varnothing$ & $\varnothing$\\
14.5.\rownumber & $P2_{1}'/c'$ & $P2_{1}'/c'$ & \#14.79 & III & Table~\ref{tab:msg14.79} & $Z_{2}$$\times$$Z_{4}$ & $Z_{2}$$\times$$Z_{4}$\\
14.6.\rownumber & $P_{2a}2_{1}/c$ & $P_{a}2_{1}/c$ & \#14.80 & IV & Table~\ref{tab:msg14.80} & $Z_{2}$ & $Z_{2}$\\
\hline
&&$P_{b}2_{1}/c$&\#14.81&IV&OG~\ref{sup:sg13}.9.85&\\
&&$P_{c}2_{1}/c$&\#14.82&IV&OG~\ref{sup:sg11}.7.65&\\
&&$P_{A}2_{1}/c$&\#14.83&IV&OG~\ref{sup:sg12}.11.76&\\
&&$P_{C}2_{1}/c$&\#14.84&IV&OG~\ref{sup:sg15}.7.98&\\
\end{longtable}
\input{src_tex_mtable/msg14.75.tex}
\input{src_tex_mtable/msg14.76.tex}
\input{src_tex_mtable/msg14.77.tex}
\input{src_tex_mtable/msg14.78.tex}
\input{src_tex_mtable/msg14.79.tex}
\input{src_tex_mtable/msg14.80.tex}
\subsection{SG 15}  \label{sup:sg15}
TABLE SG15. The crystalline space group is what the crystal has if the magnetic order is neglected. Once condidering magnetic order, the MSGs, magnetic type, and the symmetry-indicator classifications are given below. For each MSG, the detailed information is given in the corresponding MSG table. \hyperref[tabindex]{See the list of tables.}
\setcounter{table}{90}
\begin{longtable}{p{1.5cm}p{2cm}|p{1.5cm}p{2cm}|c|c|c|c}
\hline
\hline
   \multicolumn{2}{c|}{OG setting} & \multicolumn{2}{c|}{BNS setting} & MSG type & Detailed Inf. & integer spin & half-integer spin\\
\hline
\endfirsthead
\multicolumn{8}{c} {{ TABLE SG\ref{sup:sg15} -- continued}}\\
\hline
   \multicolumn{2}{c|}{OG setting} & \multicolumn{2}{c|}{BNS setting} & MSG type & Detailed Inf. & integer spin & half-integer spin\\
\hline
\endhead
\hline
\endfoot
\hline
\hline
\endlastfoot
15.1.\rownumber & $C2/c$ & $C2/c$ & \#15.85 & I & Table~\ref{tab:msg15.85} & $Z_{2}$$\times$$Z_{2}$ & $Z_{2}$$\times$$Z_{2}$\\
15.2.\rownumber & $C2/c1'$ & $C2/c1'$ & \#15.86 & II & Table~\ref{tab:msg15.86} & $Z_{2}$$\times$$Z_{2}$ & $Z_{2}$$\times$$Z_{4}$\\
15.3.\rownumber & $C2'/c$ & $C2'/c$ & \#15.87 & III & Table~\ref{tab:msg15.87} & $\varnothing$ & $\varnothing$\\
15.4.\rownumber & $C2/c'$ & $C2/c'$ & \#15.88 & III & Table~\ref{tab:msg15.88} & $\varnothing$ & $\varnothing$\\
15.5.\rownumber & $C2'/c'$ & $C2'/c'$ & \#15.89 & III & Table~\ref{tab:msg15.89} & $Z_{2}$$\times$$Z_{4}$ & $Z_{2}$$\times$$Z_{4}$\\
15.6.\rownumber & $C_{P}2/c$ & $P_{C}2/c$ & \#13.74 & IV & Table~\ref{tab:msg13.74} & $Z_{2}$ & $Z_{2}$\\
15.7.\rownumber & $C_{P}2'/c$ & $P_{C}2_{1}/c$ & \#14.84 & IV & Table~\ref{tab:msg14.84} & $Z_{2}$ & $Z_{2}$\\
\hline
&&$C_{c}2/c$&\#15.90&IV&OG~\ref{sup:sg12}.8.73&\\
&&$C_{a}2/c$&\#15.91&IV&OG~\ref{sup:sg13}.8.84&\\
\end{longtable}
\input{src_tex_mtable/msg15.85.tex}
\input{src_tex_mtable/msg15.86.tex}
\input{src_tex_mtable/msg15.87.tex}
\input{src_tex_mtable/msg15.88.tex}
\input{src_tex_mtable/msg15.89.tex}
\input{src_tex_mtable/msg13.74.tex}
\input{src_tex_mtable/msg14.84.tex}
\subsection{SG 16}  \label{sup:sg16}
TABLE SG16. The crystalline space group is what the crystal has if the magnetic order is neglected. Once condidering magnetic order, the MSGs, magnetic type, and the symmetry-indicator classifications are given below. For each MSG, the detailed information is given in the corresponding MSG table. \hyperref[tabindex]{See the list of tables.}
\setcounter{table}{97}
\begin{longtable}{p{1.5cm}p{2cm}|p{1.5cm}p{2cm}|c|c|c|c}
\hline
\hline
   \multicolumn{2}{c|}{OG setting} & \multicolumn{2}{c|}{BNS setting} & MSG type & Detailed Inf. & integer spin & half-integer spin\\
\hline
\endfirsthead
\multicolumn{8}{c} {{ TABLE SG\ref{sup:sg16} -- continued}}\\
\hline
   \multicolumn{2}{c|}{OG setting} & \multicolumn{2}{c|}{BNS setting} & MSG type & Detailed Inf. & integer spin & half-integer spin\\
\hline
\endhead
\hline
\endfoot
\hline
\hline
\endlastfoot
16.1.\rownumber & $P222$ & $P222$ & \#16.1 & I & Table~\ref{tab:msg16.1} & $\varnothing$ & $\varnothing$\\
16.2.\rownumber & $P2221'$ & $P2221'$ & \#16.2 & II & Table~\ref{tab:msg16.2} & $\varnothing$ & $\varnothing$\\
16.3.\rownumber & $P2'2'2$ & $P2'2'2$ & \#16.3 & III & Table~\ref{tab:msg16.3} & $Z_{2}$ & $Z_{2}$\\
16.4.\rownumber & $P_{2a}222$ & $P_{a}222$ & \#16.4 & IV & Table~\ref{tab:msg16.4} & $Z_{2}$ & $\varnothing$\\
16.5.\rownumber & $P_{C}222$ & $C_{a}222$ & \#21.43 & IV & Table~\ref{tab:msg21.43} & $Z_{2}$ & $\varnothing$\\
16.6.\rownumber & $P_{I}222$ & $F_{S}222$ & \#22.48 & IV & Table~\ref{tab:msg22.48} & $Z_{2}$ & $\varnothing$\\
16.7.\rownumber & $P_{2c}22'2'$ & $P_{c}222_{1}$ & \#17.12 & IV & Table~\ref{tab:msg17.12} & $\varnothing$ & $\varnothing$\\
\hline
&&$P_{C}222$&\#16.5&IV&OG~\ref{sup:sg21}.6.134&\\
&&$P_{I}222$&\#16.6&IV&OG~\ref{sup:sg23}.4.148&\\
\end{longtable}
\input{src_tex_mtable/msg16.1.tex}
\input{src_tex_mtable/msg16.2.tex}
\input{src_tex_mtable/msg16.3.tex}
\input{src_tex_mtable/msg16.4.tex}
\input{src_tex_mtable/msg21.43.tex}
\input{src_tex_mtable/msg22.48.tex}
\input{src_tex_mtable/msg17.12.tex}
\subsection{SG 17}  \label{sup:sg17}
TABLE SG17. The crystalline space group is what the crystal has if the magnetic order is neglected. Once condidering magnetic order, the MSGs, magnetic type, and the symmetry-indicator classifications are given below. For each MSG, the detailed information is given in the corresponding MSG table. \hyperref[tabindex]{See the list of tables.}
\setcounter{table}{104}
\begin{longtable}{p{1.5cm}p{2cm}|p{1.5cm}p{2cm}|c|c|c|c}
\hline
\hline
   \multicolumn{2}{c|}{OG setting} & \multicolumn{2}{c|}{BNS setting} & MSG type & Detailed Inf. & integer spin & half-integer spin\\
\hline
\endfirsthead
\multicolumn{8}{c} {{ TABLE SG\ref{sup:sg17} -- continued}}\\
\hline
   \multicolumn{2}{c|}{OG setting} & \multicolumn{2}{c|}{BNS setting} & MSG type & Detailed Inf. & integer spin & half-integer spin\\
\hline
\endhead
\hline
\endfoot
\hline
\hline
\endlastfoot
17.1.\rownumber & $P222_{1}$ & $P222_{1}$ & \#17.7 & I & Table~\ref{tab:msg17.7} & $\varnothing$ & $\varnothing$\\
17.2.\rownumber & $P222_{1}1'$ & $P222_{1}1'$ & \#17.8 & II & Table~\ref{tab:msg17.8} & $\varnothing$ & $\varnothing$\\
17.3.\rownumber & $P2'2'2_{1}$ & $P2'2'2_{1}$ & \#17.9 & III & Table~\ref{tab:msg17.9} & $\varnothing$ & $\varnothing$\\
17.4.\rownumber & $P22'2_{1}'$ & $P22'2_{1}'$ & \#17.10 & III & Table~\ref{tab:msg17.10} & $Z_{2}$ & $Z_{2}$\\
17.5.\rownumber & $P_{2a}222_{1}$ & $P_{a}222_{1}$ & \#17.11 & IV & Table~\ref{tab:msg17.11} & $Z_{2}$ & $\varnothing$\\
17.6.\rownumber & $P_{C}222_{1}$ & $C_{a}222_{1}$ & \#20.36 & IV & Table~\ref{tab:msg20.36} & $Z_{2}$ & $\varnothing$\\
17.7.\rownumber & $P_{2a}2'2'2_{1}$ & $P_{b}2_{1}2_{1}2$ & \#18.20 & IV & Table~\ref{tab:msg18.20} & $\varnothing$ & $\varnothing$\\
\hline
&&$P_{c}222_{1}$&\#17.12&IV&OG~\ref{sup:sg16}.7.105&\\
&&$P_{B}222_{1}$&\#17.13&IV&OG~\ref{sup:sg21}.10.138&\\
&&$P_{C}222_{1}$&\#17.14&IV&OG~\ref{sup:sg20}.5.126&\\
&&$P_{I}222_{1}$&\#17.15&IV&OG~\ref{sup:sg24}.5.154&\\
\end{longtable}
\input{src_tex_mtable/msg17.7.tex}
\input{src_tex_mtable/msg17.8.tex}
\input{src_tex_mtable/msg17.9.tex}
\input{src_tex_mtable/msg17.10.tex}
\input{src_tex_mtable/msg17.11.tex}
\input{src_tex_mtable/msg20.36.tex}
\input{src_tex_mtable/msg18.20.tex}
\subsection{SG 18}  \label{sup:sg18}
TABLE SG18. The crystalline space group is what the crystal has if the magnetic order is neglected. Once condidering magnetic order, the MSGs, magnetic type, and the symmetry-indicator classifications are given below. For each MSG, the detailed information is given in the corresponding MSG table. \hyperref[tabindex]{See the list of tables.}
\setcounter{table}{111}
\begin{longtable}{p{1.5cm}p{2cm}|p{1.5cm}p{2cm}|c|c|c|c}
\hline
\hline
   \multicolumn{2}{c|}{OG setting} & \multicolumn{2}{c|}{BNS setting} & MSG type & Detailed Inf. & integer spin & half-integer spin\\
\hline
\endfirsthead
\multicolumn{8}{c} {{ TABLE SG\ref{sup:sg18} -- continued}}\\
\hline
   \multicolumn{2}{c|}{OG setting} & \multicolumn{2}{c|}{BNS setting} & MSG type & Detailed Inf. & integer spin & half-integer spin\\
\hline
\endhead
\hline
\endfoot
\hline
\hline
\endlastfoot
18.1.\rownumber & $P2_{1}2_{1}2$ & $P2_{1}2_{1}2$ & \#18.16 & I & Table~\ref{tab:msg18.16} & $\varnothing$ & $\varnothing$\\
18.2.\rownumber & $P2_{1}2_{1}21'$ & $P2_{1}2_{1}21'$ & \#18.17 & II & Table~\ref{tab:msg18.17} & $\varnothing$ & $\varnothing$\\
18.3.\rownumber & $P2_{1}'2_{1}'2$ & $P2_{1}'2_{1}'2$ & \#18.18 & III & Table~\ref{tab:msg18.18} & $Z_{2}$ & $Z_{2}$\\
18.4.\rownumber & $P2_{1}2_{1}'2'$ & $P2_{1}2_{1}'2'$ & \#18.19 & III & Table~\ref{tab:msg18.19} & $\varnothing$ & $\varnothing$\\
18.5.\rownumber & $P_{2c}2_{1}2_{1}2$ & $P_{c}2_{1}2_{1}2$ & \#18.21 & IV & Table~\ref{tab:msg18.21} & $Z_{2}$ & $\varnothing$\\
18.6.\rownumber & $P_{2c}2_{1}2_{1}'2'$ & $P_{c}2_{1}2_{1}2_{1}$ & \#19.28 & IV & Table~\ref{tab:msg19.28} & $\varnothing$ & $\varnothing$\\
\hline
&&$P_{b}2_{1}2_{1}2$&\#18.20&IV&OG~\ref{sup:sg17}.7.112&\\
&&$P_{B}2_{1}2_{1}2$&\#18.22&IV&OG~\ref{sup:sg20}.7.128&\\
&&$P_{C}2_{1}2_{1}2$&\#18.23&IV&OG~\ref{sup:sg21}.9.137&\\
&&$P_{I}2_{1}2_{1}2$&\#18.24&IV&OG~\ref{sup:sg23}.5.149&\\
\end{longtable}
\input{src_tex_mtable/msg18.16.tex}
\input{src_tex_mtable/msg18.17.tex}
\input{src_tex_mtable/msg18.18.tex}
\input{src_tex_mtable/msg18.19.tex}
\input{src_tex_mtable/msg18.21.tex}
\input{src_tex_mtable/msg19.28.tex}
\subsection{SG 19}  \label{sup:sg19}
TABLE SG19. The crystalline space group is what the crystal has if the magnetic order is neglected. Once condidering magnetic order, the MSGs, magnetic type, and the symmetry-indicator classifications are given below. For each MSG, the detailed information is given in the corresponding MSG table. \hyperref[tabindex]{See the list of tables.}
\setcounter{table}{117}
\begin{longtable}{p{1.5cm}p{2cm}|p{1.5cm}p{2cm}|c|c|c|c}
\hline
\hline
   \multicolumn{2}{c|}{OG setting} & \multicolumn{2}{c|}{BNS setting} & MSG type & Detailed Inf. & integer spin & half-integer spin\\
\hline
\endfirsthead
\multicolumn{8}{c} {{ TABLE SG\ref{sup:sg19} -- continued}}\\
\hline
   \multicolumn{2}{c|}{OG setting} & \multicolumn{2}{c|}{BNS setting} & MSG type & Detailed Inf. & integer spin & half-integer spin\\
\hline
\endhead
\hline
\endfoot
\hline
\hline
\endlastfoot
19.1.\rownumber & $P2_{1}2_{1}2_{1}$ & $P2_{1}2_{1}2_{1}$ & \#19.25 & I & Table~\ref{tab:msg19.25} & $\varnothing$ & $\varnothing$\\
19.2.\rownumber & $P2_{1}2_{1}2_{1}1'$ & $P2_{1}2_{1}2_{1}1'$ & \#19.26 & II & Table~\ref{tab:msg19.26} & $\varnothing$ & $\varnothing$\\
19.3.\rownumber & $P2_{1}'2_{1}'2_{1}$ & $P2_{1}'2_{1}'2_{1}$ & \#19.27 & III & Table~\ref{tab:msg19.27} & $\varnothing$ & $\varnothing$\\
\hline
&&$P_{c}2_{1}2_{1}2_{1}$&\#19.28&IV&OG~\ref{sup:sg18}.6.118&\\
&&$P_{C}2_{1}2_{1}2_{1}$&\#19.29&IV&OG~\ref{sup:sg20}.6.127&\\
&&$P_{I}2_{1}2_{1}2_{1}$&\#19.30&IV&OG~\ref{sup:sg24}.4.153&\\
\end{longtable}
\input{src_tex_mtable/msg19.25.tex}
\input{src_tex_mtable/msg19.26.tex}
\input{src_tex_mtable/msg19.27.tex}
\subsection{SG 20}  \label{sup:sg20}
TABLE SG20. The crystalline space group is what the crystal has if the magnetic order is neglected. Once condidering magnetic order, the MSGs, magnetic type, and the symmetry-indicator classifications are given below. For each MSG, the detailed information is given in the corresponding MSG table. \hyperref[tabindex]{See the list of tables.}
\setcounter{table}{120}
\begin{longtable}{p{1.5cm}p{2cm}|p{1.5cm}p{2cm}|c|c|c|c}
\hline
\hline
   \multicolumn{2}{c|}{OG setting} & \multicolumn{2}{c|}{BNS setting} & MSG type & Detailed Inf. & integer spin & half-integer spin\\
\hline
\endfirsthead
\multicolumn{8}{c} {{ TABLE SG\ref{sup:sg20} -- continued}}\\
\hline
   \multicolumn{2}{c|}{OG setting} & \multicolumn{2}{c|}{BNS setting} & MSG type & Detailed Inf. & integer spin & half-integer spin\\
\hline
\endhead
\hline
\endfoot
\hline
\hline
\endlastfoot
20.1.\rownumber & $C222_{1}$ & $C222_{1}$ & \#20.31 & I & Table~\ref{tab:msg20.31} & $\varnothing$ & $\varnothing$\\
20.2.\rownumber & $C222_{1}1'$ & $C222_{1}1'$ & \#20.32 & II & Table~\ref{tab:msg20.32} & $\varnothing$ & $\varnothing$\\
20.3.\rownumber & $C2'2'2_{1}$ & $C2'2'2_{1}$ & \#20.33 & III & Table~\ref{tab:msg20.33} & $\varnothing$ & $\varnothing$\\
20.4.\rownumber & $C22'2_{1}'$ & $C22'2_{1}'$ & \#20.34 & III & Table~\ref{tab:msg20.34} & $\varnothing$ & $\varnothing$\\
20.5.\rownumber & $C_{P}222_{1}$ & $P_{C}222_{1}$ & \#17.14 & IV & Table~\ref{tab:msg17.14} & $Z_{2}$ & $\varnothing$\\
20.6.\rownumber & $C_{P}2'2'2_{1}$ & $P_{C}2_{1}2_{1}2_{1}$ & \#19.29 & IV & Table~\ref{tab:msg19.29} & $\varnothing$ & $\varnothing$\\
20.7.\rownumber & $C_{P}22'2_{1}'$ & $P_{B}2_{1}2_{1}2$ & \#18.22 & IV & Table~\ref{tab:msg18.22} & $\varnothing$ & $\varnothing$\\
\hline
&&$C_{c}222_{1}$&\#20.35&IV&OG~\ref{sup:sg21}.8.136&\\
&&$C_{a}222_{1}$&\#20.36&IV&OG~\ref{sup:sg17}.6.111&\\
&&$C_{A}222_{1}$&\#20.37&IV&OG~\ref{sup:sg22}.5.144&\\
\end{longtable}
\input{src_tex_mtable/msg20.31.tex}
\input{src_tex_mtable/msg20.32.tex}
\input{src_tex_mtable/msg20.33.tex}
\input{src_tex_mtable/msg20.34.tex}
\input{src_tex_mtable/msg17.14.tex}
\input{src_tex_mtable/msg19.29.tex}
\input{src_tex_mtable/msg18.22.tex}
\subsection{SG 21}  \label{sup:sg21}
TABLE SG21. The crystalline space group is what the crystal has if the magnetic order is neglected. Once condidering magnetic order, the MSGs, magnetic type, and the symmetry-indicator classifications are given below. For each MSG, the detailed information is given in the corresponding MSG table. \hyperref[tabindex]{See the list of tables.}
\setcounter{table}{127}
\begin{longtable}{p{1.5cm}p{2cm}|p{1.5cm}p{2cm}|c|c|c|c}
\hline
\hline
   \multicolumn{2}{c|}{OG setting} & \multicolumn{2}{c|}{BNS setting} & MSG type & Detailed Inf. & integer spin & half-integer spin\\
\hline
\endfirsthead
\multicolumn{8}{c} {{ TABLE SG\ref{sup:sg21} -- continued}}\\
\hline
   \multicolumn{2}{c|}{OG setting} & \multicolumn{2}{c|}{BNS setting} & MSG type & Detailed Inf. & integer spin & half-integer spin\\
\hline
\endhead
\hline
\endfoot
\hline
\hline
\endlastfoot
21.1.\rownumber & $C222$ & $C222$ & \#21.38 & I & Table~\ref{tab:msg21.38} & $\varnothing$ & $\varnothing$\\
21.2.\rownumber & $C2221'$ & $C2221'$ & \#21.39 & II & Table~\ref{tab:msg21.39} & $\varnothing$ & $\varnothing$\\
21.3.\rownumber & $C2'2'2$ & $C2'2'2$ & \#21.40 & III & Table~\ref{tab:msg21.40} & $Z_{2}$ & $Z_{2}$\\
21.4.\rownumber & $C22'2'$ & $C22'2'$ & \#21.41 & III & Table~\ref{tab:msg21.41} & $\varnothing$ & $\varnothing$\\
21.5.\rownumber & $C_{2c}222$ & $C_{c}222$ & \#21.42 & IV & Table~\ref{tab:msg21.42} & $Z_{2}$ & $\varnothing$\\
21.6.\rownumber & $C_{P}222$ & $P_{C}222$ & \#16.5 & IV & Table~\ref{tab:msg16.5} & $Z_{2}$ & $\varnothing$\\
21.7.\rownumber & $C_{I}222$ & $I_{c}222$ & \#23.52 & IV & Table~\ref{tab:msg23.52} & $Z_{2}$ & $\varnothing$\\
21.8.\rownumber & $C_{2c}22'2'$ & $C_{c}222_{1}$ & \#20.35 & IV & Table~\ref{tab:msg20.35} & $\varnothing$ & $\varnothing$\\
21.9.\rownumber & $C_{P}2'2'2$ & $P_{C}2_{1}2_{1}2$ & \#18.23 & IV & Table~\ref{tab:msg18.23} & $\varnothing$ & $\varnothing$\\
21.10.\rownumber & $C_{P}22'2'$ & $P_{B}222_{1}$ & \#17.13 & IV & Table~\ref{tab:msg17.13} & $\varnothing$ & $\varnothing$\\
21.11.\rownumber & $C_{I}2'22'$ & $I_{c}2_{1}2_{1}2_{1}$ & \#24.56 & IV & Table~\ref{tab:msg24.56} & $Z_{2}$ & $\varnothing$\\
\hline
&&$C_{a}222$&\#21.43&IV&OG~\ref{sup:sg16}.5.103&\\
&&$C_{A}222$&\#21.44&IV&OG~\ref{sup:sg22}.4.143&\\
\end{longtable}
\input{src_tex_mtable/msg21.38.tex}
\input{src_tex_mtable/msg21.39.tex}
\input{src_tex_mtable/msg21.40.tex}
\input{src_tex_mtable/msg21.41.tex}
\input{src_tex_mtable/msg21.42.tex}
\input{src_tex_mtable/msg16.5.tex}
\input{src_tex_mtable/msg23.52.tex}
\input{src_tex_mtable/msg20.35.tex}
\input{src_tex_mtable/msg18.23.tex}
\input{src_tex_mtable/msg17.13.tex}
\input{src_tex_mtable/msg24.56.tex}
\subsection{SG 22}  \label{sup:sg22}
TABLE SG22. The crystalline space group is what the crystal has if the magnetic order is neglected. Once condidering magnetic order, the MSGs, magnetic type, and the symmetry-indicator classifications are given below. For each MSG, the detailed information is given in the corresponding MSG table. \hyperref[tabindex]{See the list of tables.}
\setcounter{table}{138}
\begin{longtable}{p{1.5cm}p{2cm}|p{1.5cm}p{2cm}|c|c|c|c}
\hline
\hline
   \multicolumn{2}{c|}{OG setting} & \multicolumn{2}{c|}{BNS setting} & MSG type & Detailed Inf. & integer spin & half-integer spin\\
\hline
\endfirsthead
\multicolumn{8}{c} {{ TABLE SG\ref{sup:sg22} -- continued}}\\
\hline
   \multicolumn{2}{c|}{OG setting} & \multicolumn{2}{c|}{BNS setting} & MSG type & Detailed Inf. & integer spin & half-integer spin\\
\hline
\endhead
\hline
\endfoot
\hline
\hline
\endlastfoot
22.1.\rownumber & $F222$ & $F222$ & \#22.45 & I & Table~\ref{tab:msg22.45} & $\varnothing$ & $\varnothing$\\
22.2.\rownumber & $F2221'$ & $F2221'$ & \#22.46 & II & Table~\ref{tab:msg22.46} & $\varnothing$ & $\varnothing$\\
22.3.\rownumber & $F2'2'2$ & $F2'2'2$ & \#22.47 & III & Table~\ref{tab:msg22.47} & $\varnothing$ & $\varnothing$\\
22.4.\rownumber & $F_{C}222$ & $C_{A}222$ & \#21.44 & IV & Table~\ref{tab:msg21.44} & $Z_{2}$ & $\varnothing$\\
22.5.\rownumber & $F_{C}22'2'$ & $C_{A}222_{1}$ & \#20.37 & IV & Table~\ref{tab:msg20.37} & $\varnothing$ & $\varnothing$\\
\hline
&&$F_{S}222$&\#22.48&IV&OG~\ref{sup:sg16}.6.104&\\
\end{longtable}
\input{src_tex_mtable/msg22.45.tex}
\input{src_tex_mtable/msg22.46.tex}
\input{src_tex_mtable/msg22.47.tex}
\input{src_tex_mtable/msg21.44.tex}
\input{src_tex_mtable/msg20.37.tex}
\subsection{SG 23}  \label{sup:sg23}
TABLE SG23. The crystalline space group is what the crystal has if the magnetic order is neglected. Once condidering magnetic order, the MSGs, magnetic type, and the symmetry-indicator classifications are given below. For each MSG, the detailed information is given in the corresponding MSG table. \hyperref[tabindex]{See the list of tables.}
\setcounter{table}{143}
\begin{longtable}{p{1.5cm}p{2cm}|p{1.5cm}p{2cm}|c|c|c|c}
\hline
\hline
   \multicolumn{2}{c|}{OG setting} & \multicolumn{2}{c|}{BNS setting} & MSG type & Detailed Inf. & integer spin & half-integer spin\\
\hline
\endfirsthead
\multicolumn{8}{c} {{ TABLE SG\ref{sup:sg23} -- continued}}\\
\hline
   \multicolumn{2}{c|}{OG setting} & \multicolumn{2}{c|}{BNS setting} & MSG type & Detailed Inf. & integer spin & half-integer spin\\
\hline
\endhead
\hline
\endfoot
\hline
\hline
\endlastfoot
23.1.\rownumber & $I222$ & $I222$ & \#23.49 & I & Table~\ref{tab:msg23.49} & $\varnothing$ & $\varnothing$\\
23.2.\rownumber & $I2221'$ & $I2221'$ & \#23.50 & II & Table~\ref{tab:msg23.50} & $\varnothing$ & $\varnothing$\\
23.3.\rownumber & $I2'2'2$ & $I2'2'2$ & \#23.51 & III & Table~\ref{tab:msg23.51} & $\varnothing$ & $\varnothing$\\
23.4.\rownumber & $I_{P}222$ & $P_{I}222$ & \#16.6 & IV & Table~\ref{tab:msg16.6} & $Z_{2}$ & $\varnothing$\\
23.5.\rownumber & $I_{P}2'2'2$ & $P_{I}2_{1}2_{1}2$ & \#18.24 & IV & Table~\ref{tab:msg18.24} & $\varnothing$ & $\varnothing$\\
\hline
&&$I_{c}222$&\#23.52&IV&OG~\ref{sup:sg21}.7.135&\\
\end{longtable}
\input{src_tex_mtable/msg23.49.tex}
\input{src_tex_mtable/msg23.50.tex}
\input{src_tex_mtable/msg23.51.tex}
\input{src_tex_mtable/msg16.6.tex}
\input{src_tex_mtable/msg18.24.tex}
\subsection{SG 24}  \label{sup:sg24}
TABLE SG24. The crystalline space group is what the crystal has if the magnetic order is neglected. Once condidering magnetic order, the MSGs, magnetic type, and the symmetry-indicator classifications are given below. For each MSG, the detailed information is given in the corresponding MSG table. \hyperref[tabindex]{See the list of tables.}
\setcounter{table}{148}
\begin{longtable}{p{1.5cm}p{2cm}|p{1.5cm}p{2cm}|c|c|c|c}
\hline
\hline
   \multicolumn{2}{c|}{OG setting} & \multicolumn{2}{c|}{BNS setting} & MSG type & Detailed Inf. & integer spin & half-integer spin\\
\hline
\endfirsthead
\multicolumn{8}{c} {{ TABLE SG\ref{sup:sg24} -- continued}}\\
\hline
   \multicolumn{2}{c|}{OG setting} & \multicolumn{2}{c|}{BNS setting} & MSG type & Detailed Inf. & integer spin & half-integer spin\\
\hline
\endhead
\hline
\endfoot
\hline
\hline
\endlastfoot
24.1.\rownumber & $I2_{1}2_{1}2_{1}$ & $I2_{1}2_{1}2_{1}$ & \#24.53 & I & Table~\ref{tab:msg24.53} & $\varnothing$ & $\varnothing$\\
24.2.\rownumber & $I2_{1}2_{1}2_{1}1'$ & $I2_{1}2_{1}2_{1}1'$ & \#24.54 & II & Table~\ref{tab:msg24.54} & $\varnothing$ & $\varnothing$\\
24.3.\rownumber & $I2_{1}'2_{1}'2_{1}$ & $I2_{1}'2_{1}'2_{1}$ & \#24.55 & III & Table~\ref{tab:msg24.55} & $\varnothing$ & $\varnothing$\\
24.4.\rownumber & $I_{P}2_{1}2_{1}2_{1}$ & $P_{I}2_{1}2_{1}2_{1}$ & \#19.30 & IV & Table~\ref{tab:msg19.30} & $\varnothing$ & $\varnothing$\\
24.5.\rownumber & $I_{P}2_{1}'2_{1}'2_{1}$ & $P_{I}222_{1}$ & \#17.15 & IV & Table~\ref{tab:msg17.15} & $\varnothing$ & $\varnothing$\\
\hline
&&$I_{c}2_{1}2_{1}2_{1}$&\#24.56&IV&OG~\ref{sup:sg21}.11.139&\\
\end{longtable}
\input{src_tex_mtable/msg24.53.tex}
\input{src_tex_mtable/msg24.54.tex}
\input{src_tex_mtable/msg24.55.tex}
\input{src_tex_mtable/msg19.30.tex}
\input{src_tex_mtable/msg17.15.tex}
\subsection{SG 25}  \label{sup:sg25}
TABLE SG25. The crystalline space group is what the crystal has if the magnetic order is neglected. Once condidering magnetic order, the MSGs, magnetic type, and the symmetry-indicator classifications are given below. For each MSG, the detailed information is given in the corresponding MSG table. \hyperref[tabindex]{See the list of tables.}
\setcounter{table}{153}
\begin{longtable}{p{1.5cm}p{2cm}|p{1.5cm}p{2cm}|c|c|c|c}
\hline
\hline
   \multicolumn{2}{c|}{OG setting} & \multicolumn{2}{c|}{BNS setting} & MSG type & Detailed Inf. & integer spin & half-integer spin\\
\hline
\endfirsthead
\multicolumn{8}{c} {{ TABLE SG\ref{sup:sg25} -- continued}}\\
\hline
   \multicolumn{2}{c|}{OG setting} & \multicolumn{2}{c|}{BNS setting} & MSG type & Detailed Inf. & integer spin & half-integer spin\\
\hline
\endhead
\hline
\endfoot
\hline
\hline
\endlastfoot
25.1.\rownumber & $Pmm2$ & $Pmm2$ & \#25.57 & I & Table~\ref{tab:msg25.57} & $\varnothing$ & $\varnothing$\\
25.2.\rownumber & $Pmm21'$ & $Pmm21'$ & \#25.58 & II & Table~\ref{tab:msg25.58} & $\varnothing$ & $\varnothing$\\
25.3.\rownumber & $Pm'm2'$ & $Pm'm2'$ & \#25.59 & III & Table~\ref{tab:msg25.59} & $\varnothing$ & $\varnothing$\\
25.4.\rownumber & $Pm'm'2$ & $Pm'm'2$ & \#25.60 & III & Table~\ref{tab:msg25.60} & $Z_{2}$ & $Z_{2}$\\
25.5.\rownumber & $P_{2c}mm2$ & $P_{c}mm2$ & \#25.61 & IV & Table~\ref{tab:msg25.61} & $\varnothing$ & $\varnothing$\\
25.6.\rownumber & $P_{2a}mm2$ & $P_{a}mm2$ & \#25.62 & IV & Table~\ref{tab:msg25.62} & $\varnothing$ & $\varnothing$\\
25.7.\rownumber & $P_{C}mm2$ & $C_{a}mm2$ & \#35.170 & IV & Table~\ref{tab:msg35.170} & $\varnothing$ & $\varnothing$\\
25.8.\rownumber & $P_{A}mm2$ & $A_{b}mm2$ & \#38.193 & IV & Table~\ref{tab:msg38.193} & $\varnothing$ & $\varnothing$\\
\end{longtable}
\input{src_tex_mtable/msg25.57.tex}
\input{src_tex_mtable/msg25.58.tex}
\input{src_tex_mtable/msg25.59.tex}
\input{src_tex_mtable/msg25.60.tex}
\input{src_tex_mtable/msg25.61.tex}
\input{src_tex_mtable/msg25.62.tex}
\input{src_tex_mtable/msg35.170.tex}
\input{src_tex_mtable/msg38.193.tex}
\input{src_tex_mtable/msg42.223.tex}
\input{src_tex_mtable/msg26.73.tex}
\input{src_tex_mtable/msg27.82.tex}
\input{src_tex_mtable/msg28.92.tex}
\input{src_tex_mtable/msg39.201.tex}
\subsection{SG 26}  \label{sup:sg26}
TABLE SG26. The crystalline space group is what the crystal has if the magnetic order is neglected. Once condidering magnetic order, the MSGs, magnetic type, and the symmetry-indicator classifications are given below. For each MSG, the detailed information is given in the corresponding MSG table. \hyperref[tabindex]{See the list of tables.}
\setcounter{table}{166}
\begin{longtable}{p{1.5cm}p{2cm}|p{1.5cm}p{2cm}|c|c|c|c}
\hline
\hline
   \multicolumn{2}{c|}{OG setting} & \multicolumn{2}{c|}{BNS setting} & MSG type & Detailed Inf. & integer spin & half-integer spin\\
\hline
\endfirsthead
\multicolumn{8}{c} {{ TABLE SG\ref{sup:sg26} -- continued}}\\
\hline
   \multicolumn{2}{c|}{OG setting} & \multicolumn{2}{c|}{BNS setting} & MSG type & Detailed Inf. & integer spin & half-integer spin\\
\hline
\endhead
\hline
\endfoot
\hline
\hline
\endlastfoot
26.1.\rownumber & $Pmc2_{1}$ & $Pmc2_{1}$ & \#26.66 & I & Table~\ref{tab:msg26.66} & $\varnothing$ & $\varnothing$\\
26.2.\rownumber & $Pmc2_{1}1'$ & $Pmc2_{1}1'$ & \#26.67 & II & Table~\ref{tab:msg26.67} & $\varnothing$ & $\varnothing$\\
26.3.\rownumber & $Pm'c2_{1}'$ & $Pm'c2_{1}'$ & \#26.68 & III & Table~\ref{tab:msg26.68} & $\varnothing$ & $\varnothing$\\
26.4.\rownumber & $Pmc'2_{1}'$ & $Pmc'2_{1}'$ & \#26.69 & III & Table~\ref{tab:msg26.69} & $\varnothing$ & $\varnothing$\\
26.5.\rownumber & $Pm'c'2_{1}$ & $Pm'c'2_{1}$ & \#26.70 & III & Table~\ref{tab:msg26.70} & $\varnothing$ & $\varnothing$\\
26.6.\rownumber & $P_{2a}mc2_{1}$ & $P_{a}mc2_{1}$ & \#26.71 & IV & Table~\ref{tab:msg26.71} & $\varnothing$ & $\varnothing$\\
26.7.\rownumber & $P_{2b}mc2_{1}$ & $P_{b}mc2_{1}$ & \#26.72 & IV & Table~\ref{tab:msg26.72} & $\varnothing$ & $\varnothing$\\
26.8.\rownumber & $P_{C}mc2_{1}$ & $C_{a}mc2_{1}$ & \#36.178 & IV & Table~\ref{tab:msg36.178} & $\varnothing$ & $\varnothing$\\
\end{longtable}
\input{src_tex_mtable/msg26.66.tex}
\input{src_tex_mtable/msg26.67.tex}
\input{src_tex_mtable/msg26.68.tex}
\input{src_tex_mtable/msg26.69.tex}
\input{src_tex_mtable/msg26.70.tex}
\input{src_tex_mtable/msg26.71.tex}
\input{src_tex_mtable/msg26.72.tex}
\input{src_tex_mtable/msg36.178.tex}
\input{src_tex_mtable/msg31.128.tex}
\input{src_tex_mtable/msg29.104.tex}
\subsection{SG 27}  \label{sup:sg27}
TABLE SG27. The crystalline space group is what the crystal has if the magnetic order is neglected. Once condidering magnetic order, the MSGs, magnetic type, and the symmetry-indicator classifications are given below. For each MSG, the detailed information is given in the corresponding MSG table. \hyperref[tabindex]{See the list of tables.}
\setcounter{table}{176}
\begin{longtable}{p{1.5cm}p{2cm}|p{1.5cm}p{2cm}|c|c|c|c}
\hline
\hline
   \multicolumn{2}{c|}{OG setting} & \multicolumn{2}{c|}{BNS setting} & MSG type & Detailed Inf. & integer spin & half-integer spin\\
\hline
\endfirsthead
\multicolumn{8}{c} {{ TABLE SG\ref{sup:sg27} -- continued}}\\
\hline
   \multicolumn{2}{c|}{OG setting} & \multicolumn{2}{c|}{BNS setting} & MSG type & Detailed Inf. & integer spin & half-integer spin\\
\hline
\endhead
\hline
\endfoot
\hline
\hline
\endlastfoot
27.1.\rownumber & $Pcc2$ & $Pcc2$ & \#27.78 & I & Table~\ref{tab:msg27.78} & $Z_{2}$ & $\varnothing$\\
27.2.\rownumber & $Pcc21'$ & $Pcc21'$ & \#27.79 & II & Table~\ref{tab:msg27.79} & $Z_{2}$ & $\varnothing$\\
27.3.\rownumber & $Pc'c2'$ & $Pc'c2'$ & \#27.80 & III & Table~\ref{tab:msg27.80} & $\varnothing$ & $\varnothing$\\
27.4.\rownumber & $Pc'c'2$ & $Pc'c'2$ & \#27.81 & III & Table~\ref{tab:msg27.81} & $Z_{2}$ & $Z_{2}$\\
27.5.\rownumber & $P_{2a}cc2$ & $P_{a}cc2$ & \#27.83 & IV & Table~\ref{tab:msg27.83} & $Z_{2}$ & $\varnothing$\\
27.6.\rownumber & $P_{C}cc2$ & $C_{a}cc2$ & \#37.185 & IV & Table~\ref{tab:msg37.185} & $Z_{2}$$\times$$Z_{2}$ & $Z_{2}$\\
27.7.\rownumber & $P_{2b}c'c2'$ & $P_{b}nc2$ & \#30.117 & IV & Table~\ref{tab:msg30.117} & $Z_{2}$ & $Z_{2}$\\
\hline
&&$P_{c}cc2$&\#27.82&IV&OG~\ref{sup:sg25}.11.165&\\
&&$P_{C}cc2$&\#27.84&IV&OG~\ref{sup:sg37}.5.262&\\
&&$P_{A}cc2$&\#27.85&IV&OG~\ref{sup:sg39}.12.289&\\
&&$P_{I}cc2$&\#27.86&IV&OG~\ref{sup:sg45}.5.335&\\
\end{longtable}
\input{src_tex_mtable/msg27.78.tex}
\input{src_tex_mtable/msg27.79.tex}
\input{src_tex_mtable/msg27.80.tex}
\input{src_tex_mtable/msg27.81.tex}
\input{src_tex_mtable/msg27.83.tex}
\input{src_tex_mtable/msg37.185.tex}
\input{src_tex_mtable/msg30.117.tex}
\subsection{SG 28}  \label{sup:sg28}
TABLE SG28. The crystalline space group is what the crystal has if the magnetic order is neglected. Once condidering magnetic order, the MSGs, magnetic type, and the symmetry-indicator classifications are given below. For each MSG, the detailed information is given in the corresponding MSG table. \hyperref[tabindex]{See the list of tables.}
\setcounter{table}{183}
\begin{longtable}{p{1.5cm}p{2cm}|p{1.5cm}p{2cm}|c|c|c|c}
\hline
\hline
   \multicolumn{2}{c|}{OG setting} & \multicolumn{2}{c|}{BNS setting} & MSG type & Detailed Inf. & integer spin & half-integer spin\\
\hline
\endfirsthead
\multicolumn{8}{c} {{ TABLE SG\ref{sup:sg28} -- continued}}\\
\hline
   \multicolumn{2}{c|}{OG setting} & \multicolumn{2}{c|}{BNS setting} & MSG type & Detailed Inf. & integer spin & half-integer spin\\
\hline
\endhead
\hline
\endfoot
\hline
\hline
\endlastfoot
28.1.\rownumber & $Pma2$ & $Pma2$ & \#28.87 & I & Table~\ref{tab:msg28.87} & $\varnothing$ & $\varnothing$\\
28.2.\rownumber & $Pma21'$ & $Pma21'$ & \#28.88 & II & Table~\ref{tab:msg28.88} & $\varnothing$ & $\varnothing$\\
28.3.\rownumber & $Pm'a2'$ & $Pm'a2'$ & \#28.89 & III & Table~\ref{tab:msg28.89} & $\varnothing$ & $\varnothing$\\
28.4.\rownumber & $Pma'2'$ & $Pma'2'$ & \#28.90 & III & Table~\ref{tab:msg28.90} & $\varnothing$ & $\varnothing$\\
28.5.\rownumber & $Pm'a'2$ & $Pm'a'2$ & \#28.91 & III & Table~\ref{tab:msg28.91} & $Z_{2}$ & $Z_{2}$\\
28.6.\rownumber & $P_{2b}ma2$ & $P_{b}ma2$ & \#28.93 & IV & Table~\ref{tab:msg28.93} & $\varnothing$ & $\varnothing$\\
28.7.\rownumber & $P_{2c}ma2$ & $P_{c}ma2$ & \#28.94 & IV & Table~\ref{tab:msg28.94} & $\varnothing$ & $\varnothing$\\
28.8.\rownumber & $P_{A}ma2$ & $A_{b}ma2$ & \#40.209 & IV & Table~\ref{tab:msg40.209} & $\varnothing$ & $\varnothing$\\
\end{longtable}
\input{src_tex_mtable/msg28.87.tex}
\input{src_tex_mtable/msg28.88.tex}
\input{src_tex_mtable/msg28.89.tex}
\input{src_tex_mtable/msg28.90.tex}
\input{src_tex_mtable/msg28.91.tex}
\input{src_tex_mtable/msg28.93.tex}
\input{src_tex_mtable/msg28.94.tex}
\input{src_tex_mtable/msg40.209.tex}
\input{src_tex_mtable/msg32.140.tex}
\input{src_tex_mtable/msg29.106.tex}
\input{src_tex_mtable/msg31.130.tex}
\input{src_tex_mtable/msg30.118.tex}
\input{src_tex_mtable/msg41.217.tex}
\subsection{SG 29}  \label{sup:sg29}
TABLE SG29. The crystalline space group is what the crystal has if the magnetic order is neglected. Once condidering magnetic order, the MSGs, magnetic type, and the symmetry-indicator classifications are given below. For each MSG, the detailed information is given in the corresponding MSG table. \hyperref[tabindex]{See the list of tables.}
\setcounter{table}{196}
\begin{longtable}{p{1.5cm}p{2cm}|p{1.5cm}p{2cm}|c|c|c|c}
\hline
\hline
   \multicolumn{2}{c|}{OG setting} & \multicolumn{2}{c|}{BNS setting} & MSG type & Detailed Inf. & integer spin & half-integer spin\\
\hline
\endfirsthead
\multicolumn{8}{c} {{ TABLE SG\ref{sup:sg29} -- continued}}\\
\hline
   \multicolumn{2}{c|}{OG setting} & \multicolumn{2}{c|}{BNS setting} & MSG type & Detailed Inf. & integer spin & half-integer spin\\
\hline
\endhead
\hline
\endfoot
\hline
\hline
\endlastfoot
29.1.\rownumber & $Pca2_{1}$ & $Pca2_{1}$ & \#29.99 & I & Table~\ref{tab:msg29.99} & $\varnothing$ & $\varnothing$\\
29.2.\rownumber & $Pca2_{1}1'$ & $Pca2_{1}1'$ & \#29.100 & II & Table~\ref{tab:msg29.100} & $\varnothing$ & $\varnothing$\\
29.3.\rownumber & $Pc'a2_{1}'$ & $Pc'a2_{1}'$ & \#29.101 & III & Table~\ref{tab:msg29.101} & $\varnothing$ & $\varnothing$\\
29.4.\rownumber & $Pca'2_{1}'$ & $Pca'2_{1}'$ & \#29.102 & III & Table~\ref{tab:msg29.102} & $\varnothing$ & $\varnothing$\\
29.5.\rownumber & $Pc'a'2_{1}$ & $Pc'a'2_{1}$ & \#29.103 & III & Table~\ref{tab:msg29.103} & $\varnothing$ & $\varnothing$\\
29.6.\rownumber & $P_{2b}ca2_{1}$ & $P_{b}ca2_{1}$ & \#29.105 & IV & Table~\ref{tab:msg29.105} & $Z_{2}$ & $\varnothing$\\
29.7.\rownumber & $P_{2b}c'a'2_{1}$ & $P_{b}na2_{1}$ & \#33.150 & IV & Table~\ref{tab:msg33.150} & $\varnothing$ & $\varnothing$\\
\hline
&&$P_{a}ca2_{1}$&\#29.104&IV&OG~\ref{sup:sg26}.10.177&\\
&&$P_{c}ca2_{1}$&\#29.106&IV&OG~\ref{sup:sg28}.10.194&\\
&&$P_{A}ca2_{1}$&\#29.107&IV&OG~\ref{sup:sg41}.7.306&\\
&&$P_{B}ca2_{1}$&\#29.108&IV&OG~\ref{sup:sg39}.11.288&\\
&&$P_{C}ca2_{1}$&\#29.109&IV&OG~\ref{sup:sg36}.7.255&\\
&&$P_{I}ca2_{1}$&\#29.110&IV&OG~\ref{sup:sg45}.6.336&\\
\end{longtable}
\input{src_tex_mtable/msg29.99.tex}
\input{src_tex_mtable/msg29.100.tex}
\input{src_tex_mtable/msg29.101.tex}
\input{src_tex_mtable/msg29.102.tex}
\input{src_tex_mtable/msg29.103.tex}
\input{src_tex_mtable/msg29.105.tex}
\input{src_tex_mtable/msg33.150.tex}
\subsection{SG 30}  \label{sup:sg30}
TABLE SG30. The crystalline space group is what the crystal has if the magnetic order is neglected. Once condidering magnetic order, the MSGs, magnetic type, and the symmetry-indicator classifications are given below. For each MSG, the detailed information is given in the corresponding MSG table. \hyperref[tabindex]{See the list of tables.}
\input{src_tex/sg30}
\subsection{SG 31}  \label{sup:sg31}
TABLE SG31. The crystalline space group is what the crystal has if the magnetic order is neglected. Once condidering magnetic order, the MSGs, magnetic type, and the symmetry-indicator classifications are given below. For each MSG, the detailed information is given in the corresponding MSG table. \hyperref[tabindex]{See the list of tables.}
\input{src_tex/sg31}
\subsection{SG 32}  \label{sup:sg32}
TABLE SG32. The crystalline space group is what the crystal has if the magnetic order is neglected. Once condidering magnetic order, the MSGs, magnetic type, and the symmetry-indicator classifications are given below. For each MSG, the detailed information is given in the corresponding MSG table. \hyperref[tabindex]{See the list of tables.}
\input{src_tex/sg32}
\subsection{SG 33}  \label{sup:sg33}
TABLE SG33. The crystalline space group is what the crystal has if the magnetic order is neglected. Once condidering magnetic order, the MSGs, magnetic type, and the symmetry-indicator classifications are given below. For each MSG, the detailed information is given in the corresponding MSG table. \hyperref[tabindex]{See the list of tables.}
\input{src_tex/sg33}
\subsection{SG 34}  \label{sup:sg34}
TABLE SG34. The crystalline space group is what the crystal has if the magnetic order is neglected. Once condidering magnetic order, the MSGs, magnetic type, and the symmetry-indicator classifications are given below. For each MSG, the detailed information is given in the corresponding MSG table. \hyperref[tabindex]{See the list of tables.}
\input{src_tex/sg34}
\subsection{SG 35}  \label{sup:sg35}
TABLE SG35. The crystalline space group is what the crystal has if the magnetic order is neglected. Once condidering magnetic order, the MSGs, magnetic type, and the symmetry-indicator classifications are given below. For each MSG, the detailed information is given in the corresponding MSG table. \hyperref[tabindex]{See the list of tables.}
\input{src_tex/sg35}
\subsection{SG 36}  \label{sup:sg36}
TABLE SG36. The crystalline space group is what the crystal has if the magnetic order is neglected. Once condidering magnetic order, the MSGs, magnetic type, and the symmetry-indicator classifications are given below. For each MSG, the detailed information is given in the corresponding MSG table. \hyperref[tabindex]{See the list of tables.}
\input{src_tex/sg36}
\subsection{SG 37}  \label{sup:sg37}
TABLE SG37. The crystalline space group is what the crystal has if the magnetic order is neglected. Once condidering magnetic order, the MSGs, magnetic type, and the symmetry-indicator classifications are given below. For each MSG, the detailed information is given in the corresponding MSG table. \hyperref[tabindex]{See the list of tables.}
\input{src_tex/sg37}
\subsection{SG 38}  \label{sup:sg38}
TABLE SG38. The crystalline space group is what the crystal has if the magnetic order is neglected. Once condidering magnetic order, the MSGs, magnetic type, and the symmetry-indicator classifications are given below. For each MSG, the detailed information is given in the corresponding MSG table. \hyperref[tabindex]{See the list of tables.}
\input{src_tex/sg38}
\subsection{SG 39}  \label{sup:sg39}
TABLE SG39. The crystalline space group is what the crystal has if the magnetic order is neglected. Once condidering magnetic order, the MSGs, magnetic type, and the symmetry-indicator classifications are given below. For each MSG, the detailed information is given in the corresponding MSG table. \hyperref[tabindex]{See the list of tables.}
\input{src_tex/sg39}
\subsection{SG 40}  \label{sup:sg40}
TABLE SG40. The crystalline space group is what the crystal has if the magnetic order is neglected. Once condidering magnetic order, the MSGs, magnetic type, and the symmetry-indicator classifications are given below. For each MSG, the detailed information is given in the corresponding MSG table. \hyperref[tabindex]{See the list of tables.}
\input{src_tex/sg40}
\subsection{SG 41}  \label{sup:sg41}
TABLE SG41. The crystalline space group is what the crystal has if the magnetic order is neglected. Once condidering magnetic order, the MSGs, magnetic type, and the symmetry-indicator classifications are given below. For each MSG, the detailed information is given in the corresponding MSG table. \hyperref[tabindex]{See the list of tables.}
\input{src_tex/sg41}
\subsection{SG 42}  \label{sup:sg42}
TABLE SG42. The crystalline space group is what the crystal has if the magnetic order is neglected. Once condidering magnetic order, the MSGs, magnetic type, and the symmetry-indicator classifications are given below. For each MSG, the detailed information is given in the corresponding MSG table. \hyperref[tabindex]{See the list of tables.}
\input{src_tex/sg42}
\subsection{SG 43}  \label{sup:sg43}
TABLE SG43. The crystalline space group is what the crystal has if the magnetic order is neglected. Once condidering magnetic order, the MSGs, magnetic type, and the symmetry-indicator classifications are given below. For each MSG, the detailed information is given in the corresponding MSG table. \hyperref[tabindex]{See the list of tables.}
\setcounter{table}{318}
\begin{longtable}{p{1.5cm}p{2cm}|p{1.5cm}p{2cm}|c|c|c|c}
\hline
\hline
   \multicolumn{2}{c|}{OG setting} & \multicolumn{2}{c|}{BNS setting} & MSG type & Detailed Inf. & integer spin & half-integer spin\\
\hline
\endfirsthead
\multicolumn{8}{c} {{ TABLE SG\ref{sup:sg43} -- continued}}\\
\hline
   \multicolumn{2}{c|}{OG setting} & \multicolumn{2}{c|}{BNS setting} & MSG type & Detailed Inf. & integer spin & half-integer spin\\
\hline
\endhead
\hline
\endfoot
\hline
\hline
\endlastfoot
43.1.\rownumber & $Fdd2$ & $Fdd2$ & \#43.224 & I & Table~\ref{tab:msg43.224} & $\varnothing$ & $\varnothing$\\
43.2.\rownumber & $Fdd21'$ & $Fdd21'$ & \#43.225 & II & Table~\ref{tab:msg43.225} & $\varnothing$ & $\varnothing$\\
43.3.\rownumber & $Fd'd2'$ & $Fd'd2'$ & \#43.226 & III & Table~\ref{tab:msg43.226} & $\varnothing$ & $\varnothing$\\
43.4.\rownumber & $Fd'd'2$ & $Fd'd'2$ & \#43.227 & III & Table~\ref{tab:msg43.227} & $\varnothing$ & $\varnothing$\\
\hline
&&$F_{S}dd2$&\#43.228&IV&OG~\ref{sup:sg34}.5.235&\\
\end{longtable}
\input{src_tex_mtable/msg43.224.tex}
\input{src_tex_mtable/msg43.225.tex}
\input{src_tex_mtable/msg43.226.tex}
\input{src_tex_mtable/msg43.227.tex}
\subsection{SG 44}  \label{sup:sg44}
TABLE SG44. The crystalline space group is what the crystal has if the magnetic order is neglected. Once condidering magnetic order, the MSGs, magnetic type, and the symmetry-indicator classifications are given below. For each MSG, the detailed information is given in the corresponding MSG table. \hyperref[tabindex]{See the list of tables.}
\input{src_tex/sg44}
\subsection{SG 45}  \label{sup:sg45}
TABLE SG45. The crystalline space group is what the crystal has if the magnetic order is neglected. Once condidering magnetic order, the MSGs, magnetic type, and the symmetry-indicator classifications are given below. For each MSG, the detailed information is given in the corresponding MSG table. \hyperref[tabindex]{See the list of tables.}
\input{src_tex/sg45}
\subsection{SG 46}  \label{sup:sg46}
TABLE SG46. The crystalline space group is what the crystal has if the magnetic order is neglected. Once condidering magnetic order, the MSGs, magnetic type, and the symmetry-indicator classifications are given below. For each MSG, the detailed information is given in the corresponding MSG table. \hyperref[tabindex]{See the list of tables.}
\input{src_tex/sg46}
\subsection{SG 47}  \label{sup:sg47}
TABLE SG47. The crystalline space group is what the crystal has if the magnetic order is neglected. Once condidering magnetic order, the MSGs, magnetic type, and the symmetry-indicator classifications are given below. For each MSG, the detailed information is given in the corresponding MSG table. \hyperref[tabindex]{See the list of tables.}
\input{src_tex/sg47}
\subsection{SG 48}  \label{sup:sg48}
TABLE SG48. The crystalline space group is what the crystal has if the magnetic order is neglected. Once condidering magnetic order, the MSGs, magnetic type, and the symmetry-indicator classifications are given below. For each MSG, the detailed information is given in the corresponding MSG table. \hyperref[tabindex]{See the list of tables.}
\input{src_tex/sg48}
\subsection{SG 49}  \label{sup:sg49}
TABLE SG49. The crystalline space group is what the crystal has if the magnetic order is neglected. Once condidering magnetic order, the MSGs, magnetic type, and the symmetry-indicator classifications are given below. For each MSG, the detailed information is given in the corresponding MSG table. \hyperref[tabindex]{See the list of tables.}
\input{src_tex/sg49}
\subsection{SG 50}  \label{sup:sg50}
TABLE SG50. The crystalline space group is what the crystal has if the magnetic order is neglected. Once condidering magnetic order, the MSGs, magnetic type, and the symmetry-indicator classifications are given below. For each MSG, the detailed information is given in the corresponding MSG table. \hyperref[tabindex]{See the list of tables.}
\input{src_tex/sg50}
\subsection{SG 51}  \label{sup:sg51}
TABLE SG51. The crystalline space group is what the crystal has if the magnetic order is neglected. Once condidering magnetic order, the MSGs, magnetic type, and the symmetry-indicator classifications are given below. For each MSG, the detailed information is given in the corresponding MSG table. \hyperref[tabindex]{See the list of tables.}
\input{src_tex/sg51}
\subsection{SG 52}  \label{sup:sg52}
TABLE SG52. The crystalline space group is what the crystal has if the magnetic order is neglected. Once condidering magnetic order, the MSGs, magnetic type, and the symmetry-indicator classifications are given below. For each MSG, the detailed information is given in the corresponding MSG table. \hyperref[tabindex]{See the list of tables.}
\input{src_tex/sg52}
\subsection{SG 53}  \label{sup:sg53}
TABLE SG53. The crystalline space group is what the crystal has if the magnetic order is neglected. Once condidering magnetic order, the MSGs, magnetic type, and the symmetry-indicator classifications are given below. For each MSG, the detailed information is given in the corresponding MSG table. \hyperref[tabindex]{See the list of tables.}
\input{src_tex/sg53}
\subsection{SG 54}  \label{sup:sg54}
TABLE SG54. The crystalline space group is what the crystal has if the magnetic order is neglected. Once condidering magnetic order, the MSGs, magnetic type, and the symmetry-indicator classifications are given below. For each MSG, the detailed information is given in the corresponding MSG table. \hyperref[tabindex]{See the list of tables.}
\input{src_tex/sg54}
\subsection{SG 55}  \label{sup:sg55}
TABLE SG55. The crystalline space group is what the crystal has if the magnetic order is neglected. Once condidering magnetic order, the MSGs, magnetic type, and the symmetry-indicator classifications are given below. For each MSG, the detailed information is given in the corresponding MSG table. \hyperref[tabindex]{See the list of tables.}
\input{src_tex/sg55}
\subsection{SG 56}  \label{sup:sg56}
TABLE SG56. The crystalline space group is what the crystal has if the magnetic order is neglected. Once condidering magnetic order, the MSGs, magnetic type, and the symmetry-indicator classifications are given below. For each MSG, the detailed information is given in the corresponding MSG table. \hyperref[tabindex]{See the list of tables.}
\input{src_tex/sg56}
\subsection{SG 57}  \label{sup:sg57}
TABLE SG57. The crystalline space group is what the crystal has if the magnetic order is neglected. Once condidering magnetic order, the MSGs, magnetic type, and the symmetry-indicator classifications are given below. For each MSG, the detailed information is given in the corresponding MSG table. \hyperref[tabindex]{See the list of tables.}
\input{src_tex/sg57}
\subsection{SG 58}  \label{sup:sg58}
TABLE SG58. The crystalline space group is what the crystal has if the magnetic order is neglected. Once condidering magnetic order, the MSGs, magnetic type, and the symmetry-indicator classifications are given below. For each MSG, the detailed information is given in the corresponding MSG table. \hyperref[tabindex]{See the list of tables.}
\input{src_tex/sg58}
\subsection{SG 59}  \label{sup:sg59}
TABLE SG59. The crystalline space group is what the crystal has if the magnetic order is neglected. Once condidering magnetic order, the MSGs, magnetic type, and the symmetry-indicator classifications are given below. For each MSG, the detailed information is given in the corresponding MSG table. \hyperref[tabindex]{See the list of tables.}
\input{src_tex/sg59}
\subsection{SG 60}  \label{sup:sg60}
TABLE SG60. The crystalline space group is what the crystal has if the magnetic order is neglected. Once condidering magnetic order, the MSGs, magnetic type, and the symmetry-indicator classifications are given below. For each MSG, the detailed information is given in the corresponding MSG table. \hyperref[tabindex]{See the list of tables.}
\input{src_tex/sg60}
\subsection{SG 61}  \label{sup:sg61}
TABLE SG61. The crystalline space group is what the crystal has if the magnetic order is neglected. Once condidering magnetic order, the MSGs, magnetic type, and the symmetry-indicator classifications are given below. For each MSG, the detailed information is given in the corresponding MSG table. \hyperref[tabindex]{See the list of tables.}
\input{src_tex/sg61}
\subsection{SG 62}  \label{sup:sg62}
TABLE SG62. The crystalline space group is what the crystal has if the magnetic order is neglected. Once condidering magnetic order, the MSGs, magnetic type, and the symmetry-indicator classifications are given below. For each MSG, the detailed information is given in the corresponding MSG table. \hyperref[tabindex]{See the list of tables.}
\input{src_tex/sg62}
\subsection{SG 63}  \label{sup:sg63}
TABLE SG63. The crystalline space group is what the crystal has if the magnetic order is neglected. Once condidering magnetic order, the MSGs, magnetic type, and the symmetry-indicator classifications are given below. For each MSG, the detailed information is given in the corresponding MSG table. \hyperref[tabindex]{See the list of tables.}
\input{src_tex/sg63}
\subsection{SG 64}  \label{sup:sg64}
TABLE SG64. The crystalline space group is what the crystal has if the magnetic order is neglected. Once condidering magnetic order, the MSGs, magnetic type, and the symmetry-indicator classifications are given below. For each MSG, the detailed information is given in the corresponding MSG table. \hyperref[tabindex]{See the list of tables.}
\input{src_tex/sg64}
\subsection{SG 65}  \label{sup:sg65}
TABLE SG65. The crystalline space group is what the crystal has if the magnetic order is neglected. Once condidering magnetic order, the MSGs, magnetic type, and the symmetry-indicator classifications are given below. For each MSG, the detailed information is given in the corresponding MSG table. \hyperref[tabindex]{See the list of tables.}
\input{src_tex/sg65}
\subsection{SG 66}  \label{sup:sg66}
TABLE SG66. The crystalline space group is what the crystal has if the magnetic order is neglected. Once condidering magnetic order, the MSGs, magnetic type, and the symmetry-indicator classifications are given below. For each MSG, the detailed information is given in the corresponding MSG table. \hyperref[tabindex]{See the list of tables.}
\input{src_tex/sg66}
\subsection{SG 67}  \label{sup:sg67}
TABLE SG67. The crystalline space group is what the crystal has if the magnetic order is neglected. Once condidering magnetic order, the MSGs, magnetic type, and the symmetry-indicator classifications are given below. For each MSG, the detailed information is given in the corresponding MSG table. \hyperref[tabindex]{See the list of tables.}
\input{src_tex/sg67}
\subsection{SG 68}  \label{sup:sg68}
TABLE SG68. The crystalline space group is what the crystal has if the magnetic order is neglected. Once condidering magnetic order, the MSGs, magnetic type, and the symmetry-indicator classifications are given below. For each MSG, the detailed information is given in the corresponding MSG table. \hyperref[tabindex]{See the list of tables.}
\input{src_tex/sg68}
\subsection{SG 69}  \label{sup:sg69}
TABLE SG69. The crystalline space group is what the crystal has if the magnetic order is neglected. Once condidering magnetic order, the MSGs, magnetic type, and the symmetry-indicator classifications are given below. For each MSG, the detailed information is given in the corresponding MSG table. \hyperref[tabindex]{See the list of tables.}
\input{src_tex/sg69}
\subsection{SG 70}  \label{sup:sg70}
TABLE SG70. The crystalline space group is what the crystal has if the magnetic order is neglected. Once condidering magnetic order, the MSGs, magnetic type, and the symmetry-indicator classifications are given below. For each MSG, the detailed information is given in the corresponding MSG table. \hyperref[tabindex]{See the list of tables.}
\input{src_tex/sg70}
\subsection{SG 71}  \label{sup:sg71}
TABLE SG71. The crystalline space group is what the crystal has if the magnetic order is neglected. Once condidering magnetic order, the MSGs, magnetic type, and the symmetry-indicator classifications are given below. For each MSG, the detailed information is given in the corresponding MSG table. \hyperref[tabindex]{See the list of tables.}
\input{src_tex/sg71}
\subsection{SG 72}  \label{sup:sg72}
TABLE SG72. The crystalline space group is what the crystal has if the magnetic order is neglected. Once condidering magnetic order, the MSGs, magnetic type, and the symmetry-indicator classifications are given below. For each MSG, the detailed information is given in the corresponding MSG table. \hyperref[tabindex]{See the list of tables.}
\input{src_tex/sg72}
\subsection{SG 73}  \label{sup:sg73}
TABLE SG73. The crystalline space group is what the crystal has if the magnetic order is neglected. Once condidering magnetic order, the MSGs, magnetic type, and the symmetry-indicator classifications are given below. For each MSG, the detailed information is given in the corresponding MSG table. \hyperref[tabindex]{See the list of tables.}
\input{src_tex/sg73}
\subsection{SG 74}  \label{sup:sg74}
TABLE SG74. The crystalline space group is what the crystal has if the magnetic order is neglected. Once condidering magnetic order, the MSGs, magnetic type, and the symmetry-indicator classifications are given below. For each MSG, the detailed information is given in the corresponding MSG table. \hyperref[tabindex]{See the list of tables.}
\input{src_tex/sg74}
\subsection{SG 75}  \label{sup:sg75}
TABLE SG75. The crystalline space group is what the crystal has if the magnetic order is neglected. Once condidering magnetic order, the MSGs, magnetic type, and the symmetry-indicator classifications are given below. For each MSG, the detailed information is given in the corresponding MSG table. \hyperref[tabindex]{See the list of tables.}
\input{src_tex/sg75}
\subsection{SG 76}  \label{sup:sg76}
TABLE SG76. The crystalline space group is what the crystal has if the magnetic order is neglected. Once condidering magnetic order, the MSGs, magnetic type, and the symmetry-indicator classifications are given below. For each MSG, the detailed information is given in the corresponding MSG table. \hyperref[tabindex]{See the list of tables.}
\input{src_tex/sg76}
\subsection{SG 77}  \label{sup:sg77}
TABLE SG77. The crystalline space group is what the crystal has if the magnetic order is neglected. Once condidering magnetic order, the MSGs, magnetic type, and the symmetry-indicator classifications are given below. For each MSG, the detailed information is given in the corresponding MSG table. \hyperref[tabindex]{See the list of tables.}
\input{src_tex/sg77}
\subsection{SG 78}  \label{sup:sg78}
TABLE SG78. The crystalline space group is what the crystal has if the magnetic order is neglected. Once condidering magnetic order, the MSGs, magnetic type, and the symmetry-indicator classifications are given below. For each MSG, the detailed information is given in the corresponding MSG table. \hyperref[tabindex]{See the list of tables.}
\input{src_tex/sg78}
\subsection{SG 79}  \label{sup:sg79}
TABLE SG79. The crystalline space group is what the crystal has if the magnetic order is neglected. Once condidering magnetic order, the MSGs, magnetic type, and the symmetry-indicator classifications are given below. For each MSG, the detailed information is given in the corresponding MSG table. \hyperref[tabindex]{See the list of tables.}
\input{src_tex/sg79}
\subsection{SG 80}  \label{sup:sg80}
TABLE SG80. The crystalline space group is what the crystal has if the magnetic order is neglected. Once condidering magnetic order, the MSGs, magnetic type, and the symmetry-indicator classifications are given below. For each MSG, the detailed information is given in the corresponding MSG table. \hyperref[tabindex]{See the list of tables.}
\input{src_tex/sg80}
\subsection{SG 81}  \label{sup:sg81}
TABLE SG81. The crystalline space group is what the crystal has if the magnetic order is neglected. Once condidering magnetic order, the MSGs, magnetic type, and the symmetry-indicator classifications are given below. For each MSG, the detailed information is given in the corresponding MSG table. \hyperref[tabindex]{See the list of tables.}
\input{src_tex/sg81}
\subsection{SG 82}  \label{sup:sg82}
TABLE SG82. The crystalline space group is what the crystal has if the magnetic order is neglected. Once condidering magnetic order, the MSGs, magnetic type, and the symmetry-indicator classifications are given below. For each MSG, the detailed information is given in the corresponding MSG table. \hyperref[tabindex]{See the list of tables.}
\input{src_tex/sg82}
\subsection{SG 83}  \label{sup:sg83}
TABLE SG83. The crystalline space group is what the crystal has if the magnetic order is neglected. Once condidering magnetic order, the MSGs, magnetic type, and the symmetry-indicator classifications are given below. For each MSG, the detailed information is given in the corresponding MSG table. \hyperref[tabindex]{See the list of tables.}
\input{src_tex/sg83}
\subsection{SG 84}  \label{sup:sg84}
TABLE SG84. The crystalline space group is what the crystal has if the magnetic order is neglected. Once condidering magnetic order, the MSGs, magnetic type, and the symmetry-indicator classifications are given below. For each MSG, the detailed information is given in the corresponding MSG table. \hyperref[tabindex]{See the list of tables.}
\input{src_tex/sg84}
\subsection{SG 85}  \label{sup:sg85}
TABLE SG85. The crystalline space group is what the crystal has if the magnetic order is neglected. Once condidering magnetic order, the MSGs, magnetic type, and the symmetry-indicator classifications are given below. For each MSG, the detailed information is given in the corresponding MSG table. \hyperref[tabindex]{See the list of tables.}
\input{src_tex/sg85}
\subsection{SG 86}  \label{sup:sg86}
TABLE SG86. The crystalline space group is what the crystal has if the magnetic order is neglected. Once condidering magnetic order, the MSGs, magnetic type, and the symmetry-indicator classifications are given below. For each MSG, the detailed information is given in the corresponding MSG table. \hyperref[tabindex]{See the list of tables.}
\input{src_tex/sg86}
\subsection{SG 87}  \label{sup:sg87}
TABLE SG87. The crystalline space group is what the crystal has if the magnetic order is neglected. Once condidering magnetic order, the MSGs, magnetic type, and the symmetry-indicator classifications are given below. For each MSG, the detailed information is given in the corresponding MSG table. \hyperref[tabindex]{See the list of tables.}
\input{src_tex/sg87}
\subsection{SG 88}  \label{sup:sg88}
TABLE SG88. The crystalline space group is what the crystal has if the magnetic order is neglected. Once condidering magnetic order, the MSGs, magnetic type, and the symmetry-indicator classifications are given below. For each MSG, the detailed information is given in the corresponding MSG table. \hyperref[tabindex]{See the list of tables.}
\input{src_tex/sg88}
\subsection{SG 89}  \label{sup:sg89}
TABLE SG89. The crystalline space group is what the crystal has if the magnetic order is neglected. Once condidering magnetic order, the MSGs, magnetic type, and the symmetry-indicator classifications are given below. For each MSG, the detailed information is given in the corresponding MSG table. \hyperref[tabindex]{See the list of tables.}
\input{src_tex/sg89}
\subsection{SG 90}  \label{sup:sg90}
TABLE SG90. The crystalline space group is what the crystal has if the magnetic order is neglected. Once condidering magnetic order, the MSGs, magnetic type, and the symmetry-indicator classifications are given below. For each MSG, the detailed information is given in the corresponding MSG table. \hyperref[tabindex]{See the list of tables.}
\input{src_tex/sg90}
\subsection{SG 91}  \label{sup:sg91}
TABLE SG91. The crystalline space group is what the crystal has if the magnetic order is neglected. Once condidering magnetic order, the MSGs, magnetic type, and the symmetry-indicator classifications are given below. For each MSG, the detailed information is given in the corresponding MSG table. \hyperref[tabindex]{See the list of tables.}
\input{src_tex/sg91}
\subsection{SG 92}  \label{sup:sg92}
TABLE SG92. The crystalline space group is what the crystal has if the magnetic order is neglected. Once condidering magnetic order, the MSGs, magnetic type, and the symmetry-indicator classifications are given below. For each MSG, the detailed information is given in the corresponding MSG table. \hyperref[tabindex]{See the list of tables.}
\input{src_tex/sg92}
\subsection{SG 93}  \label{sup:sg93}
TABLE SG93. The crystalline space group is what the crystal has if the magnetic order is neglected. Once condidering magnetic order, the MSGs, magnetic type, and the symmetry-indicator classifications are given below. For each MSG, the detailed information is given in the corresponding MSG table. \hyperref[tabindex]{See the list of tables.}
\input{src_tex/sg93}
\subsection{SG 94}  \label{sup:sg94}
TABLE SG94. The crystalline space group is what the crystal has if the magnetic order is neglected. Once condidering magnetic order, the MSGs, magnetic type, and the symmetry-indicator classifications are given below. For each MSG, the detailed information is given in the corresponding MSG table. \hyperref[tabindex]{See the list of tables.}
\input{src_tex/sg94}
\subsection{SG 95}  \label{sup:sg95}
TABLE SG95. The crystalline space group is what the crystal has if the magnetic order is neglected. Once condidering magnetic order, the MSGs, magnetic type, and the symmetry-indicator classifications are given below. For each MSG, the detailed information is given in the corresponding MSG table. \hyperref[tabindex]{See the list of tables.}
\input{src_tex/sg95}
\subsection{SG 96}  \label{sup:sg96}
TABLE SG96. The crystalline space group is what the crystal has if the magnetic order is neglected. Once condidering magnetic order, the MSGs, magnetic type, and the symmetry-indicator classifications are given below. For each MSG, the detailed information is given in the corresponding MSG table. \hyperref[tabindex]{See the list of tables.}
\input{src_tex/sg96}
\subsection{SG 97}  \label{sup:sg97}
TABLE SG97. The crystalline space group is what the crystal has if the magnetic order is neglected. Once condidering magnetic order, the MSGs, magnetic type, and the symmetry-indicator classifications are given below. For each MSG, the detailed information is given in the corresponding MSG table. \hyperref[tabindex]{See the list of tables.}
\input{src_tex/sg97}
\subsection{SG 98}  \label{sup:sg98}
TABLE SG98. The crystalline space group is what the crystal has if the magnetic order is neglected. Once condidering magnetic order, the MSGs, magnetic type, and the symmetry-indicator classifications are given below. For each MSG, the detailed information is given in the corresponding MSG table. \hyperref[tabindex]{See the list of tables.}
\input{src_tex/sg98}
\subsection{SG 99}  \label{sup:sg99}
TABLE SG99. The crystalline space group is what the crystal has if the magnetic order is neglected. Once condidering magnetic order, the MSGs, magnetic type, and the symmetry-indicator classifications are given below. For each MSG, the detailed information is given in the corresponding MSG table. \hyperref[tabindex]{See the list of tables.}
\input{src_tex/sg99}
\subsection{SG 100}  \label{sup:sg100}
TABLE SG100. The crystalline space group is what the crystal has if the magnetic order is neglected. Once condidering magnetic order, the MSGs, magnetic type, and the symmetry-indicator classifications are given below. For each MSG, the detailed information is given in the corresponding MSG table. \hyperref[tabindex]{See the list of tables.}
\input{src_tex/sg100}
\subsection{SG 101}  \label{sup:sg101}
TABLE SG101. The crystalline space group is what the crystal has if the magnetic order is neglected. Once condidering magnetic order, the MSGs, magnetic type, and the symmetry-indicator classifications are given below. For each MSG, the detailed information is given in the corresponding MSG table. \hyperref[tabindex]{See the list of tables.}
\input{src_tex/sg101}
\subsection{SG 102}  \label{sup:sg102}
TABLE SG102. The crystalline space group is what the crystal has if the magnetic order is neglected. Once condidering magnetic order, the MSGs, magnetic type, and the symmetry-indicator classifications are given below. For each MSG, the detailed information is given in the corresponding MSG table. \hyperref[tabindex]{See the list of tables.}
\input{src_tex/sg102}
\subsection{SG 103}  \label{sup:sg103}
TABLE SG103. The crystalline space group is what the crystal has if the magnetic order is neglected. Once condidering magnetic order, the MSGs, magnetic type, and the symmetry-indicator classifications are given below. For each MSG, the detailed information is given in the corresponding MSG table. \hyperref[tabindex]{See the list of tables.}
\input{src_tex/sg103}
\subsection{SG 104}  \label{sup:sg104}
TABLE SG104. The crystalline space group is what the crystal has if the magnetic order is neglected. Once condidering magnetic order, the MSGs, magnetic type, and the symmetry-indicator classifications are given below. For each MSG, the detailed information is given in the corresponding MSG table. \hyperref[tabindex]{See the list of tables.}
\input{src_tex/sg104}
\subsection{SG 105}  \label{sup:sg105}
TABLE SG105. The crystalline space group is what the crystal has if the magnetic order is neglected. Once condidering magnetic order, the MSGs, magnetic type, and the symmetry-indicator classifications are given below. For each MSG, the detailed information is given in the corresponding MSG table. \hyperref[tabindex]{See the list of tables.}
\input{src_tex/sg105}
\subsection{SG 106}  \label{sup:sg106}
TABLE SG106. The crystalline space group is what the crystal has if the magnetic order is neglected. Once condidering magnetic order, the MSGs, magnetic type, and the symmetry-indicator classifications are given below. For each MSG, the detailed information is given in the corresponding MSG table. \hyperref[tabindex]{See the list of tables.}
\input{src_tex/sg106}
\subsection{SG 107}  \label{sup:sg107}
TABLE SG107. The crystalline space group is what the crystal has if the magnetic order is neglected. Once condidering magnetic order, the MSGs, magnetic type, and the symmetry-indicator classifications are given below. For each MSG, the detailed information is given in the corresponding MSG table. \hyperref[tabindex]{See the list of tables.}
\input{src_tex/sg107}
\subsection{SG 108}  \label{sup:sg108}
TABLE SG108. The crystalline space group is what the crystal has if the magnetic order is neglected. Once condidering magnetic order, the MSGs, magnetic type, and the symmetry-indicator classifications are given below. For each MSG, the detailed information is given in the corresponding MSG table. \hyperref[tabindex]{See the list of tables.}
\input{src_tex/sg108}
\subsection{SG 109}  \label{sup:sg109}
TABLE SG109. The crystalline space group is what the crystal has if the magnetic order is neglected. Once condidering magnetic order, the MSGs, magnetic type, and the symmetry-indicator classifications are given below. For each MSG, the detailed information is given in the corresponding MSG table. \hyperref[tabindex]{See the list of tables.}
\input{src_tex/sg109}
\subsection{SG 110}  \label{sup:sg110}
TABLE SG110. The crystalline space group is what the crystal has if the magnetic order is neglected. Once condidering magnetic order, the MSGs, magnetic type, and the symmetry-indicator classifications are given below. For each MSG, the detailed information is given in the corresponding MSG table. \hyperref[tabindex]{See the list of tables.}
\input{src_tex/sg110}
\subsection{SG 111}  \label{sup:sg111}
TABLE SG111. The crystalline space group is what the crystal has if the magnetic order is neglected. Once condidering magnetic order, the MSGs, magnetic type, and the symmetry-indicator classifications are given below. For each MSG, the detailed information is given in the corresponding MSG table. \hyperref[tabindex]{See the list of tables.}
\input{src_tex/sg111}
\subsection{SG 112}  \label{sup:sg112}
TABLE SG112. The crystalline space group is what the crystal has if the magnetic order is neglected. Once condidering magnetic order, the MSGs, magnetic type, and the symmetry-indicator classifications are given below. For each MSG, the detailed information is given in the corresponding MSG table. \hyperref[tabindex]{See the list of tables.}
\input{src_tex/sg112}
\subsection{SG 113}  \label{sup:sg113}
TABLE SG113. The crystalline space group is what the crystal has if the magnetic order is neglected. Once condidering magnetic order, the MSGs, magnetic type, and the symmetry-indicator classifications are given below. For each MSG, the detailed information is given in the corresponding MSG table. \hyperref[tabindex]{See the list of tables.}
\input{src_tex/sg113}
\subsection{SG 114}  \label{sup:sg114}
TABLE SG114. The crystalline space group is what the crystal has if the magnetic order is neglected. Once condidering magnetic order, the MSGs, magnetic type, and the symmetry-indicator classifications are given below. For each MSG, the detailed information is given in the corresponding MSG table. \hyperref[tabindex]{See the list of tables.}
\input{src_tex/sg114}
\subsection{SG 115}  \label{sup:sg115}
TABLE SG115. The crystalline space group is what the crystal has if the magnetic order is neglected. Once condidering magnetic order, the MSGs, magnetic type, and the symmetry-indicator classifications are given below. For each MSG, the detailed information is given in the corresponding MSG table. \hyperref[tabindex]{See the list of tables.}
\input{src_tex/sg115}
\subsection{SG 116}  \label{sup:sg116}
TABLE SG116. The crystalline space group is what the crystal has if the magnetic order is neglected. Once condidering magnetic order, the MSGs, magnetic type, and the symmetry-indicator classifications are given below. For each MSG, the detailed information is given in the corresponding MSG table. \hyperref[tabindex]{See the list of tables.}
\input{src_tex/sg116}
\subsection{SG 117}  \label{sup:sg117}
TABLE SG117. The crystalline space group is what the crystal has if the magnetic order is neglected. Once condidering magnetic order, the MSGs, magnetic type, and the symmetry-indicator classifications are given below. For each MSG, the detailed information is given in the corresponding MSG table. \hyperref[tabindex]{See the list of tables.}
\input{src_tex/sg117}
\subsection{SG 118}  \label{sup:sg118}
TABLE SG118. The crystalline space group is what the crystal has if the magnetic order is neglected. Once condidering magnetic order, the MSGs, magnetic type, and the symmetry-indicator classifications are given below. For each MSG, the detailed information is given in the corresponding MSG table. \hyperref[tabindex]{See the list of tables.}
\input{src_tex/sg118}
\subsection{SG 119}  \label{sup:sg119}
TABLE SG119. The crystalline space group is what the crystal has if the magnetic order is neglected. Once condidering magnetic order, the MSGs, magnetic type, and the symmetry-indicator classifications are given below. For each MSG, the detailed information is given in the corresponding MSG table. \hyperref[tabindex]{See the list of tables.}
\input{src_tex/sg119}
\subsection{SG 120}  \label{sup:sg120}
TABLE SG120. The crystalline space group is what the crystal has if the magnetic order is neglected. Once condidering magnetic order, the MSGs, magnetic type, and the symmetry-indicator classifications are given below. For each MSG, the detailed information is given in the corresponding MSG table. \hyperref[tabindex]{See the list of tables.}
\input{src_tex/sg120}
\subsection{SG 121}  \label{sup:sg121}
TABLE SG121. The crystalline space group is what the crystal has if the magnetic order is neglected. Once condidering magnetic order, the MSGs, magnetic type, and the symmetry-indicator classifications are given below. For each MSG, the detailed information is given in the corresponding MSG table. \hyperref[tabindex]{See the list of tables.}
\input{src_tex/sg121}
\subsection{SG 122}  \label{sup:sg122}
TABLE SG122. The crystalline space group is what the crystal has if the magnetic order is neglected. Once condidering magnetic order, the MSGs, magnetic type, and the symmetry-indicator classifications are given below. For each MSG, the detailed information is given in the corresponding MSG table. \hyperref[tabindex]{See the list of tables.}
\input{src_tex/sg122}
\subsection{SG 123}  \label{sup:sg123}
TABLE SG123. The crystalline space group is what the crystal has if the magnetic order is neglected. Once condidering magnetic order, the MSGs, magnetic type, and the symmetry-indicator classifications are given below. For each MSG, the detailed information is given in the corresponding MSG table. \hyperref[tabindex]{See the list of tables.}
\input{src_tex/sg123}
\subsection{SG 124}  \label{sup:sg124}
TABLE SG124. The crystalline space group is what the crystal has if the magnetic order is neglected. Once condidering magnetic order, the MSGs, magnetic type, and the symmetry-indicator classifications are given below. For each MSG, the detailed information is given in the corresponding MSG table. \hyperref[tabindex]{See the list of tables.}
\input{src_tex/sg124}
\subsection{SG 125}  \label{sup:sg125}
TABLE SG125. The crystalline space group is what the crystal has if the magnetic order is neglected. Once condidering magnetic order, the MSGs, magnetic type, and the symmetry-indicator classifications are given below. For each MSG, the detailed information is given in the corresponding MSG table. \hyperref[tabindex]{See the list of tables.}
\input{src_tex/sg125}
\subsection{SG 126}  \label{sup:sg126}
TABLE SG126. The crystalline space group is what the crystal has if the magnetic order is neglected. Once condidering magnetic order, the MSGs, magnetic type, and the symmetry-indicator classifications are given below. For each MSG, the detailed information is given in the corresponding MSG table. \hyperref[tabindex]{See the list of tables.}
\input{src_tex/sg126}
\subsection{SG 127}  \label{sup:sg127}
TABLE SG127. The crystalline space group is what the crystal has if the magnetic order is neglected. Once condidering magnetic order, the MSGs, magnetic type, and the symmetry-indicator classifications are given below. For each MSG, the detailed information is given in the corresponding MSG table. \hyperref[tabindex]{See the list of tables.}
\input{src_tex/sg127}
\subsection{SG 128}  \label{sup:sg128}
TABLE SG128. The crystalline space group is what the crystal has if the magnetic order is neglected. Once condidering magnetic order, the MSGs, magnetic type, and the symmetry-indicator classifications are given below. For each MSG, the detailed information is given in the corresponding MSG table. \hyperref[tabindex]{See the list of tables.}
\input{src_tex/sg128}
\subsection{SG 129}  \label{sup:sg129}
TABLE SG129. The crystalline space group is what the crystal has if the magnetic order is neglected. Once condidering magnetic order, the MSGs, magnetic type, and the symmetry-indicator classifications are given below. For each MSG, the detailed information is given in the corresponding MSG table. \hyperref[tabindex]{See the list of tables.}
\input{src_tex/sg129}
\subsection{SG 130}  \label{sup:sg130}
TABLE SG130. The crystalline space group is what the crystal has if the magnetic order is neglected. Once condidering magnetic order, the MSGs, magnetic type, and the symmetry-indicator classifications are given below. For each MSG, the detailed information is given in the corresponding MSG table. \hyperref[tabindex]{See the list of tables.}
\input{src_tex/sg130}
\subsection{SG 131}  \label{sup:sg131}
TABLE SG131. The crystalline space group is what the crystal has if the magnetic order is neglected. Once condidering magnetic order, the MSGs, magnetic type, and the symmetry-indicator classifications are given below. For each MSG, the detailed information is given in the corresponding MSG table. \hyperref[tabindex]{See the list of tables.}
\input{src_tex/sg131}
\subsection{SG 132}  \label{sup:sg132}
TABLE SG132. The crystalline space group is what the crystal has if the magnetic order is neglected. Once condidering magnetic order, the MSGs, magnetic type, and the symmetry-indicator classifications are given below. For each MSG, the detailed information is given in the corresponding MSG table. \hyperref[tabindex]{See the list of tables.}
\input{src_tex/sg132}
\subsection{SG 133}  \label{sup:sg133}
TABLE SG133. The crystalline space group is what the crystal has if the magnetic order is neglected. Once condidering magnetic order, the MSGs, magnetic type, and the symmetry-indicator classifications are given below. For each MSG, the detailed information is given in the corresponding MSG table. \hyperref[tabindex]{See the list of tables.}
\input{src_tex/sg133}
\subsection{SG 134}  \label{sup:sg134}
TABLE SG134. The crystalline space group is what the crystal has if the magnetic order is neglected. Once condidering magnetic order, the MSGs, magnetic type, and the symmetry-indicator classifications are given below. For each MSG, the detailed information is given in the corresponding MSG table. \hyperref[tabindex]{See the list of tables.}
\input{src_tex/sg134}
\subsection{SG 135}  \label{sup:sg135}
TABLE SG135. The crystalline space group is what the crystal has if the magnetic order is neglected. Once condidering magnetic order, the MSGs, magnetic type, and the symmetry-indicator classifications are given below. For each MSG, the detailed information is given in the corresponding MSG table. \hyperref[tabindex]{See the list of tables.}
\input{src_tex/sg135}
\subsection{SG 136}  \label{sup:sg136}
TABLE SG136. The crystalline space group is what the crystal has if the magnetic order is neglected. Once condidering magnetic order, the MSGs, magnetic type, and the symmetry-indicator classifications are given below. For each MSG, the detailed information is given in the corresponding MSG table. \hyperref[tabindex]{See the list of tables.}
\input{src_tex/sg136}
\subsection{SG 137}  \label{sup:sg137}
TABLE SG137. The crystalline space group is what the crystal has if the magnetic order is neglected. Once condidering magnetic order, the MSGs, magnetic type, and the symmetry-indicator classifications are given below. For each MSG, the detailed information is given in the corresponding MSG table. \hyperref[tabindex]{See the list of tables.}
\input{src_tex/sg137}
\subsection{SG 138}  \label{sup:sg138}
TABLE SG138. The crystalline space group is what the crystal has if the magnetic order is neglected. Once condidering magnetic order, the MSGs, magnetic type, and the symmetry-indicator classifications are given below. For each MSG, the detailed information is given in the corresponding MSG table. \hyperref[tabindex]{See the list of tables.}
\input{src_tex/sg138}
\subsection{SG 139}  \label{sup:sg139}
TABLE SG139. The crystalline space group is what the crystal has if the magnetic order is neglected. Once condidering magnetic order, the MSGs, magnetic type, and the symmetry-indicator classifications are given below. For each MSG, the detailed information is given in the corresponding MSG table. \hyperref[tabindex]{See the list of tables.}
\input{src_tex/sg139}
\subsection{SG 140}  \label{sup:sg140}
TABLE SG140. The crystalline space group is what the crystal has if the magnetic order is neglected. Once condidering magnetic order, the MSGs, magnetic type, and the symmetry-indicator classifications are given below. For each MSG, the detailed information is given in the corresponding MSG table. \hyperref[tabindex]{See the list of tables.}
\input{src_tex/sg140}
\subsection{SG 141}  \label{sup:sg141}
TABLE SG141. The crystalline space group is what the crystal has if the magnetic order is neglected. Once condidering magnetic order, the MSGs, magnetic type, and the symmetry-indicator classifications are given below. For each MSG, the detailed information is given in the corresponding MSG table. \hyperref[tabindex]{See the list of tables.}
\input{src_tex/sg141}
\subsection{SG 142}  \label{sup:sg142}
TABLE SG142. The crystalline space group is what the crystal has if the magnetic order is neglected. Once condidering magnetic order, the MSGs, magnetic type, and the symmetry-indicator classifications are given below. For each MSG, the detailed information is given in the corresponding MSG table. \hyperref[tabindex]{See the list of tables.}
\input{src_tex/sg142}
\subsection{SG 143}  \label{sup:sg143}
TABLE SG143. The crystalline space group is what the crystal has if the magnetic order is neglected. Once condidering magnetic order, the MSGs, magnetic type, and the symmetry-indicator classifications are given below. For each MSG, the detailed information is given in the corresponding MSG table. \hyperref[tabindex]{See the list of tables.}
\input{src_tex/sg143}
\subsection{SG 144}  \label{sup:sg144}
TABLE SG144. The crystalline space group is what the crystal has if the magnetic order is neglected. Once condidering magnetic order, the MSGs, magnetic type, and the symmetry-indicator classifications are given below. For each MSG, the detailed information is given in the corresponding MSG table. \hyperref[tabindex]{See the list of tables.}
\input{src_tex/sg144}
\subsection{SG 145}  \label{sup:sg145}
TABLE SG145. The crystalline space group is what the crystal has if the magnetic order is neglected. Once condidering magnetic order, the MSGs, magnetic type, and the symmetry-indicator classifications are given below. For each MSG, the detailed information is given in the corresponding MSG table. \hyperref[tabindex]{See the list of tables.}
\input{src_tex/sg145}
\subsection{SG 146}  \label{sup:sg146}
TABLE SG146. The crystalline space group is what the crystal has if the magnetic order is neglected. Once condidering magnetic order, the MSGs, magnetic type, and the symmetry-indicator classifications are given below. For each MSG, the detailed information is given in the corresponding MSG table. \hyperref[tabindex]{See the list of tables.}
\input{src_tex/sg146}
\subsection{SG 147}  \label{sup:sg147}
TABLE SG147. The crystalline space group is what the crystal has if the magnetic order is neglected. Once condidering magnetic order, the MSGs, magnetic type, and the symmetry-indicator classifications are given below. For each MSG, the detailed information is given in the corresponding MSG table. \hyperref[tabindex]{See the list of tables.}
\input{src_tex/sg147}
\subsection{SG 148}  \label{sup:sg148}
TABLE SG148. The crystalline space group is what the crystal has if the magnetic order is neglected. Once condidering magnetic order, the MSGs, magnetic type, and the symmetry-indicator classifications are given below. For each MSG, the detailed information is given in the corresponding MSG table. \hyperref[tabindex]{See the list of tables.}
\input{src_tex/sg148}
\subsection{SG 149}  \label{sup:sg149}
TABLE SG149. The crystalline space group is what the crystal has if the magnetic order is neglected. Once condidering magnetic order, the MSGs, magnetic type, and the symmetry-indicator classifications are given below. For each MSG, the detailed information is given in the corresponding MSG table. \hyperref[tabindex]{See the list of tables.}
\input{src_tex/sg149}
\subsection{SG 150}  \label{sup:sg150}
TABLE SG150. The crystalline space group is what the crystal has if the magnetic order is neglected. Once condidering magnetic order, the MSGs, magnetic type, and the symmetry-indicator classifications are given below. For each MSG, the detailed information is given in the corresponding MSG table. \hyperref[tabindex]{See the list of tables.}
\input{src_tex/sg150}
\subsection{SG 151}  \label{sup:sg151}
TABLE SG151. The crystalline space group is what the crystal has if the magnetic order is neglected. Once condidering magnetic order, the MSGs, magnetic type, and the symmetry-indicator classifications are given below. For each MSG, the detailed information is given in the corresponding MSG table. \hyperref[tabindex]{See the list of tables.}
\input{src_tex/sg151}
\subsection{SG 152}  \label{sup:sg152}
TABLE SG152. The crystalline space group is what the crystal has if the magnetic order is neglected. Once condidering magnetic order, the MSGs, magnetic type, and the symmetry-indicator classifications are given below. For each MSG, the detailed information is given in the corresponding MSG table. \hyperref[tabindex]{See the list of tables.}
\input{src_tex/sg152}
\subsection{SG 153}  \label{sup:sg153}
TABLE SG153. The crystalline space group is what the crystal has if the magnetic order is neglected. Once condidering magnetic order, the MSGs, magnetic type, and the symmetry-indicator classifications are given below. For each MSG, the detailed information is given in the corresponding MSG table. \hyperref[tabindex]{See the list of tables.}
\input{src_tex/sg153}
\subsection{SG 154}  \label{sup:sg154}
TABLE SG154. The crystalline space group is what the crystal has if the magnetic order is neglected. Once condidering magnetic order, the MSGs, magnetic type, and the symmetry-indicator classifications are given below. For each MSG, the detailed information is given in the corresponding MSG table. \hyperref[tabindex]{See the list of tables.}
\input{src_tex/sg154}
\subsection{SG 155}  \label{sup:sg155}
TABLE SG155. The crystalline space group is what the crystal has if the magnetic order is neglected. Once condidering magnetic order, the MSGs, magnetic type, and the symmetry-indicator classifications are given below. For each MSG, the detailed information is given in the corresponding MSG table. \hyperref[tabindex]{See the list of tables.}
\input{src_tex/sg155}
\subsection{SG 156}  \label{sup:sg156}
TABLE SG156. The crystalline space group is what the crystal has if the magnetic order is neglected. Once condidering magnetic order, the MSGs, magnetic type, and the symmetry-indicator classifications are given below. For each MSG, the detailed information is given in the corresponding MSG table. \hyperref[tabindex]{See the list of tables.}
\input{src_tex/sg156}
\subsection{SG 157}  \label{sup:sg157}
TABLE SG157. The crystalline space group is what the crystal has if the magnetic order is neglected. Once condidering magnetic order, the MSGs, magnetic type, and the symmetry-indicator classifications are given below. For each MSG, the detailed information is given in the corresponding MSG table. \hyperref[tabindex]{See the list of tables.}
\input{src_tex/sg157}
\subsection{SG 158}  \label{sup:sg158}
TABLE SG158. The crystalline space group is what the crystal has if the magnetic order is neglected. Once condidering magnetic order, the MSGs, magnetic type, and the symmetry-indicator classifications are given below. For each MSG, the detailed information is given in the corresponding MSG table. \hyperref[tabindex]{See the list of tables.}
\input{src_tex/sg158}
\subsection{SG 159}  \label{sup:sg159}
TABLE SG159. The crystalline space group is what the crystal has if the magnetic order is neglected. Once condidering magnetic order, the MSGs, magnetic type, and the symmetry-indicator classifications are given below. For each MSG, the detailed information is given in the corresponding MSG table. \hyperref[tabindex]{See the list of tables.}
\input{src_tex/sg159}
\subsection{SG 160}  \label{sup:sg160}
TABLE SG160. The crystalline space group is what the crystal has if the magnetic order is neglected. Once condidering magnetic order, the MSGs, magnetic type, and the symmetry-indicator classifications are given below. For each MSG, the detailed information is given in the corresponding MSG table. \hyperref[tabindex]{See the list of tables.}
\input{src_tex/sg160}
\subsection{SG 161}  \label{sup:sg161}
TABLE SG161. The crystalline space group is what the crystal has if the magnetic order is neglected. Once condidering magnetic order, the MSGs, magnetic type, and the symmetry-indicator classifications are given below. For each MSG, the detailed information is given in the corresponding MSG table. \hyperref[tabindex]{See the list of tables.}
\input{src_tex/sg161}
\subsection{SG 162}  \label{sup:sg162}
TABLE SG162. The crystalline space group is what the crystal has if the magnetic order is neglected. Once condidering magnetic order, the MSGs, magnetic type, and the symmetry-indicator classifications are given below. For each MSG, the detailed information is given in the corresponding MSG table. \hyperref[tabindex]{See the list of tables.}
\input{src_tex/sg162}
\subsection{SG 163}  \label{sup:sg163}
TABLE SG163. The crystalline space group is what the crystal has if the magnetic order is neglected. Once condidering magnetic order, the MSGs, magnetic type, and the symmetry-indicator classifications are given below. For each MSG, the detailed information is given in the corresponding MSG table. \hyperref[tabindex]{See the list of tables.}
\input{src_tex/sg163}
\subsection{SG 164}  \label{sup:sg164}
TABLE SG164. The crystalline space group is what the crystal has if the magnetic order is neglected. Once condidering magnetic order, the MSGs, magnetic type, and the symmetry-indicator classifications are given below. For each MSG, the detailed information is given in the corresponding MSG table. \hyperref[tabindex]{See the list of tables.}
\input{src_tex/sg164}
\subsection{SG 165}  \label{sup:sg165}
TABLE SG165. The crystalline space group is what the crystal has if the magnetic order is neglected. Once condidering magnetic order, the MSGs, magnetic type, and the symmetry-indicator classifications are given below. For each MSG, the detailed information is given in the corresponding MSG table. \hyperref[tabindex]{See the list of tables.}
\input{src_tex/sg165}
\subsection{SG 166}  \label{sup:sg166}
TABLE SG166. The crystalline space group is what the crystal has if the magnetic order is neglected. Once condidering magnetic order, the MSGs, magnetic type, and the symmetry-indicator classifications are given below. For each MSG, the detailed information is given in the corresponding MSG table. \hyperref[tabindex]{See the list of tables.}
\input{src_tex/sg166}
\subsection{SG 167}  \label{sup:sg167}
TABLE SG167. The crystalline space group is what the crystal has if the magnetic order is neglected. Once condidering magnetic order, the MSGs, magnetic type, and the symmetry-indicator classifications are given below. For each MSG, the detailed information is given in the corresponding MSG table. \hyperref[tabindex]{See the list of tables.}
\input{src_tex/sg167}
\subsection{SG 168}  \label{sup:sg168}
TABLE SG168. The crystalline space group is what the crystal has if the magnetic order is neglected. Once condidering magnetic order, the MSGs, magnetic type, and the symmetry-indicator classifications are given below. For each MSG, the detailed information is given in the corresponding MSG table. \hyperref[tabindex]{See the list of tables.}
\input{src_tex/sg168}
\subsection{SG 169}  \label{sup:sg169}
TABLE SG169. The crystalline space group is what the crystal has if the magnetic order is neglected. Once condidering magnetic order, the MSGs, magnetic type, and the symmetry-indicator classifications are given below. For each MSG, the detailed information is given in the corresponding MSG table. \hyperref[tabindex]{See the list of tables.}
\input{src_tex/sg169}
\subsection{SG 170}  \label{sup:sg170}
TABLE SG170. The crystalline space group is what the crystal has if the magnetic order is neglected. Once condidering magnetic order, the MSGs, magnetic type, and the symmetry-indicator classifications are given below. For each MSG, the detailed information is given in the corresponding MSG table. \hyperref[tabindex]{See the list of tables.}
\input{src_tex/sg170}
\subsection{SG 171}  \label{sup:sg171}
TABLE SG171. The crystalline space group is what the crystal has if the magnetic order is neglected. Once condidering magnetic order, the MSGs, magnetic type, and the symmetry-indicator classifications are given below. For each MSG, the detailed information is given in the corresponding MSG table. \hyperref[tabindex]{See the list of tables.}
\input{src_tex/sg171}
\subsection{SG 172}  \label{sup:sg172}
TABLE SG172. The crystalline space group is what the crystal has if the magnetic order is neglected. Once condidering magnetic order, the MSGs, magnetic type, and the symmetry-indicator classifications are given below. For each MSG, the detailed information is given in the corresponding MSG table. \hyperref[tabindex]{See the list of tables.}
\input{src_tex/sg172}
\subsection{SG 173}  \label{sup:sg173}
TABLE SG173. The crystalline space group is what the crystal has if the magnetic order is neglected. Once condidering magnetic order, the MSGs, magnetic type, and the symmetry-indicator classifications are given below. For each MSG, the detailed information is given in the corresponding MSG table. \hyperref[tabindex]{See the list of tables.}
\input{src_tex/sg173}
\subsection{SG 174}  \label{sup:sg174}
TABLE SG174. The crystalline space group is what the crystal has if the magnetic order is neglected. Once condidering magnetic order, the MSGs, magnetic type, and the symmetry-indicator classifications are given below. For each MSG, the detailed information is given in the corresponding MSG table. \hyperref[tabindex]{See the list of tables.}
\input{src_tex/sg174}
\subsection{SG 175}  \label{sup:sg175}
TABLE SG175. The crystalline space group is what the crystal has if the magnetic order is neglected. Once condidering magnetic order, the MSGs, magnetic type, and the symmetry-indicator classifications are given below. For each MSG, the detailed information is given in the corresponding MSG table. \hyperref[tabindex]{See the list of tables.}
\input{src_tex/sg175}
\subsection{SG 176}  \label{sup:sg176}
TABLE SG176. The crystalline space group is what the crystal has if the magnetic order is neglected. Once condidering magnetic order, the MSGs, magnetic type, and the symmetry-indicator classifications are given below. For each MSG, the detailed information is given in the corresponding MSG table. \hyperref[tabindex]{See the list of tables.}
\input{src_tex/sg176}
\subsection{SG 177}  \label{sup:sg177}
TABLE SG177. The crystalline space group is what the crystal has if the magnetic order is neglected. Once condidering magnetic order, the MSGs, magnetic type, and the symmetry-indicator classifications are given below. For each MSG, the detailed information is given in the corresponding MSG table. \hyperref[tabindex]{See the list of tables.}
\input{src_tex/sg177}
\subsection{SG 178}  \label{sup:sg178}
TABLE SG178. The crystalline space group is what the crystal has if the magnetic order is neglected. Once condidering magnetic order, the MSGs, magnetic type, and the symmetry-indicator classifications are given below. For each MSG, the detailed information is given in the corresponding MSG table. \hyperref[tabindex]{See the list of tables.}
\input{src_tex/sg178}
\subsection{SG 179}  \label{sup:sg179}
TABLE SG179. The crystalline space group is what the crystal has if the magnetic order is neglected. Once condidering magnetic order, the MSGs, magnetic type, and the symmetry-indicator classifications are given below. For each MSG, the detailed information is given in the corresponding MSG table. \hyperref[tabindex]{See the list of tables.}
\input{src_tex/sg179}
\subsection{SG 180}  \label{sup:sg180}
TABLE SG180. The crystalline space group is what the crystal has if the magnetic order is neglected. Once condidering magnetic order, the MSGs, magnetic type, and the symmetry-indicator classifications are given below. For each MSG, the detailed information is given in the corresponding MSG table. \hyperref[tabindex]{See the list of tables.}
\input{src_tex/sg180}
\subsection{SG 181}  \label{sup:sg181}
TABLE SG181. The crystalline space group is what the crystal has if the magnetic order is neglected. Once condidering magnetic order, the MSGs, magnetic type, and the symmetry-indicator classifications are given below. For each MSG, the detailed information is given in the corresponding MSG table. \hyperref[tabindex]{See the list of tables.}
\input{src_tex/sg181}
\subsection{SG 182}  \label{sup:sg182}
TABLE SG182. The crystalline space group is what the crystal has if the magnetic order is neglected. Once condidering magnetic order, the MSGs, magnetic type, and the symmetry-indicator classifications are given below. For each MSG, the detailed information is given in the corresponding MSG table. \hyperref[tabindex]{See the list of tables.}
\input{src_tex/sg182}
\subsection{SG 183}  \label{sup:sg183}
TABLE SG183. The crystalline space group is what the crystal has if the magnetic order is neglected. Once condidering magnetic order, the MSGs, magnetic type, and the symmetry-indicator classifications are given below. For each MSG, the detailed information is given in the corresponding MSG table. \hyperref[tabindex]{See the list of tables.}
\input{src_tex/sg183}
\subsection{SG 184}  \label{sup:sg184}
TABLE SG184. The crystalline space group is what the crystal has if the magnetic order is neglected. Once condidering magnetic order, the MSGs, magnetic type, and the symmetry-indicator classifications are given below. For each MSG, the detailed information is given in the corresponding MSG table. \hyperref[tabindex]{See the list of tables.}
\input{src_tex/sg184}
\subsection{SG 185}  \label{sup:sg185}
TABLE SG185. The crystalline space group is what the crystal has if the magnetic order is neglected. Once condidering magnetic order, the MSGs, magnetic type, and the symmetry-indicator classifications are given below. For each MSG, the detailed information is given in the corresponding MSG table. \hyperref[tabindex]{See the list of tables.}
\input{src_tex/sg185}
\subsection{SG 186}  \label{sup:sg186}
TABLE SG186. The crystalline space group is what the crystal has if the magnetic order is neglected. Once condidering magnetic order, the MSGs, magnetic type, and the symmetry-indicator classifications are given below. For each MSG, the detailed information is given in the corresponding MSG table. \hyperref[tabindex]{See the list of tables.}
\input{src_tex/sg186}
\subsection{SG 187}  \label{sup:sg187}
TABLE SG187. The crystalline space group is what the crystal has if the magnetic order is neglected. Once condidering magnetic order, the MSGs, magnetic type, and the symmetry-indicator classifications are given below. For each MSG, the detailed information is given in the corresponding MSG table. \hyperref[tabindex]{See the list of tables.}
\input{src_tex/sg187}
\subsection{SG 188}  \label{sup:sg188}
TABLE SG188. The crystalline space group is what the crystal has if the magnetic order is neglected. Once condidering magnetic order, the MSGs, magnetic type, and the symmetry-indicator classifications are given below. For each MSG, the detailed information is given in the corresponding MSG table. \hyperref[tabindex]{See the list of tables.}
\input{src_tex/sg188}
\subsection{SG 189}  \label{sup:sg189}
TABLE SG189. The crystalline space group is what the crystal has if the magnetic order is neglected. Once condidering magnetic order, the MSGs, magnetic type, and the symmetry-indicator classifications are given below. For each MSG, the detailed information is given in the corresponding MSG table. \hyperref[tabindex]{See the list of tables.}
\input{src_tex/sg189}
\subsection{SG 190}  \label{sup:sg190}
TABLE SG190. The crystalline space group is what the crystal has if the magnetic order is neglected. Once condidering magnetic order, the MSGs, magnetic type, and the symmetry-indicator classifications are given below. For each MSG, the detailed information is given in the corresponding MSG table. \hyperref[tabindex]{See the list of tables.}
\input{src_tex/sg190}
\subsection{SG 191}  \label{sup:sg191}
TABLE SG191. The crystalline space group is what the crystal has if the magnetic order is neglected. Once condidering magnetic order, the MSGs, magnetic type, and the symmetry-indicator classifications are given below. For each MSG, the detailed information is given in the corresponding MSG table. \hyperref[tabindex]{See the list of tables.}
\input{src_tex/sg191}
\subsection{SG 192}  \label{sup:sg192}
TABLE SG192. The crystalline space group is what the crystal has if the magnetic order is neglected. Once condidering magnetic order, the MSGs, magnetic type, and the symmetry-indicator classifications are given below. For each MSG, the detailed information is given in the corresponding MSG table. \hyperref[tabindex]{See the list of tables.}
\input{src_tex/sg192}
\subsection{SG 193}  \label{sup:sg193}
TABLE SG193. The crystalline space group is what the crystal has if the magnetic order is neglected. Once condidering magnetic order, the MSGs, magnetic type, and the symmetry-indicator classifications are given below. For each MSG, the detailed information is given in the corresponding MSG table. \hyperref[tabindex]{See the list of tables.}
\input{src_tex/sg193}
\subsection{SG 194}  \label{sup:sg194}
TABLE SG194. The crystalline space group is what the crystal has if the magnetic order is neglected. Once condidering magnetic order, the MSGs, magnetic type, and the symmetry-indicator classifications are given below. For each MSG, the detailed information is given in the corresponding MSG table. \hyperref[tabindex]{See the list of tables.}
\input{src_tex/sg194}
\subsection{SG 195}  \label{sup:sg195}
TABLE SG195. The crystalline space group is what the crystal has if the magnetic order is neglected. Once condidering magnetic order, the MSGs, magnetic type, and the symmetry-indicator classifications are given below. For each MSG, the detailed information is given in the corresponding MSG table. \hyperref[tabindex]{See the list of tables.}
\input{src_tex/sg195}
\subsection{SG 196}  \label{sup:sg196}
TABLE SG196. The crystalline space group is what the crystal has if the magnetic order is neglected. Once condidering magnetic order, the MSGs, magnetic type, and the symmetry-indicator classifications are given below. For each MSG, the detailed information is given in the corresponding MSG table. \hyperref[tabindex]{See the list of tables.}
\input{src_tex/sg196}
\subsection{SG 197}  \label{sup:sg197}
TABLE SG197. The crystalline space group is what the crystal has if the magnetic order is neglected. Once condidering magnetic order, the MSGs, magnetic type, and the symmetry-indicator classifications are given below. For each MSG, the detailed information is given in the corresponding MSG table. \hyperref[tabindex]{See the list of tables.}
\input{src_tex/sg197}
\subsection{SG 198}  \label{sup:sg198}
TABLE SG198. The crystalline space group is what the crystal has if the magnetic order is neglected. Once condidering magnetic order, the MSGs, magnetic type, and the symmetry-indicator classifications are given below. For each MSG, the detailed information is given in the corresponding MSG table. \hyperref[tabindex]{See the list of tables.}
\input{src_tex/sg198}
\subsection{SG 199}  \label{sup:sg199}
TABLE SG199. The crystalline space group is what the crystal has if the magnetic order is neglected. Once condidering magnetic order, the MSGs, magnetic type, and the symmetry-indicator classifications are given below. For each MSG, the detailed information is given in the corresponding MSG table. \hyperref[tabindex]{See the list of tables.}
\input{src_tex/sg199}
\subsection{SG 200}  \label{sup:sg200}
TABLE SG200. The crystalline space group is what the crystal has if the magnetic order is neglected. Once condidering magnetic order, the MSGs, magnetic type, and the symmetry-indicator classifications are given below. For each MSG, the detailed information is given in the corresponding MSG table. \hyperref[tabindex]{See the list of tables.}
\input{src_tex/sg200}
\subsection{SG 201}  \label{sup:sg201}
TABLE SG201. The crystalline space group is what the crystal has if the magnetic order is neglected. Once condidering magnetic order, the MSGs, magnetic type, and the symmetry-indicator classifications are given below. For each MSG, the detailed information is given in the corresponding MSG table. \hyperref[tabindex]{See the list of tables.}
\input{src_tex/sg201}
\subsection{SG 202}  \label{sup:sg202}
TABLE SG202. The crystalline space group is what the crystal has if the magnetic order is neglected. Once condidering magnetic order, the MSGs, magnetic type, and the symmetry-indicator classifications are given below. For each MSG, the detailed information is given in the corresponding MSG table. \hyperref[tabindex]{See the list of tables.}
\input{src_tex/sg202}
\subsection{SG 203}  \label{sup:sg203}
TABLE SG203. The crystalline space group is what the crystal has if the magnetic order is neglected. Once condidering magnetic order, the MSGs, magnetic type, and the symmetry-indicator classifications are given below. For each MSG, the detailed information is given in the corresponding MSG table. \hyperref[tabindex]{See the list of tables.}
\input{src_tex/sg203}
\subsection{SG 204}  \label{sup:sg204}
TABLE SG204. The crystalline space group is what the crystal has if the magnetic order is neglected. Once condidering magnetic order, the MSGs, magnetic type, and the symmetry-indicator classifications are given below. For each MSG, the detailed information is given in the corresponding MSG table. \hyperref[tabindex]{See the list of tables.}
\input{src_tex/sg204}
\subsection{SG 205}  \label{sup:sg205}
TABLE SG205. The crystalline space group is what the crystal has if the magnetic order is neglected. Once condidering magnetic order, the MSGs, magnetic type, and the symmetry-indicator classifications are given below. For each MSG, the detailed information is given in the corresponding MSG table. \hyperref[tabindex]{See the list of tables.}
\input{src_tex/sg205}
\subsection{SG 206}  \label{sup:sg206}
TABLE SG206. The crystalline space group is what the crystal has if the magnetic order is neglected. Once condidering magnetic order, the MSGs, magnetic type, and the symmetry-indicator classifications are given below. For each MSG, the detailed information is given in the corresponding MSG table. \hyperref[tabindex]{See the list of tables.}
\input{src_tex/sg206}
\subsection{SG 207}  \label{sup:sg207}
TABLE SG207. The crystalline space group is what the crystal has if the magnetic order is neglected. Once condidering magnetic order, the MSGs, magnetic type, and the symmetry-indicator classifications are given below. For each MSG, the detailed information is given in the corresponding MSG table. \hyperref[tabindex]{See the list of tables.}
\input{src_tex/sg207}
\subsection{SG 208}  \label{sup:sg208}
TABLE SG208. The crystalline space group is what the crystal has if the magnetic order is neglected. Once condidering magnetic order, the MSGs, magnetic type, and the symmetry-indicator classifications are given below. For each MSG, the detailed information is given in the corresponding MSG table. \hyperref[tabindex]{See the list of tables.}
\input{src_tex/sg208}
\subsection{SG 209}  \label{sup:sg209}
TABLE SG209. The crystalline space group is what the crystal has if the magnetic order is neglected. Once condidering magnetic order, the MSGs, magnetic type, and the symmetry-indicator classifications are given below. For each MSG, the detailed information is given in the corresponding MSG table. \hyperref[tabindex]{See the list of tables.}
\input{src_tex/sg209}
\subsection{SG 210}  \label{sup:sg210}
TABLE SG210. The crystalline space group is what the crystal has if the magnetic order is neglected. Once condidering magnetic order, the MSGs, magnetic type, and the symmetry-indicator classifications are given below. For each MSG, the detailed information is given in the corresponding MSG table. \hyperref[tabindex]{See the list of tables.}
\input{src_tex/sg210}
\subsection{SG 211}  \label{sup:sg211}
TABLE SG211. The crystalline space group is what the crystal has if the magnetic order is neglected. Once condidering magnetic order, the MSGs, magnetic type, and the symmetry-indicator classifications are given below. For each MSG, the detailed information is given in the corresponding MSG table. \hyperref[tabindex]{See the list of tables.}
\input{src_tex/sg211}
\subsection{SG 212}  \label{sup:sg212}
TABLE SG212. The crystalline space group is what the crystal has if the magnetic order is neglected. Once condidering magnetic order, the MSGs, magnetic type, and the symmetry-indicator classifications are given below. For each MSG, the detailed information is given in the corresponding MSG table. \hyperref[tabindex]{See the list of tables.}
\input{src_tex/sg212}
\subsection{SG 213}  \label{sup:sg213}
TABLE SG213. The crystalline space group is what the crystal has if the magnetic order is neglected. Once condidering magnetic order, the MSGs, magnetic type, and the symmetry-indicator classifications are given below. For each MSG, the detailed information is given in the corresponding MSG table. \hyperref[tabindex]{See the list of tables.}
\input{src_tex/sg213}
\subsection{SG 214}  \label{sup:sg214}
TABLE SG214. The crystalline space group is what the crystal has if the magnetic order is neglected. Once condidering magnetic order, the MSGs, magnetic type, and the symmetry-indicator classifications are given below. For each MSG, the detailed information is given in the corresponding MSG table. \hyperref[tabindex]{See the list of tables.}
\input{src_tex/sg214}
\subsection{SG 215}  \label{sup:sg215}
TABLE SG215. The crystalline space group is what the crystal has if the magnetic order is neglected. Once condidering magnetic order, the MSGs, magnetic type, and the symmetry-indicator classifications are given below. For each MSG, the detailed information is given in the corresponding MSG table. \hyperref[tabindex]{See the list of tables.}
\input{src_tex/sg215}
\subsection{SG 216}  \label{sup:sg216}
TABLE SG216. The crystalline space group is what the crystal has if the magnetic order is neglected. Once condidering magnetic order, the MSGs, magnetic type, and the symmetry-indicator classifications are given below. For each MSG, the detailed information is given in the corresponding MSG table. \hyperref[tabindex]{See the list of tables.}
\input{src_tex/sg216}
\subsection{SG 217}  \label{sup:sg217}
TABLE SG217. The crystalline space group is what the crystal has if the magnetic order is neglected. Once condidering magnetic order, the MSGs, magnetic type, and the symmetry-indicator classifications are given below. For each MSG, the detailed information is given in the corresponding MSG table. \hyperref[tabindex]{See the list of tables.}
\input{src_tex/sg217}
\subsection{SG 218}  \label{sup:sg218}
TABLE SG218. The crystalline space group is what the crystal has if the magnetic order is neglected. Once condidering magnetic order, the MSGs, magnetic type, and the symmetry-indicator classifications are given below. For each MSG, the detailed information is given in the corresponding MSG table. \hyperref[tabindex]{See the list of tables.}
\input{src_tex/sg218}
\subsection{SG 219}  \label{sup:sg219}
TABLE SG219. The crystalline space group is what the crystal has if the magnetic order is neglected. Once condidering magnetic order, the MSGs, magnetic type, and the symmetry-indicator classifications are given below. For each MSG, the detailed information is given in the corresponding MSG table. \hyperref[tabindex]{See the list of tables.}
\input{src_tex/sg219}
\subsection{SG 220}  \label{sup:sg220}
TABLE SG220. The crystalline space group is what the crystal has if the magnetic order is neglected. Once condidering magnetic order, the MSGs, magnetic type, and the symmetry-indicator classifications are given below. For each MSG, the detailed information is given in the corresponding MSG table. \hyperref[tabindex]{See the list of tables.}
\input{src_tex/sg220}
\subsection{SG 221}  \label{sup:sg221}
TABLE SG221. The crystalline space group is what the crystal has if the magnetic order is neglected. Once condidering magnetic order, the MSGs, magnetic type, and the symmetry-indicator classifications are given below. For each MSG, the detailed information is given in the corresponding MSG table. \hyperref[tabindex]{See the list of tables.}
\input{src_tex/sg221}
\subsection{SG 222}  \label{sup:sg222}
TABLE SG222. The crystalline space group is what the crystal has if the magnetic order is neglected. Once condidering magnetic order, the MSGs, magnetic type, and the symmetry-indicator classifications are given below. For each MSG, the detailed information is given in the corresponding MSG table. \hyperref[tabindex]{See the list of tables.}
\input{src_tex/sg222}
\subsection{SG 223}  \label{sup:sg223}
TABLE SG223. The crystalline space group is what the crystal has if the magnetic order is neglected. Once condidering magnetic order, the MSGs, magnetic type, and the symmetry-indicator classifications are given below. For each MSG, the detailed information is given in the corresponding MSG table. \hyperref[tabindex]{See the list of tables.}
\input{src_tex/sg223}
\subsection{SG 224}  \label{sup:sg224}
TABLE SG224. The crystalline space group is what the crystal has if the magnetic order is neglected. Once condidering magnetic order, the MSGs, magnetic type, and the symmetry-indicator classifications are given below. For each MSG, the detailed information is given in the corresponding MSG table. \hyperref[tabindex]{See the list of tables.}
\input{src_tex/sg224}
\subsection{SG 225}  \label{sup:sg225}
TABLE SG225. The crystalline space group is what the crystal has if the magnetic order is neglected. Once condidering magnetic order, the MSGs, magnetic type, and the symmetry-indicator classifications are given below. For each MSG, the detailed information is given in the corresponding MSG table. \hyperref[tabindex]{See the list of tables.}
\input{src_tex/sg225}
\subsection{SG 226}  \label{sup:sg226}
TABLE SG226. The crystalline space group is what the crystal has if the magnetic order is neglected. Once condidering magnetic order, the MSGs, magnetic type, and the symmetry-indicator classifications are given below. For each MSG, the detailed information is given in the corresponding MSG table. \hyperref[tabindex]{See the list of tables.}
\input{src_tex/sg226}
\subsection{SG 227}  \label{sup:sg227}
TABLE SG227. The crystalline space group is what the crystal has if the magnetic order is neglected. Once condidering magnetic order, the MSGs, magnetic type, and the symmetry-indicator classifications are given below. For each MSG, the detailed information is given in the corresponding MSG table. \hyperref[tabindex]{See the list of tables.}
\input{src_tex/sg227}
\subsection{SG 228}  \label{sup:sg228}
TABLE SG228. The crystalline space group is what the crystal has if the magnetic order is neglected. Once condidering magnetic order, the MSGs, magnetic type, and the symmetry-indicator classifications are given below. For each MSG, the detailed information is given in the corresponding MSG table. \hyperref[tabindex]{See the list of tables.}
\input{src_tex/sg228}
\subsection{SG 229}  \label{sup:sg229}
TABLE SG229. The crystalline space group is what the crystal has if the magnetic order is neglected. Once condidering magnetic order, the MSGs, magnetic type, and the symmetry-indicator classifications are given below. For each MSG, the detailed information is given in the corresponding MSG table. \hyperref[tabindex]{See the list of tables.}
\input{src_tex/sg229}
\subsection{SG 230}  \label{sup:sg230}
TABLE SG230. The crystalline space group is what the crystal has if the magnetic order is neglected. Once condidering magnetic order, the MSGs, magnetic type, and the symmetry-indicator classifications are given below. For each MSG, the detailed information is given in the corresponding MSG table. \hyperref[tabindex]{See the list of tables.}
\input{src_tex/sg230}

 \clearpage

 \end{widetext}

\end{document}